\documentclass[traditabstract]{aa} %
\usepackage{graphicx}

\usepackage{txfonts}
\usepackage{natbib}

\newcommand{\hMpc}{{\rm h^{-1}Mpc}}
\newcommand{\Lya}{Ly$\alpha$~}
\newcommand{\Lyb}{Ly$\beta$~}
\newcommand{\om}{\Omega_M}
\newcommand{\ol}{\Omega_\Lambda}
\newcommand{\da}{D_A}
\newcommand{\dv}{D_V}
\newcommand{\dl}{D_L}

\begin{document}

\title{Baryon Acoustic Oscillations in the \Lya forest of BOSS quasars}

   \author{
Nicol\'as G. Busca\inst{1}, 
Timoth\'ee Delubac\inst{2}, 
James Rich\inst{2},
Stephen~Bailey\inst{3},
Andreu Font-Ribera\inst{3,27},
David~Kirkby\inst{4},
J.-M. Le Goff\inst{2}, 
Matthew M. Pieri\inst{5},
An\v{z}e Slosar\inst{6},
\'{E}ric Aubourg\inst{1},
Julian E. Bautista\inst{1},
Dmitry Bizyaev\inst{7}, 
Michael~Blomqvist\inst{4},
Adam S. Bolton\inst{8},  
Jo Bovy\inst{9},
Howard Brewington\inst{7}, 
Arnaud Borde\inst{2},
J. Brinkmann\inst{7},
Bill Carithers\inst{3},
Rupert A.C. Croft\inst{10},
Kyle S. Dawson\inst{8},
Garrett Ebelke\inst{7}, 
Daniel J. Eisenstein\inst{11},
Jean-Christophe Hamilton\inst{1},
Shirley Ho\inst{10},
David W. Hogg\inst{12},
Klaus Honscheid\inst{13},
Khee-Gan Lee\inst{14},
Britt Lundgren\inst{15},
Elena  Malanushenko\inst{7},  
Viktor Malanushenko\inst{7},
Daniel~Margala\inst{4},
Claudia~Maraston\inst{5},
Kushal Mehta\inst{16},
Jordi Miralda-Escud\'{e}\inst{17,18},
Adam D. Myers\inst{19},
Robert C. Nichol\inst{5},
Pasquier Noterdaeme\inst{20},
Matthew D Olmstead\inst{8},  
Daniel Oravetz\inst{7}, 
Nathalie Palanque-Delabrouille\inst{2}, 
Kaike Pan\inst{7}, 
Isabelle P\^aris\inst{20,28},
Will J. Percival\inst{5},
Patrick Petitjean\inst{20},
N. A. Roe\inst{3}, 
Emmanuel Rollinde\inst{20},
Nicholas P. Ross\inst{3},
Graziano Rossi\inst{2},
David J. Schlegel\inst{3},
Donald P. Schneider\inst{21,22},
Alaina Shelden\inst{7},   
Erin S. Sheldon\inst{6},
Audrey Simmons\inst{7},   
Stephanie Snedden\inst{6}, 
Jeremy L. Tinker\inst{12},
Matteo Viel\inst{23,24},
Benjamin A. Weaver\inst{12},
David H. Weinberg\inst{25},
Martin White\inst{3},
Christophe Y\`eche\inst{2},
Donald G. York\inst{26}
                          }
\institute{APC, Universit\'{e} Paris Diderot-Paris 7, CNRS/IN2P3, CEA,
       Observatoire de Paris, 10, rueA. Domon \& L. Duquet,  Paris, France
   \and
CEA, Centre de Saclay, IRFU,  F-91191 Gif-sur-Yvette, France
   \and
Lawrence Berkeley National Laboratory, 1 Cyclotron Road, Berkeley, CA 94720, USA
   \and
Department of Physics and Astronomy, University of California, Irvine, CA 92697, USA
\and
Institute of Cosmology and Gravitation, Dennis Sciama Building, University of Portsmouth, Portsmouth, PO1 3FX, UK
  \and
Bldg 510 Brookhaven National Laboratory, Upton, NY 11973, USA
  \and
Apache Point Observatory, P.O. Box 59, Sunspot, NM 88349, USA
\and
Department of Physics and Astronomy, University of Utah, 115 S 1400 E, Salt Lake City, UT 84112, USA
\and
Institute for Advanced Study, Einstein Drive, Princeton, NJ 08540, USA
\and
Bruce and Astrid McWilliams Center for Cosmology, Carnegie Mellon University, Pittsburgh, PA 15213, USA
\and
Harvard-Smithsonian Center for Astrophysics, Harvard University, 60 Garden St., Cambridge MA 02138, USA
\and
Center for Cosmology and Particle Physics, New York University, New York, NY 10003, USA
\and
Department of Physics and Center for Cosmology and Astro-Particle Physics, Ohio State University, Columbus, OH 43210, USA
\and
Max-Planck-Institut f\"ur Astronomie, K\"onigstuhl 17, D69117 Heidelberg, Germany
\and
Department of Astronomy, University of Wisconsin, 475 North Charter Street, Madison, WI 53706, USA
\and
Steward Observatory, University of Arizona, 933 N. Cherry Ave., Tucson, AZ 85121, USA
\and
Instituci\'{o} Catalana de Recerca i Estudis  Avan\c{c}ats, Barcelona, Catalonia
\and
Institut de Ci\`{e}ncies del Cosmos, Universitat de Barcelona/IEEC, Barcelona 08028, Catalonia
\and
Department of Physics and Astronomy, University of Wyoming, Laramie, WY 82071, USA
\and
Universit\'e Paris 6 et CNRS, Institut d'Astrophysique de Paris, 98bis blvd. Arago, 75014 Paris, France
\and
Department of Astronomy and Astrophysics, The Pennsylvania State University, University Park, PA 16802, USA
\and
Institute for Gravitation and the Cosmos, The Pennsylvania State University, University Park, PA 16802, USA
\and
INAF, Osservatorio Astronomico di Trieste, Via G. B. Tiepolo 11, 34131 Trieste, Italy
\and
INFN/National Institute for Nuclear Physics, Via Valerio 2, I-34127 Trieste, Italy.
\and
Department of Astronomy, Ohio State University, 140 West 18th Avenue, Columbus, OH 43210, USA
\and
Department of Astronomy and Astrophysics and the Enrico Fermi Institute, The University of Chicago, 5640 South Ellis Avenue, Chicago, Illinois, 60615, USA
\and
Institute of Theoretical Physics, University of Zurich, 8057 Zurich, Switzerland 
\and
Departamento de Astronom\'ia, Universidad de Chile, Casilla 36-D, Santiago, Chile
}

   \date{Received November 13th, 2012; accepted February 4th, 2013}

\authorrunning{N.G. Busca et al.}
\titlerunning{BAO in the \Lya forest of BOSS quasars}

  \abstract{
We report a detection of the baryon acoustic oscillation (BAO)
feature in the three-dimensional correlation function of the transmitted flux fraction in the \Lya forest of high-redshift quasars.
The study uses 48,640 quasars in the redshift range $2.1\le z \le 3.5$
from the
Baryon  Oscillation Spectroscopic Survey (BOSS) of the
third generation of the Sloan Digital Sky Survey (SDSS-III).
At a mean redshift $z=2.3$,
we measure the monopole and quadrupole components of the correlation
function for separations in the range $20\hMpc<r<200\hMpc$.
A peak in the correlation function is seen 
at a separation equal to $(1.01\pm0.03)$ times 
the distance expected for the BAO peak within a concordance $\Lambda$CDM
cosmology. This first detection of the BAO peak at high redshift,
when the universe was strongly matter dominated, results in 
constraints on
the angular diameter distance $\da$ and the expansion
rate $H$ at $z=2.3$
that, combined with priors on $H_0$ and the baryon density, 
require the existence of dark energy.
Combined with constraints derived from 
Cosmic Microwave Background (CMB) observations,
this result implies
$H(z=2.3)=(224\pm8){\rm km\,s^{-1}Mpc^{-1}}$,
indicating that the time derivative of the cosmological scale parameter
$\dot{a}=H(z=2.3)/(1+z)$ is significantly
greater than that measured with BAO at $z\sim0.5$.
This demonstrates that the expansion was decelerating in the range $0.7<z<2.3$, as expected from the matter domination during this epoch.
Combined with measurements of $H_0$, one sees the pattern
of deceleration followed by acceleration characteristic of a
dark-energy dominated universe. 
}


   \maketitle

\section{Introduction}
\label{introsec}

Baryon acoustic oscillations (BAO) in the pre-recombination
universe have striking effects on the anisotropies of the Cosmic Microwave Background (CMB)
and on the large scale structure (LSS) of matter (\citet{baotheory} and references therein). 
The BAO effects were first seen in the  series of peaks in the CMB
angular power spectrum \citep{boomerang}.
Subsequently, the BAO relic at redshift $z\sim0.3$ was seen \citep{sdss1bao,2dfbao} as a peak in the galaxy-galaxy correlation function at a co-moving distance
corresponding to the sound horizon at recombination.
For the WMAP7 cosmological parameters \citep{wmap7ref},
the expected comoving scale of the BAO peak is
$r_s=153~{\rm Mpc}$, with an uncertainty of $\approx 1\%$.

The BAO peak in the correlation function at a redshift $z$ appears at
an angular separation $\Delta\theta=r_s/(1+z)\da(z)$ 
and at a redshift separation
$\Delta z=r_sH(z)/c$, where $\da$ and $H$ are the angular distances and
expansion rates. Measurement of the peak position at any redshift thus
constrains the combinations of  cosmological parameters that
determine $r_sH$ and $r_s/\da$.
While the possibility of measuring both combinations is beginning to be exploited \citep{chuang2012,xu2012}, most present measurements have concentrated on the combination $\dv\equiv[(1+z)^2\da^2cz/H]^{1/3}$, which
determines the peak position
for an isotropic distribution of galaxy pairs and an isotropic clustering strength.
The ``BAO Hubble
diagram'', $\dv/r_s\,vs.\,z$,
now includes the Sloan Digital Sky Survey (SDSS) measurement
\citep{sdss1bao} updated to the DR7 \citep{abazajian09} sample
and combined with
2dF data \citep{percival2010}, the 6dF point at $z=0.1$ \citep{6df},
the WiggleZ points at $(0.4<z<0.8)$ \citep{wigglez},
and a reanalysis of the SDSS DR7 sample that uses
reconstruction \citep{eisenstein07,padmanabhan09} to sharpen
the precision of the BAO measurement \citep{padmanabhan12,mehtaref}.
Recently, the Baryon Oscillation Spectroscopic Survey (BOSS;
\citealt{bossoverview}) of SDSS-III
\citep{Eisenstein:2011sa}
has added a precise measurement at $z\sim 0.57$ \citep{bossgal}.
BOSS has also reported a measurement of $\da(z=0.55)/r_s$ based on
the BAO structure in the angular power spectrum of galaxies \citep{seo}.

In this paper, we present an observation of the BAO peak at $z\sim2.3$
found in the flux correlation function of the \Lya forest of BOSS
quasars.
This is the first such observation at a redshift 
where the expansion dynamics is matter-dominated, 
$z>0.8$. The possibility of such a measurement was suggested 
by \citet{macdo2003} and \citet{white2003} and first studied 
in detail  by \citet{McDonald:2006qs}.
While the galaxy BAO measurements are most sensitive
to $\dv\propto \da^{2/3}H^{-1/3}$, the \Lya flux transmission 
is more sensitive to peculiar velocity gradient
effects, which enhance redshift
distortions  and shift our sensitivity to the expansion rate.
As we shall show below, 
the most accurately measured combination from the
\Lya forest BAO peak is $\propto \da^{0.2}H^{-0.8}$,
and the present BOSS data set allows us to determine its value to
a precision of 3.5\%.
Combining this result  with constraints 
from CMB observations
 allows  us to deduce the value  of $H(z=2.3)$ accurate to 4\%. 
Comparing our
results with measurements of $H_0$ and of $H(0.2<z<0.8)$  reveals the
expected sequence of deceleration and acceleration in models
with dark energy.

The last decade has seen increasing use of \Lya absorption
to investigate large scale structure.
The number of quasars in early studies 
\citep{croft99,macdo2000,croft02,viel04,macdo2006} 
was enough only to determine the \Lya absorption correlation
along individual lines of sight.
With the BOSS project
the surface density of quasars is sufficient to probe the full three-dimensional
distribution of neutral hydrogen.
A study using the first 10,000 BOSS quasars was presented by  \citet{Slosar}.
This sample provided clear evidence for the expected long-range
correlations, including the redshift-space distortions
due to the gravitational growth of structure.
With the SDSS data release DR9 \citep{dr9ref},
we now have $\sim60,000$ quasars at $z\sim2.3$ \citep{dr9qso}, 
with a high enough surface density to observe the BAO peak.

The use of \Lya absorption to trace matter has certain
interesting differences from the use of galaxies.
Galaxy surveys provide a catalog of positions in redshift space
that correspond to points of high over-densities.
On the other hand, the forest region of a quasar spectrum provides
a complete mapping of the absorption over a $\sim400\,\hMpc$ (comoving) range
starting about $100\,\hMpc$ in front of the quasar (so as
to avoid the necessity of modeling the quasar's \Lya emission line). 
To the extent that quasar lines of sight are random,
a large collection of quasars can  provide a nearly unbiased
sample of points where the absorption is measured.
Cosmological simulations
\citep{cen94,petitjean95,yu95,hernquist96,miralda96,theuns98}
indicate that most of the \Lya absorption is due to cosmic filamentary
structures with overdensities of order one to ten, much lower than
the overdensities of virialized halos sampled by galaxies.
These simulations have also indicated that, on large scales, the
mean \Lya absorption is a linear tracer of the mass overdensity
\citep{croft97,croft98,weinberg98,macdo2000,macdo2003}, implying
a relation of the power spectrum of the measured absorption to that
of the underlying mass fluctuations.
Finally, the forest is observable in a redshift range
inaccessible to current large galaxy surveys and where theoretical
modeling is less dependent on non-linear effects in cosmological
structure formation.
These factors combine to make \Lya absorption a promising tracer of mass
that is complementary to galaxy tracers.

With \Lya forest measurements along multiple sightlines, one can
attempt to reconstruct the underlying 3-dimensional mass density
field \citep{nusser99,pichon01,gallerani11}, from which one can investigate
topological characteristics \citep{caucci08} or the power spectrum
\citep{kitaura12}.  However, the BOSS sample is fairly sparse, with
a typical transverse sightline separation $\sim 15~\hMpc$ (comoving),
and the signal-to-noise ratio in individual spectra is low
(see Fig. \ref{zdistfig} below), 
which makes it poorly suited to such reconstruction
techniques.  In this paper we take the more direct approach of 
measuring the
BAO feature in the correlation function of transmitted flux fraction (the ratio
of the observed flux to the flux expected in the
absence of absorption).

The use of the \Lya forest is handicapped
by the fact that not all fluctuations of the transmitted flux fraction are
due to fluctuations of the  density of hydrogen.
Because the neutral hydrogen density is believed to be determined
by photo-ionization equilibrium with the flux of UV photons from stars and quasars, variations
of the UV flux may contribute to fluctuations of the transmitted flux \citep{fluxflucref}.
In addition, hydrogen can be self-shielded to ionizing photons,
leading to much higher neutral fractions and damping wings in strong absorption systems \citep{dlasmocks}.
Metals present in the intergalactic
medium provide additional absorption superimposed on the \Lya absorption \citep{metalref}.
A further complication results from the fact that the flux correlation function uses the fluctuations
of the transmitted flux about its mean value;
this requires an estimate of the product of the mean absorption 
(as a function of absorber redshift) and
the unabsorbed flux for individual quasars \citep{legoff}.

All of these complications are  important for a complete understanding
of the statistics of the \Lya forest.
Fortunately,  they are not expected
to produce a sharp peak-like feature in the correlation function,
so  the interpretation of a peak as due to BAO  should
give robust constraints on the cosmological parameters.

The BOSS collaboration has performed three independent analyses to search for BAO in \Lya forest.
This paper presents two of them,
both of which  aim to analyze the forest using  simple
procedures at the expense of some loss of sensitivity.
The third analysis, with the goal of an optimal measurement
of the flux correlation function with a more complex method,
is described in a separate publication \citep{method3}.

Our methods are tested extensively on a set of detailed
mock catalogs of the BOSS \Lya forest data set.
These mock catalogs, which use the method presented by
\cite{mockref}, will be described in detail in a
forthcoming public release paper (S.\ Bailey et al., in preparation).
In addition, the BOSS collaboration have also released 
a fiducial version of 
the DR9 \Lya forest spectra \citep{lee13}, 
with various per-object products including masks, continua, and 
noise correction vectors designed to aid in \Lya forest analysis.
While our analyses implement their own sample selection criteria 
and continuum determination procedures, we have
also applied our measurement to this \citet{lee13} sample.

This paper is organized as follows.
Section \ref{samplesec} presents the BOSS quasar sample
used in this analysis and the procedure used to produce
the quasar spectra.
Section \ref{xisec}
describes the analysis to measure the correlation
function. 
Section \ref{fitssec}  derives the monopole and quadrupole components of the correlation function
and determines the significance of the
peak observed in these functions at the  BAO scale.
The cosmological implications of our detection of a BAO peak are discussed in Section \ref{cosmosec}.
Finally, Appendix~\ref{mocksec} provides a brief description
of the mock spectra used to test our methodology and
Appendix~\ref{kgsample} shows the result of our BAO
measurement applied to the BOSS \Lya sample of \citet{lee13}.

\section{The BOSS quasar sample and data reduction}
\label{samplesec}

The BOSS project
\citep{bossoverview}
of SDSS-III \citep{Eisenstein:2011sa}
is obtaining the spectra of
$\sim1.6\times10^6$ luminous galaxies  and  $\sim150,000$ quasars.
The project uses upgraded versions of the
SDSS spectrographs
\citep{bossspectrometer}
mounted on
the Sloan 2.5-meter telescope \citep{gunn06} at Apache Point, New Mexico.
BOSS galaxy and quasar spectroscopic targets are selected using
algorithms based primarily on photometry from the SDSS camera
\citep{gunn98,York:2000gk} in the
$ugriz$ bands \citep{fukugita96,smith02} reduced and calibrated as
described by \cite{stoughton02}, \cite{pier03}, and \cite{padmanabhan08}.
Targets are assigned to fibers
appropriately positioned in the $3^{\circ}$ diameter focal plane according to a
specially designed tiling algorithm
\citep{Blanton:2001yk}.
Fibers are fixed in place by a pierced metal plate drilled 
for each observed field and fed to one of two spectrographs.
Each exposed plate generates 1,000 spectra covering
wavelengths of
360 to 1000~nm with a resolving power ranging
from 1500 to 3000 \citep{bossspectrometer}.
A median of 631 of these
fibers are assigned to galactic targets and 204 to quasar targets.
The BOSS spectroscopic targets are observed in dark and gray time,
while the bright-time is used by other SDSS-III
surveys (see \citealt{Eisenstein:2011sa}).

The quasar spectroscopy targets are
selected from photometric data
with a combination of algorithms
(\citealt{richards09,Yeche:2009yh,Kirkpatrick:2011gq,coreref,
PalanqueDelabrouille:2010wi};
for a summary, see \citet{ross2012}).
The algorithms use SDSS fluxes and, for SDSS Stripe 82, photometric
variability.
When available, we also use
data from non-optical surveys \citep{bovy2012}:
the GALEX survey \citep{galexref} in the UV;
the UKIDSS survey \citep{ukidssref} in the NIR,
and the FIRST survey \citep{firstref} in the radio.

The quasar spectroscopy targets
are divided into two samples ``CORE'' and ``BONUS''.
The CORE sample consists of 20 quasar targets per
square degree selected from  SDSS photometry 
with a uniform algorithm,
for which
the selection efficiency for $z>2.1$ quasars is $\sim50\%$. The selection
algorithm for the CORE sample
\citep{coreref}
was fixed at the end of the first year of
the survey, thus making it useful for studies that require
a uniform target selection across the sky.
The BONUS sample was chosen from a combination of algorithms with the purpose
of increasing the  density on the sky
of observed quasars beyond that of the CORE
sample.
The combined samples yield a mean density of identified quasars
of $15~{\rm deg}^{-2}$ with
a maximum of $20~{\rm deg}^{-2}$, mostly
in zones where photometric variability,
UV, and/or NIR
data are available.
The combined BONUS plus CORE sample can be used for \Lya BAO
studies, which require the highest possible quasar density in a broad sky area
but are insensitive to the uniformity of the quasar selection
criteria because the structure being mapped is in the foreground
of these quasar back-lights.

\begin{figure}
\includegraphics[width=\columnwidth]{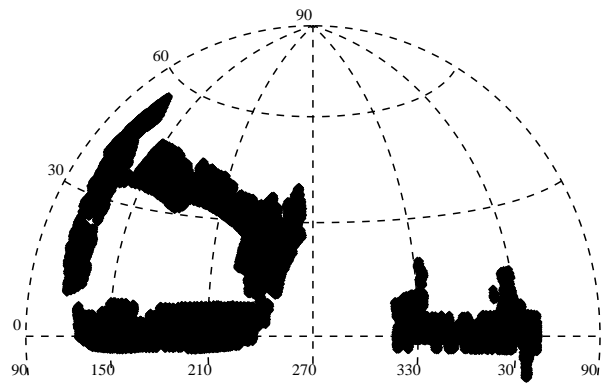}
\caption{Hammer-Aitoff projection in equatorial coordinates
of the BOSS DR9 footprint.  The observations cover $\sim 3000~{\rm deg^2}$.
}
\label{fig:skysectors}
\end{figure}

The data presented here consist of the DR9 data release \citep{dr9ref}
covering $\sim3000~{\rm deg}^2$ of the sky 
shown in figure \ref{fig:skysectors}.
These data cover about one-third of the ultimate BOSS
footprint.

The data were reduced with the SDSS-III pipeline as described in
\citet{pipeline}.
Typically four exposures of 15 minutes were co-added in 
pixels of wavelength width $\sim0.09~{\rm nm}$.
Besides providing flux calibrated spectra, the pipeline provided
preliminary object classifications (galaxy, quasar, star) and
redshift estimates.

The spectra of all quasar targets were visually inspected, as
described in \citet{dr9qso}, to correct for misidentifications or
inaccurate redshift determinations and to flag broad absorption lines (BAL).
Damped \Lya
troughs are visually flagged, but also identified and characterized
automatically \citep{noterdaeme}. 
The visual inspection of DR9 confirms
60,369 quasars with 2.1~$\leq$~$z_q$~$\leq$~3.5. 
In order to simplify the analysis of the \Lya forest, we discarded
quasars with visually identified BALs and DLAs, leaving 
48,640 quasars.

\begin{figure}[htbp]
\includegraphics[width=\columnwidth]{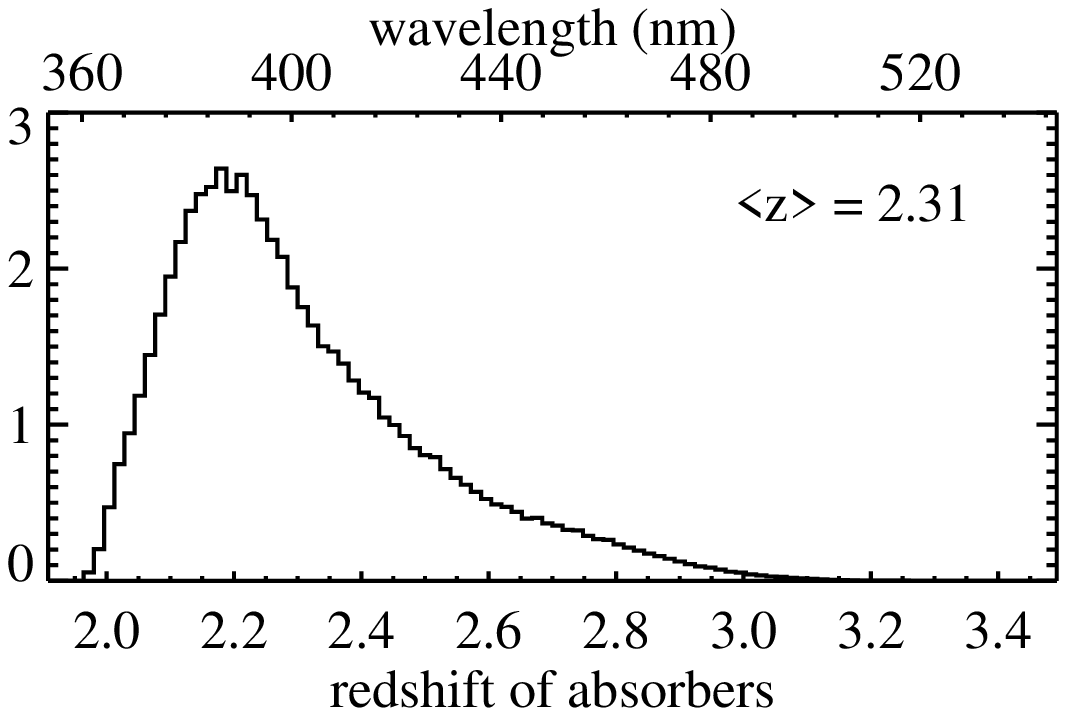}
\includegraphics[width=\columnwidth]{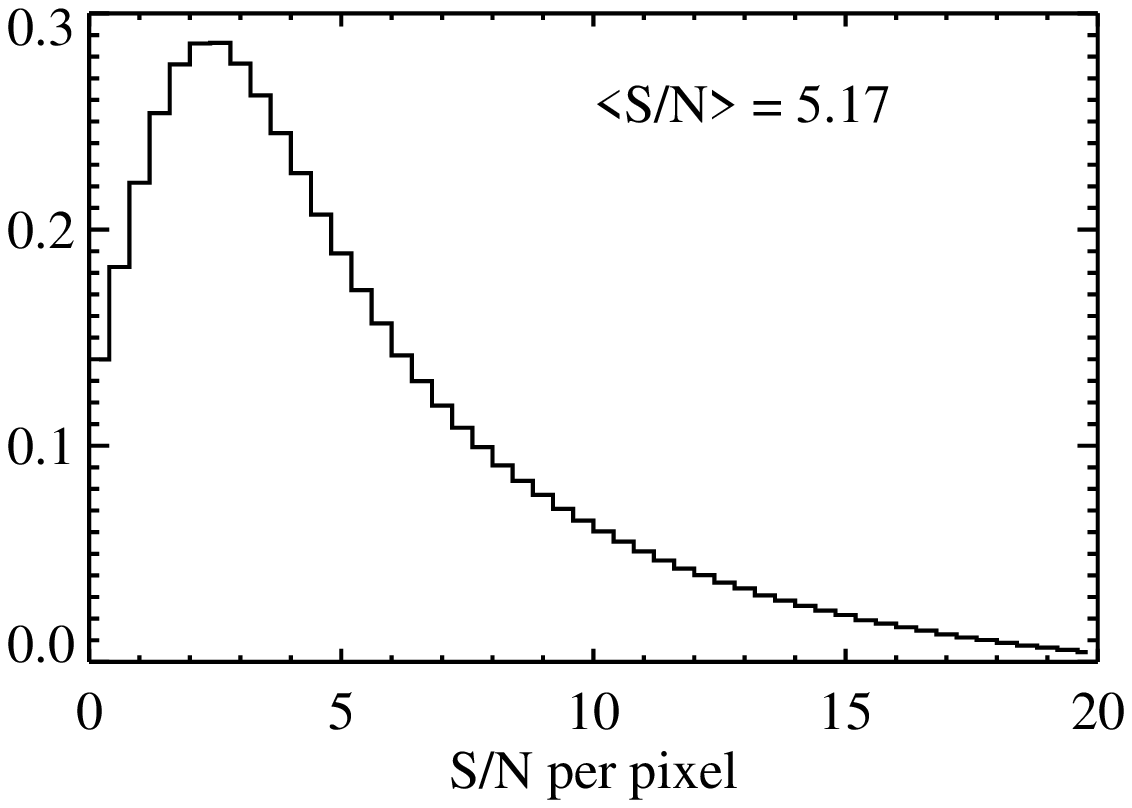}
\caption{Top:
weighted distribution of absorber redshifts used in the calculation of 
the correlation function in the distance range $80~\hMpc<r<120~\hMpc$.
Bottom: distribution of signal-to-noise ratio 
for analysis pixels
(triplets of pipeline pixels)  
averaged over the forest region.
}
\label{zdistfig}
\end{figure}

For the measurement of the flux transmission, we use the
rest-frame wavelength interval
\begin{equation}
104.5\,<\lambda_{\rm rf}<118.0~{\rm nm} \; .
\label{rflambdarange}
\end{equation}
The range is 
bracketed by the \Lya and \Lyb emission lines at 121.6 and 102.5~nm.
The limits are chosen conservatively to avoid problems of modeling
the shapes of the two emission lines and to avoid quasar proximate absorbers.
The absorber redshift, $z=\lambda/\lambda_{{\rm Ly\alpha}}-1$,
is in the range 
$1.96<z<3.38$. The lower limit
is set by the requirement that the observed wavelength be greater
than 360~nm below which the system throughput is less than 10\% its peak value. 
The upper limit comes from the maximum quasar
redshift of 3.5, beyond which the BOSS surface density of quasars
is not sufficient to be useful. 
The distribution of absorber redshift 
is shown in figure \ref{zdistfig} (top panel).  
When given the weights used for the calculation of the correlation
function (section \ref{sec:weights}), the absorbers have a mean
redshift of
$\langle z\rangle=2.31$.

For the determination of the correlation function, we
use ``analysis pixels'' 
that are the flux average over three adjacent pipeline pixels.
Throughout the rest of this paper, ``pixel'' refers to analysis
pixels unless otherwise stated.
The effective width of these pixels is $210~{\rm km~s^{-1}}$, i.e. an
observed-wavelength width $\sim0.27\,{\rm nm}\sim2~\hMpc$.
The total sample of 
48,640 quasars thus provides $\sim8\times10^6$ measurements
of \Lya absorption
over a total volume of ${\rm \sim20h^{-3}Gpc^3}$. 

Figure \ref{zdistfig} (bottom panel)  shows the distribution of the
signal-to-noise ratio for pixels 
averaged over the forest region.
The relatively modest mean value of $5.17$ reflects the
exposure times necessary to acquire such a large number of spectra.

In addition to the BOSS spectra, we analyzed 15 sets of mock
spectra that were produced by the methods described in appendix \ref{mocksec}.
These spectra do not yet reproduce all of the characteristics
of the BOSS sample, but they are nevertheless 
useful for a qualitative understanding
of the shape of the measured correlation function.
More importantly,
they are  useful for understanding the detectability
of a BAO-like peak and the precision of the measurement of
its position.

\section{Measurement of the correlation function}
\label{xisec}

The flux correlation function can be determined through
a simple two-step process.
In the first step, 
for each pixel in the forest region (equation \ref{rflambdarange}) 
of quasar $q$, 
the measured flux $f_q(\lambda)$ at observed wavelength $\lambda$
is compared with the mean expected flux, $C_q(\lambda)\overline{F}(z)$,
thus defining the ``delta field'': 
\begin{equation}
\delta_q(\lambda)={f_q(\lambda)\over C_q(\lambda)\overline{F}(z)}-1 
\; .
\label{delta:def}
\end{equation}
Here,
$C_q(\lambda)$ is the unabsorbed flux (the so-called ``continuum'') and
 $\overline{F}(z)$ is the mean transmitted fraction
 at the HI absorber redshift.
The quantities $\lambda$ and $z$ in equation \ref{delta:def} 
are not independent but related via $z=\lambda/\lambda_{{\rm Ly\alpha}}-1$.

\begin{figure}[htbp]
\begin{center}
\includegraphics[width=\columnwidth]{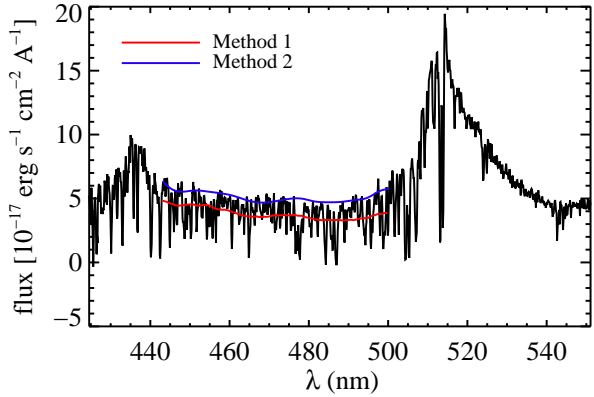}
\caption{
An example of a BOSS quasar spectrum of redshift 3.239. 
The red and blue lines cover the
forest region used here, $104.5<\lambda_{\rm rf}<118.0$.
This region is sandwiched between the quasar's \Lyb and \Lya
emission lines respectively at 435 and 515~nm.
The blue line is
an estimate of the continuum (unabsorbed flux) by method 2
and the red line is the estimate of the 
product of the continuum and the
mean absorption by method 1.
}
\label{spectrumfig}
\end{center}
\end{figure}

Figure \ref{spectrumfig} shows
an example of an estimation for $C_q(\lambda)$ (blue line) 
and $C_q\overline{F}$ (red line).
Our two methods for estimating $C_q$ and $\overline{F}$ are described in 
sections \ref{contfitsec1} and \ref{contfitsec2}. 

In the second step, the correlation function is calculated as
a weighted sum of products of the deltas:
 \begin{equation}
 \hat{\xi}_A=\sum_{ij\in A}w_{ij}\delta_i\delta_j \; /
\sum_{ij\in A}w_{ij}
\; ,
\label{eq:thexi}
 \end{equation}
where the $w_{ij}$ are weights and each $i$ or $j$ indexes
a measurement on a quasar $q$ at wavelength $\lambda$.
The sum over $(i,j)$ is understood to run over all pairs of 
pixels of all 
pairs of quasars within $A$ defining
a region in space of pixel separations, 
$\vec{r}_i-\vec{r}_j$.
The region $A$ is generally defined by a range  
$r_{\rm min}<r<r_{\rm max}$ and $\mu_{\rm min}<\mu<\mu_{\rm max}$ with:
\begin{equation}
r=|\vec{r}_i-\vec{r}_j|
\hspace*{10mm}
\mu=\frac{(\vec{r}_i-\vec{r}_j)_\parallel}{r}
\end{equation}
where $(\vec{r}_i-\vec{r}_j)_\parallel$ is the component along
the line of sight.
Separations in observational pixel coordinates (ra,dec,$z$) are transformed
to $(r,\mu)$ in units of $\hMpc$ by using a $\Lambda$CDM
fiducial cosmology with matter and vacuum densities of
\begin{equation}
(\om,\ol)=(0.27,0.73) \;.
\label{fiducialmv}
\end{equation}

In the sum (\ref{eq:thexi}), we exclude
pairs of pixels from only one quasar 
to avoid the correlated errors in $\delta_i$ and $\delta_j$ coming from
the estimate of $C_q$.
Note that the weights in eq. \ref{eq:thexi} are set to zero for 
pixels flagged by the pipeline as having problems due, e.g., 
to sky emission lines or cosmic rays.

A procedure for determining $\xi$ 
is defined by its method for estimating the expected flux 
$C_q\overline{F}$
and by its choice of weights, $w_{ij}$.
The two methods described here use the same technique to calculate
weights  but
have different approaches to estimate  $C_q\overline{F}$.
We will see that the two methods produce correlation functions
that have no significant differences.
However, the two independent codes were invaluable for
consistency checks throughout the analysis.

The two methods were ``blind'' to the extent that
many of the procedures were defined 
during tests either with mock data or with
the real data in which we masked the region 
of the peak in the correlation function.
Among those aspects fixed in 
this way were the quasar sample, the continuum determination, 
the weighting, the extraction
of the monopole and quadrupole correlation function
and the determination of the peak significance (section \ref{fitssec}).
This early freezing of procedures
resulted in some that are suboptimal
but which will be improved in future analyses.
We note, however, that
the procedures used to extract cosmological information
(section \ref{cosmosec}) were decided on only after
de-masking the data.

\subsection{Continuum fits, method 1}
\label{contfitsec1}

 Both methods for estimating the product $C_q\overline{F}$ 
 assume that 
 $C_q$  is, to first approximation, proportional to a universal
quasar spectrum that is a function of  rest-frame wavelength,
$\lambda_{\rm rf}=\lambda/(1+z_q)$  (for quasar redshift $z_q$), 
multiplied by a mean
 transmission fraction that slowly varies with absorber redshift.
Following this assumption, the universal spectrum is found by
  stacking the appropriately normalized spectra of quasars in our
  sample,
thus averaging out the fluctuating \Lya absorption.
The product  $C_q\overline{F}$ for individual quasars is then
derived from  the universal
spectrum by normalizing it to account for the quasar's mean forest flux
and then modifying its slope   to account  
for  spectral-index diversity   and/or
photo-spectroscopic miscalibration.

Method 1 estimates directly the product $C_q\overline{F}$ in equation
\ref{delta:def}.
An example is given by the red line in figure \ref{spectrumfig}.
The estimate is made
by modeling each spectrum as
\begin{equation}
C_q\overline{F}= a_q \left(
\frac{\lambda}{\langle\lambda\rangle}
\right)^{b_q} \overline{f}(\lambda_{\rm rf},z)
\label{eq:delta_model_meth1}
\end{equation}
 where $a_q$ is a normalization, $b_q$ a ``deformation parameter'', 
and $\langle\lambda\rangle$ is the mean wavelength 
in the forest
for the quasar $q$ and $\overline{f}(\lambda_\mathrm{rf},z)$ is 
the mean normalized flux  obtained by stacking spectra
in bins of width $\Delta z=0.1$: 
 \begin{equation}
 \overline{f}(\lambda_\mathrm{rf},z) 
= 
\sum_q w_q f_q(\lambda)/f_q^{128} \label{eq:univ_spectr} \,/\, \sum_q w_q
\;.
\end{equation}
Here $z$ is the redshift of the absorption line at observed wavelength 
$\lambda$ ($z = \lambda/\lambda_{{\rm Ly\alpha}}-1$), 
$f_q$ is the observed flux of quasar $q$ at wavelength $\lambda$ and 
$f_q^{128}$ is the average of the flux of quasar $q$ 
for $127.5<\lambda_{\rm rf}<128.5$~nm.
The weight
$w_q(\lambda)$ is given by 
$w_q^{-1} = 1/[\mathrm{ivar}(\lambda)\cdot(f^{128}_q)^2]+\sigma_{flux,~\mathrm{LSS}}^2$.
The quantity ivar is the pipeline estimate of the inverse 
flux variance in the 
pixel corresponding to wavelength $\lambda$.  
The quantity $\sigma_{flux,~\mathrm{LSS}}^2$ is the contribution to the variance in the flux due to the LSS. We approximate it by its value at the typical redshift of the survey, $z\sim2.3$: $\sigma_{flux,~\mathrm{LSS}}^2\sim0.035$ (section \ref{sec:weights}).

Figure \ref{calciumfig} shows
the resulting mean $\delta_i$ as a function of observed
wavelength.  
The mean fluctuates about zero with up to 2\% deviations 
with correlated features that include the H and K lines of 
singly ionized calcium 
(presumably originating from some combination of solar neighborhood, 
interstellar medium and the Milky Way halo absorption)
and features related to Balmer lines. 
These Balmer features are a by-product of imperfect masking 
of Balmer absorption lines in F-star spectroscopic standards, 
which are used to produce calibration vectors 
(in the conversion of CCD counts to flux) for DR9 quasars. 
Therefore such Balmer artifacts are constant for all fibers 
in a plate fed to one of the two spectrographs and 
so they are approximately constant for every 'half-plate'.

If unsubtracted, the artifacts in figure \ref{calciumfig}  would lead to 
spurious correlations, especially between pairs of  pixels with
separations that are purely transverse to the line of sight. 
We have made a global correction by subtracting the quantity 
$\langle\delta\rangle(\lambda)$  in figure \ref{calciumfig} (un-smoothed) 
from individual measurements of $\delta$.
This is 
justified if the variance of the artifacts from 
half-plate-to-half-plate is sufficiently small, 
as half-plate-wide deviations from our global correction could, 
in principle add spurious correlations. 

We have investigated this variance both by measuring the 
Balmer artifacts in the calibration vectors themselves and by 
studying continuum regions of all available quasars in the DR9 sample. 
Both studies yield no detection of excess variance arising from 
these artifacts, but do provide upper limits. 
The study of the calibration vectors indicate that the 
square-root of the variance is less than 20\% of the mean 
Balmer artifact deviations and the study of quasar spectra 
indicate that the square-root of the variance is less than 100\% 
of the mean Balmer artifacts (and less than 50\% of the 
mean calcium line deviations).

We then performed Monte Carlo simulations by adding a 
random sampling of our measured artifacts to our data to 
confirm that our global correction is adequate. 
We found that there is no significant effect on the 
determination of the BAO peak position, 
even if the variations are as large as that allowed in our tests.

\subsection{Continuum fits, method 2}
\label{contfitsec2}

Method 1 would be especially appropriate if the fluxes had a Gaussian
distribution  about the mean absorbed flux, $C_q\overline{F}$.
Since this is not the case, we have developed method 2 
which explicitly uses
the probability distribution function for the transmitted flux
fraction  $F$, $P(F,z)$, where
$0<F<1$.
We use the $P(F,z)$ that results from the log-normal model 
used to generate mock data (see appendix \ref{mocksec}).

Using $P(F,z)$, we can construct for each BOSS quasar
the PDF of the flux in pixel $i$, $f_i$, by assuming a continuum 
$C_q(\lambda_i)$ and  convolving with the pixel noise, $\sigma_i$:
\begin{equation}
P_i(f_i,C_q(\lambda_i),z_i) \propto \int_0^1 dF P(F,z_i) 
\exp \left[\frac{-(C_qF-f_i)^2}{2\sigma_i^2}\right] \;.
\end{equation}
The continuum  is assumed to be of the form
\begin{equation}
C_q(\lambda)=(a_q + b_q\lambda )\overline{f}(\lambda_{\rm rf})
\end{equation}
where $\overline{f}(\lambda_{\rm rf})$ is the mean flux as determined
by stacking spectra as follows:
\begin{equation}
\overline{f}(\lambda_{\rm rf}) = \sum_q w_q(\lambda_\mathrm{rf})
\left[ f_q(\lambda_\mathrm{rf})/f^{128}_q\right]\,/\, \sum_q w_q
\end{equation}
as in 
equation \ref{eq:univ_spectr}
except that here there is no redshift binning. 
The parameters $a_q$ and $b_q$ 
are then determined for each quasar by
maximizing a likelihood given by 
\begin{equation}
L(C_q) = \prod_{i} P_i[f_i,C_q(\lambda_i)] \;.
\end{equation} 
Figure \ref{spectrumfig} shows the $C_q(\lambda)$ estimated for
a typical quasar (blue line).

\begin{figure}[htbp]  
\includegraphics[width=\columnwidth]{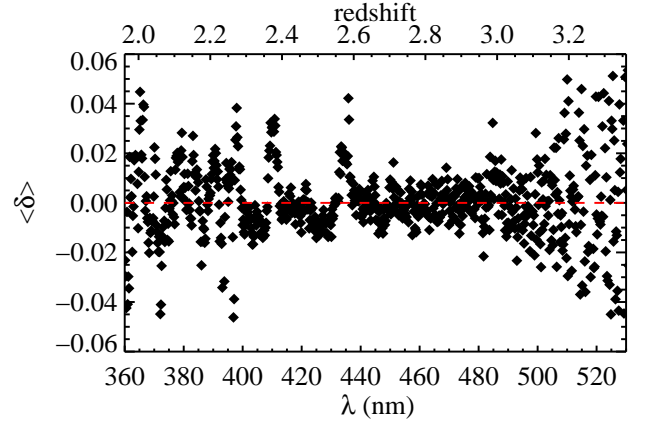}
\caption{The mean of $\delta(\lambda)$ plotted as a function
of observed wavelength (method 1).
Systematic offsets from zero are seen at the 2\% level.
The calcium lines (393.4,396.8~nm) is present.
The features around the 
hydrogen lines H$\gamma$, $\delta$ and $\epsilon$  
(434.1, 410.2, 397.0~nm) are artifacts from the use of F-stars
for the photocalibration of the spectrometer. 
}
\label{calciumfig}
\end{figure}

\begin{figure}[ht]
\includegraphics[width=\columnwidth]{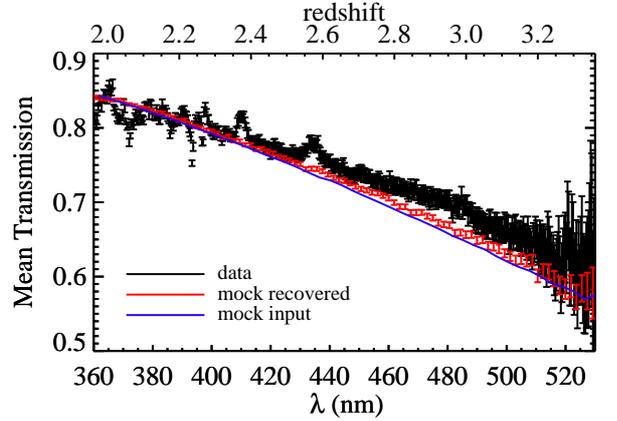}
\caption{Mean transmitted flux fraction  as a function of redshift 
obtained from the continuum fits with method 2. Data are 
shown in black, mock-000 in red and the input mean transmitted flux
fraction  in blue.}
\label{fig:meantransmission}
\end{figure}

The last element necessary to use
equation (\ref{delta:def}) is the 
mean transmitted flux fraction  $\overline{F}(z)$.
If $P(F,z)$ 
derived from the mocks were the true distribution of the
  transmitted flux fraction, then $\overline{F}(z)$ could simply be computed
  from the average of this distribution. 
Since this is not precisely true, we determine $\overline{F}(z)$ from the data by
requiring that the mean of the delta field vanish for all redshifts.
The  $\overline{F}(z)$ we obtain is shown in figure \ref{fig:meantransmission}.
The unphysical wiggles in the derived  $\overline{F}(z)$ are 
associated  with the aforementioned residuals in $\overline{\delta}(\lambda)$
for method 1 (figure \ref{calciumfig}).

There is  one inevitable effect of our two continuum estimating
procedures. 
The use of the forest data in fitting the continuum
effectively forces each quasar to have a mean absorption
near that of the mean for the entire quasar sample.
This approach introduces a spurious negative correlation between
pixels on a given quasar even when well separated in wavelength.
This negative correlation has no direct effect on our  measurement
of the flux correlation function because we do not use pixel
pairs from the same quasar.
However, the physical correlation between absorption on 
neighboring quasars causes the unphysical negative correlation
for individual quasars to generate a negative contribution
to the correlation measured with quasar pairs.
Fortunately, this distortion is a smooth function of scale
so it can be expected to have little effect on the observability or position
of the BAO peak.  This expectation is confirmed by analysis of the
mock spectra (section \ref{cosmosec}).

\subsection{Weights}
\label{sec:weights}

A discussion on the optimal use of weights 
for the \Lya correlation function is found in 
\citet{mcquinnwhiteref}.
Here 
we  simply choose the weights $w_{ij}$ 
so as to approximately  minimize the relative error on 
$\hat{\xi}_A$ 
estimated with equation (\ref{eq:thexi}).
In the  approximation of uncorrelated pixels, 
the variance of  $\hat{\xi}_A$  is
\begin{equation}
\mathrm{Var}(\hat{\xi}_A)=
\frac{\sum_{i,j\in A} w_{ij}^2 \xi_{ii}\xi_{jj}}
{\left[ \sum_{i,j\in A} w_{ij} \right]^2}
\hspace*{10mm}\xi_{ii}=\langle \delta_i^2 \rangle
\label{pixvariance}
\end{equation}
where the pixel variance, $\xi_{ii}$, includes  contributions from
both observational noise and LSS.
The signal-to-noise ratio is:
\begin{equation}
\left({S\over N}\right)^2=
\frac{\langle\hat\xi_A\rangle^2}
{\mathrm{Var}(\hat{\xi}_A)}
\simeq
\frac
{\left(\sum_{ij\in A}\xi_{ij}w_{ij}\right)^2} 
{\sum_{ij\in A} \xi_{ii}\xi_{jj}w_{ij}^2} \;.
\label{noise2signal}
\end{equation}
Because of LSS growth and redshift evolution of the mean absorption,
the $\xi_{ij}$ depend on redshift and 
we use the measured dependence of the 1d correlation
function \citep{macdo2006} 
\begin{equation}
\xi_{ij}(z)=(1+z_i)^{\gamma/2}(1+z_j)^{\gamma/2}\xi_{ij}(z_0) 
\hspace*{10mm}\gamma\sim3.8 \;.
\label{zevol}
\end{equation}
Maximizing the signal-to-noise ratio  with respect to $w_{ij}$ 
this gives:
\begin{equation}
w_{ij}\propto
{(1+z_i)^{\gamma/2}(1+z_j)^{\gamma/2}\over\xi_{ii}^2\xi_{jj}^2} \;.
\label{weights}
\end{equation}
For this expression to be used, we require
a way of estimating the $\xi_{ii}$.
We assume that it can be decomposed into a noise
term and a LSS term ($\sigma_{LSS}$):
\begin{equation}
 \xi_{ii}^2={\sigma^2_{pipeline,i}\over\eta(z_i)}+\sigma^2_{\mathrm{LSS}}(z_i)
\hspace*{10mm}z_i=\lambda_i/\lambda_{{\rm Ly\alpha}} -1 \;,
\label{twocontris}
\end{equation}  
where $\sigma_{pipeline,i}^2=[\mathrm{ivar}(C_q\overline{F})^2]^{-1}$ 
is the pipeline estimate of 
the noise-variance of pixel $i$ 
and $\eta$ is a factor that corrects  for a possible 
misestimate of the variance by the pipeline.

We then organize the data in bins of $\sigma_{pipeline,i}^2$ 
and redshift. In each such bin, 
we measure the variance of 
$\delta_i$,  which serves as an estimator 
of $\xi_{ii}$ for the bin in question. 
The two functions $\eta(z)$ and $\sigma^2_{\mathrm{LSS}}(z)$
can then be determined by fitting equation (\ref{twocontris}).

\begin{figure}
\includegraphics[width=\columnwidth]{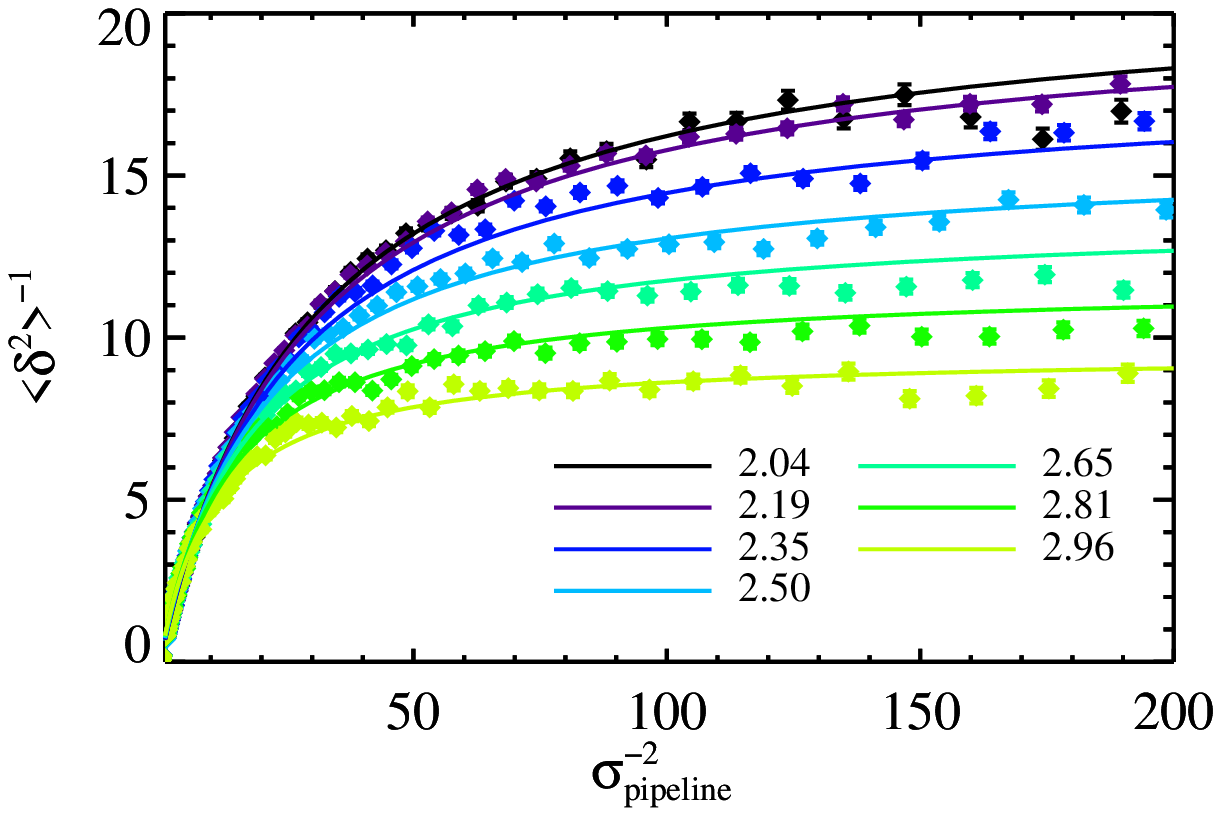}
\includegraphics[width=\columnwidth]{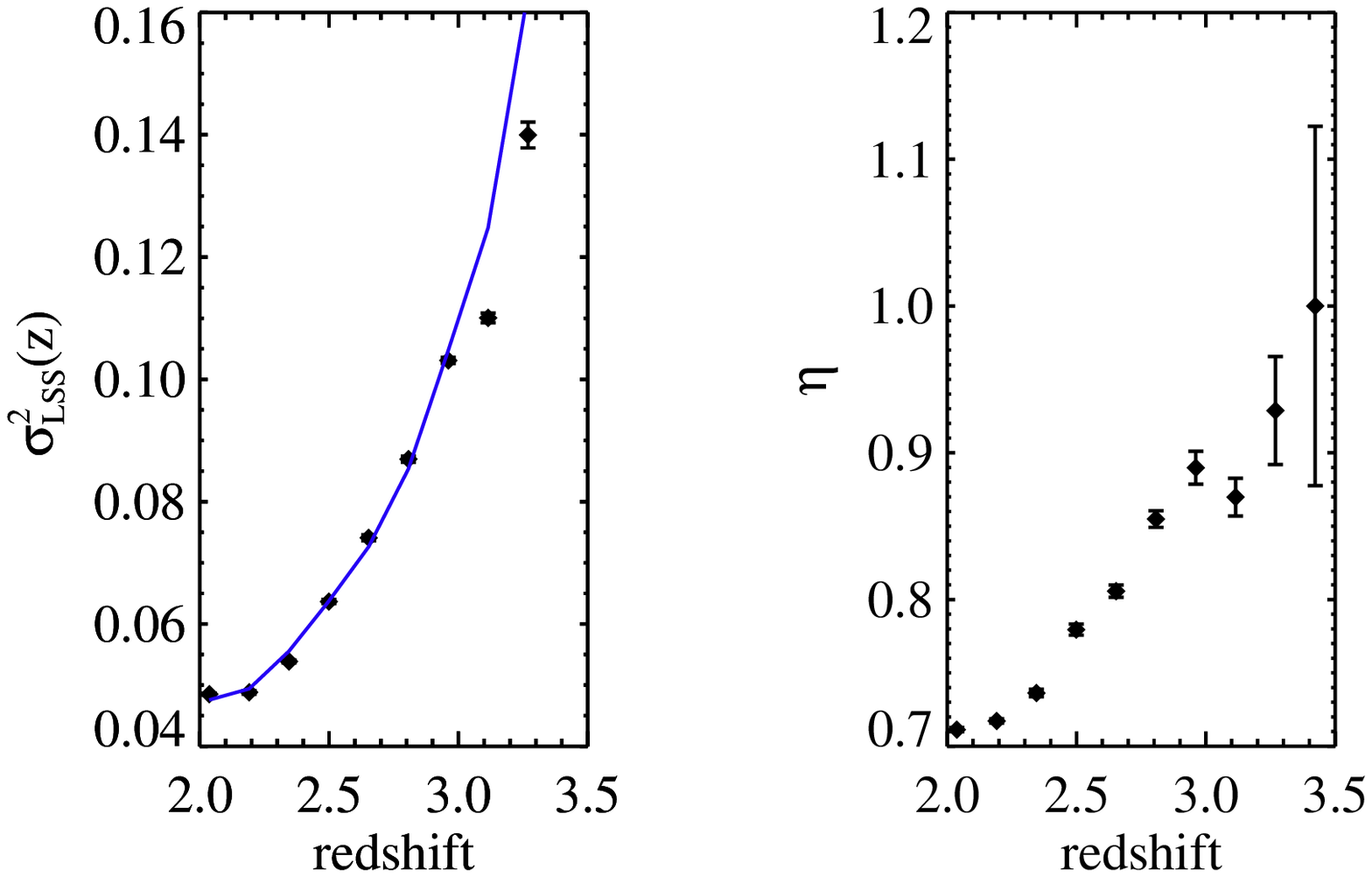}
\caption{Top panel: Inverse total variance in bins of redshift
as a function of 
the pipeline inverse variance. Bottom panel: Parameters of the fit:
the LSS contribution  
$\sigma_\mathrm{LSS}$ (left) and 
the pipeline correction factor
$\eta$ (right) as a function of redshift.
The lines show fits to the data as explained in the text.
}
\label{fig:fitsigmaLSS}
\end{figure}

These fits are shown in figure \ref{fig:fitsigmaLSS}.
The top panel shows that the measured inverse variance follows
the inverse pipeline variance until saturating at the redshift-dependent
LSS variance (shown on the bottom left panel).
For $z>3$, there are not enough pixel pairs to determine
$\eta(z)$ and  $\sigma^2_{\mathrm{LSS}}(z)$. 
In this high redshift range, we assumed $\eta=1$ and 
extrapolated $\sigma_\mathrm{LSS}(z)$
with a second-degree polynomial fit to the $z<3$ data.

\subsection{$\xi(r,\mu)$}

The procedure described above was used to
determine $\xi(r,\mu)$ through
equation \ref{eq:thexi} in
$r$-bins of width $4~\hMpc$ (centered at 2,6,..., $198~\hMpc$) and in 
$\mu$-bins of width  0.02, (centered  at 0.01, 0.03, ... 0.99).
The $50\times50$ $r-\mu$ bins have an average of 
$6\times10^6$ terms in the sum
(\ref{eq:thexi}) 
with an average nominal variance of $\xi$ for individual bins
of $(10^{-4})^2$ as given by
(eqn.\ref{pixvariance}).

\begin{figure}[ht]
\includegraphics[width=\columnwidth]{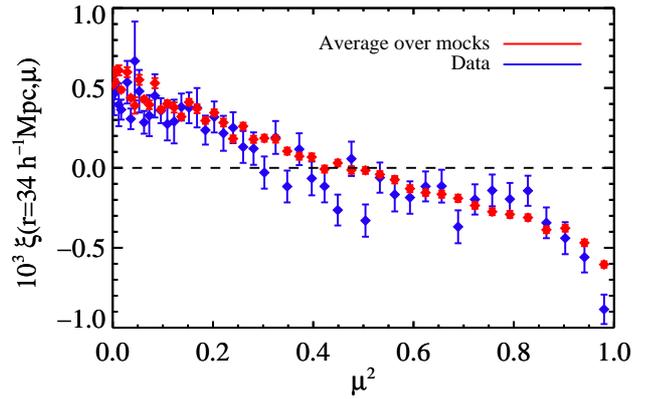}
\caption{$\xi(r,\mu)\,vs.\,\mu^2$ for the bin centered on $r=34~\hMpc$.
The red dots are the mean of the 15 mocks and the blue dots are the data.
}
\label{fig:xivsmusq}
\end{figure}

Figure \ref{fig:xivsmusq} shows an example of $\xi(r,\mu)$  
for the $r$ bin centered on $34~\hMpc$.  
The blue dots are the data and the red dots are the mean of the
15 mocks.
The function falls from positive to 
negative values with increasing $\mu$, as expected from redshift
distortions.  The effect is enhanced by the deformation due to
the continuum subtraction.

\begin{figure}[ht]
\includegraphics[width=\columnwidth]{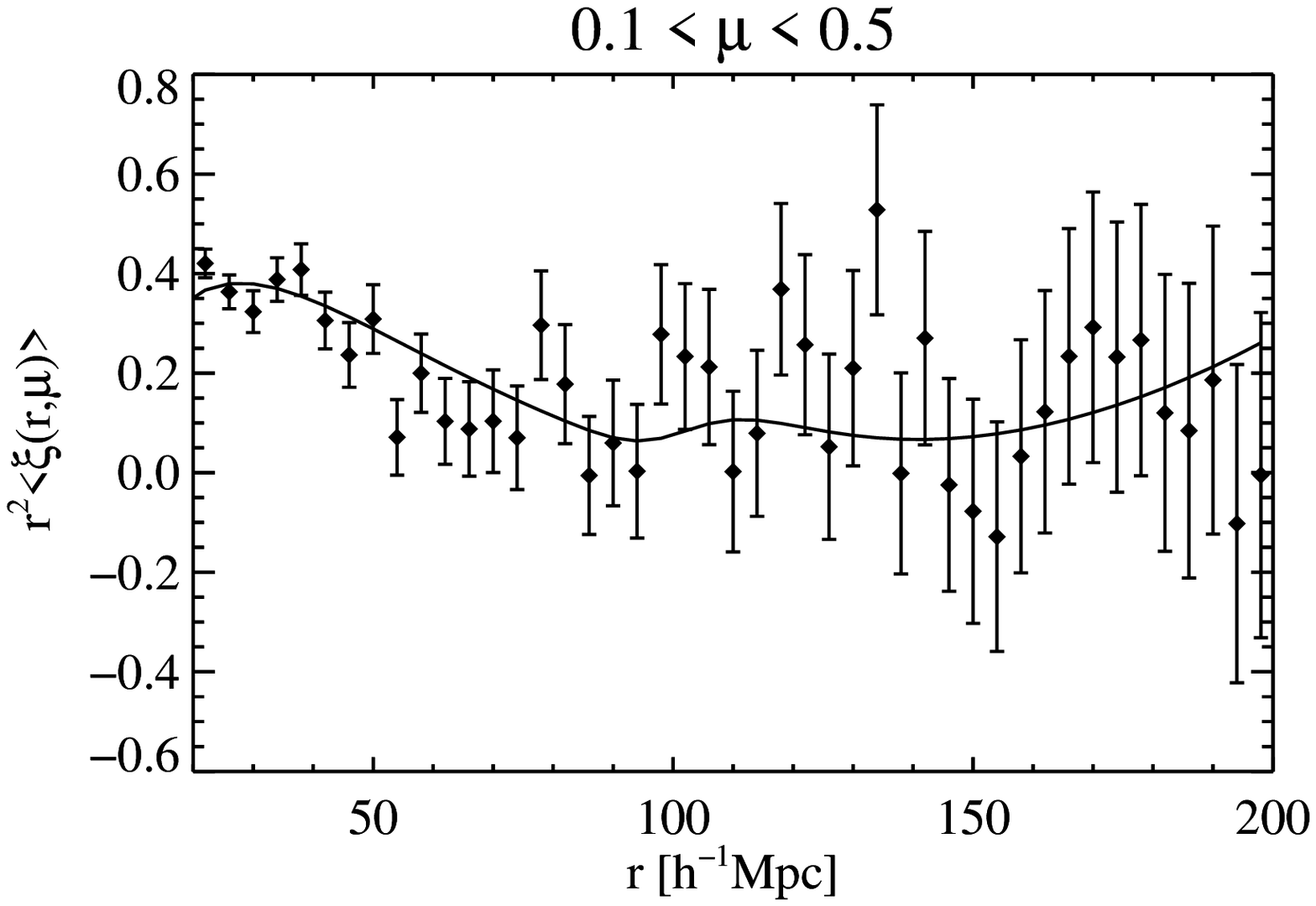}
\includegraphics[width=\columnwidth]{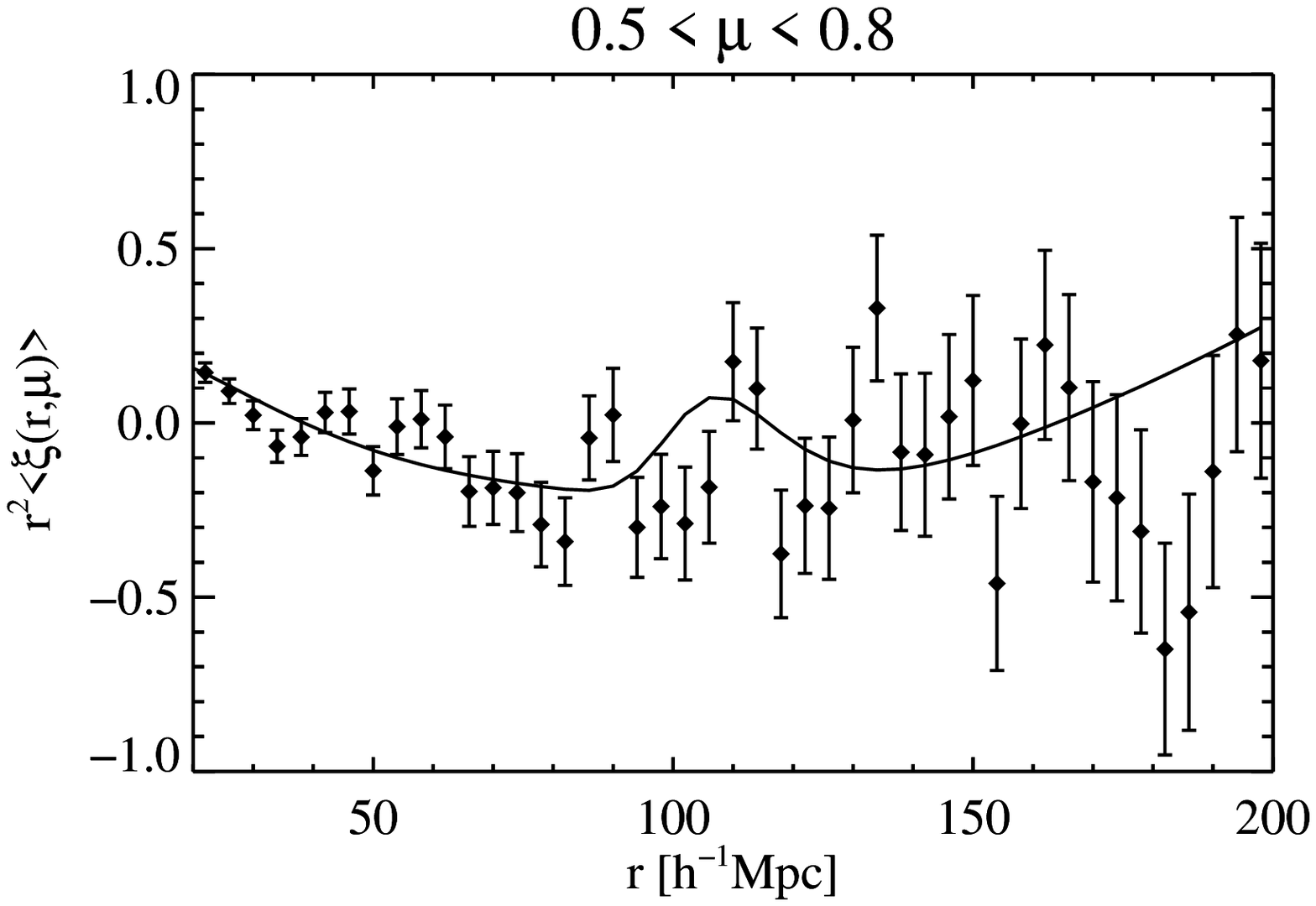}
\includegraphics[width=\columnwidth]{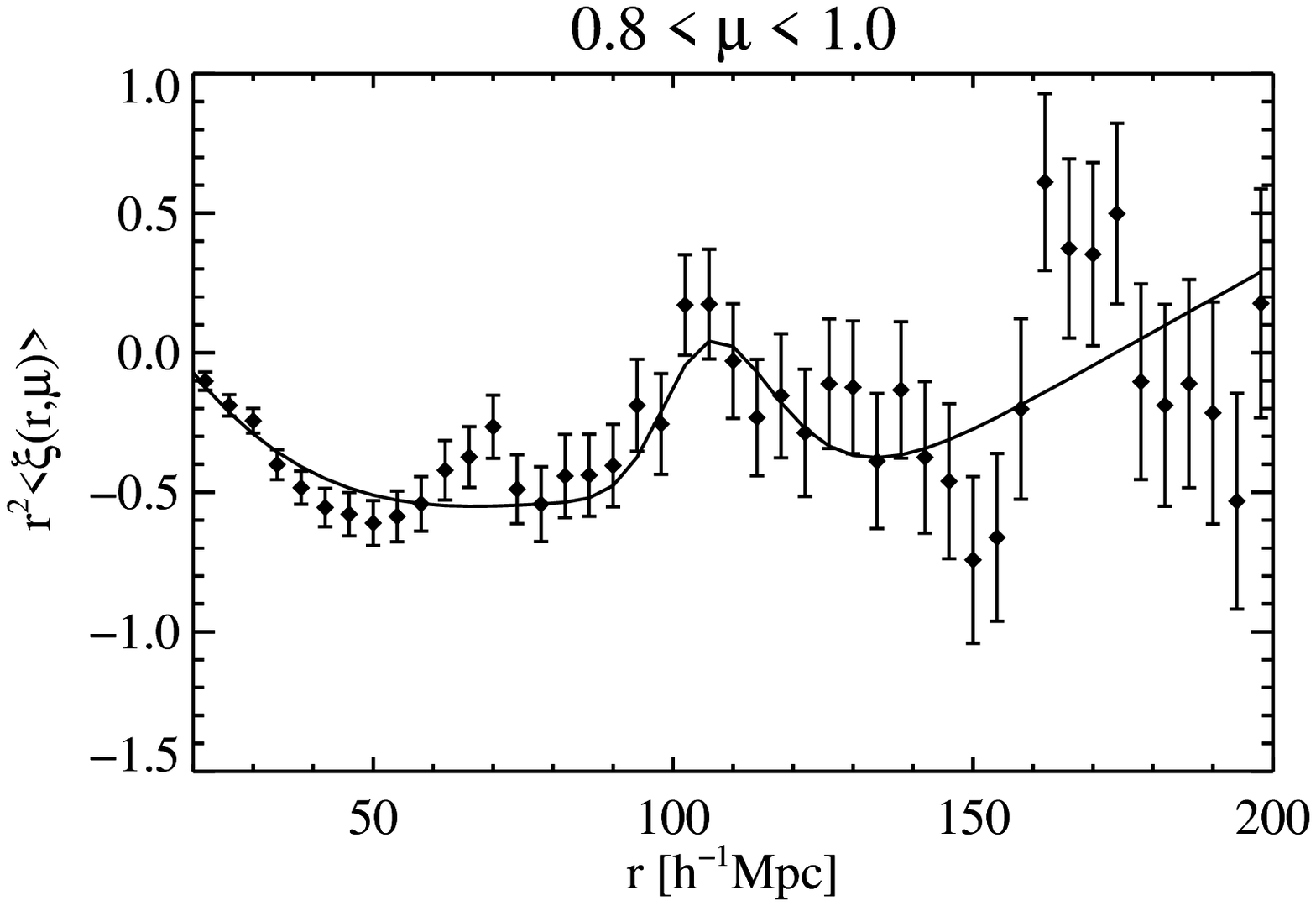}
\caption{$\xi(r,\mu)$ averaged over $0.1<\mu<0.5$, $0.5<\mu<0.8$
and $0.8<\mu<1$. 
The curves give fits (section \ref{cosmosec}) to the data
imposing concordance $\Lambda$CDM cosmology. 
The BAO peak is most clearly present in the data for $\mu>0.8$.
}
\label{ximubins1}
\end{figure}

Figure \ref{ximubins1} presents $\xi(r,\mu)$ averaged over   
three bins in $\mu$.
A clear peak at the expected BAO position, $r_s=105~\hMpc$, is present
in the bin $0.8<\mu<1.0$ corresponding
to separation vectors within $37^\circ$ of the line-of-sight.
The curves show the best fits for a $\Lambda$CDM correlation function,
as described in section \ref{cosmosec}.

The data were divided into various subsamples to search for systematic
errors in $\xi(r,\mu)$.  For example, searches were made for
differences between the northern and southern Galactic cap regions
and between higher and lower signal-to-noise ratio quasars.
No significant differences were found in
the overall shape and amplitude of the 
correlation function.
We also verified that the BAO peak position does not
change significantly when
wavelength slices of \Lya forest data are eliminated,
in particular slices centered on the Balmer features in 
figure \ref{calciumfig}.
The peak position also does not change significantly if
the subtraction of the mean $\delta$ (figure \ref{calciumfig})
is suppressed.

\section{The Monopole and quadrupole}
\label{fitssec}

The analysis of the correlation function was performed
in the framework of the standard multipole decomposition
\citep{hamilton}.
For each bin in $r$ we fit a monopole ($\ell=0$) and 
quadrupole ($\ell=2$) to the angular dependence:
\begin{equation}
\hat{\xi}(r,\mu)=\sum_{\ell=0,2}\xi_\ell(r)P_\ell(\mu)
\;=[\xi_0(r)-\xi_2(r)/2] + [3\xi_2(r)/2]\mu^2
\end{equation}
where $P_\ell$ is the $\ell$-Legendre polynomial.
We ignore the small and poorly determined $\ell=4$ term. 
This fit is performed using a simple $\chi^2$ 
minimization with 
the nominal variance (equation \ref{pixvariance})
 and ignoring the  correlations between bins.
This approach makes the fit slightly sub-optimal.
(Later, we will correctly take into account correlations
between $r$-bins of the monopole and quadrupole.)
We also exclude from this fit the portion $\mu<0.1$ to avoid residual
biases due to correlated sky subtraction across quasars;
this has
a negligible impact on the fits and, 
at any rate, there is little BAO signal at low $\mu$.

\begin{figure}[ht]
\includegraphics[width=\columnwidth]{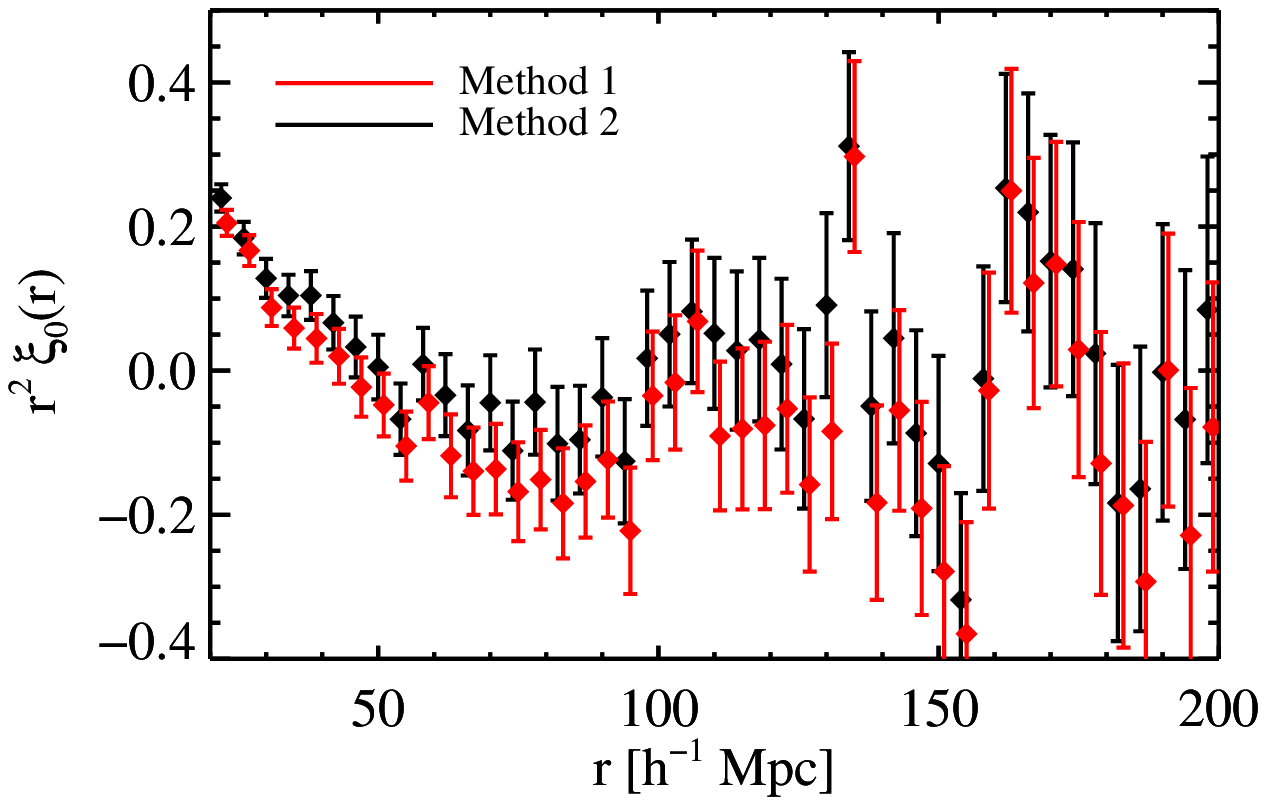}
\includegraphics[width=\columnwidth]{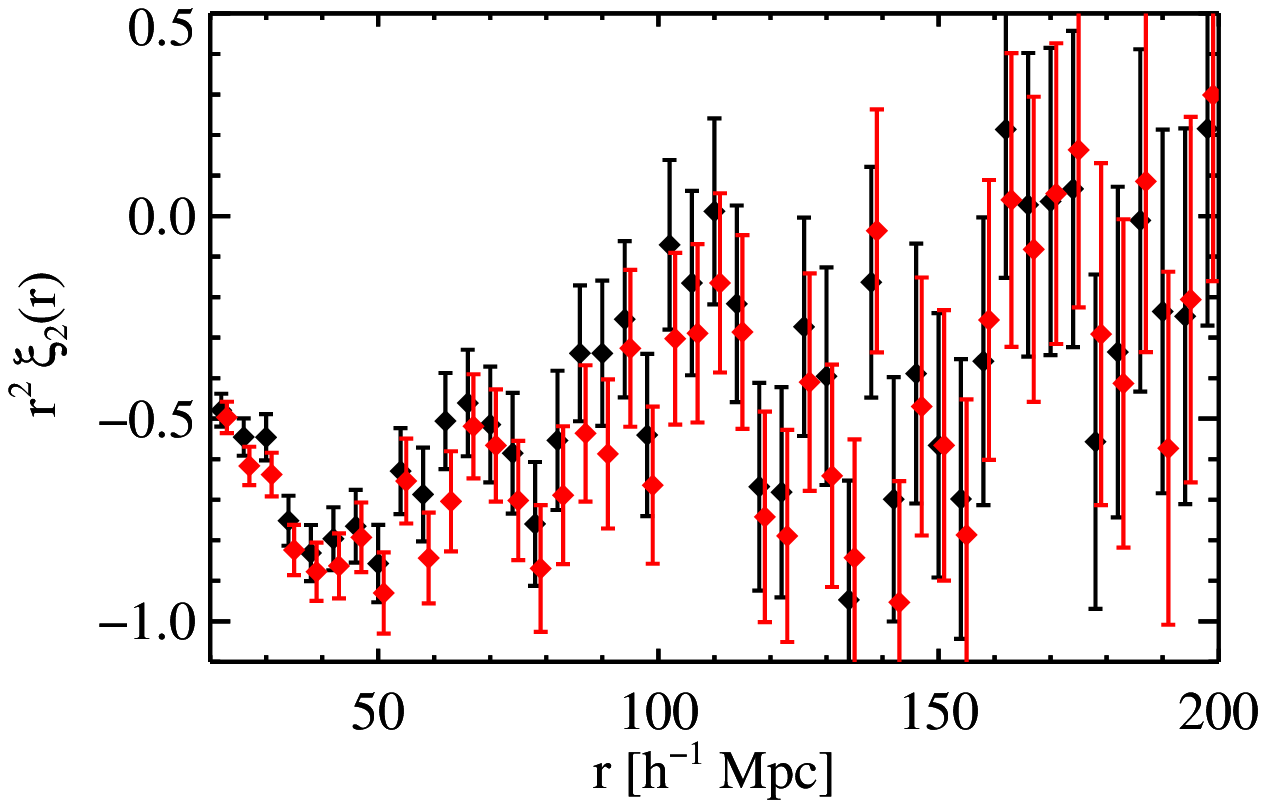}
\includegraphics[width=\columnwidth]{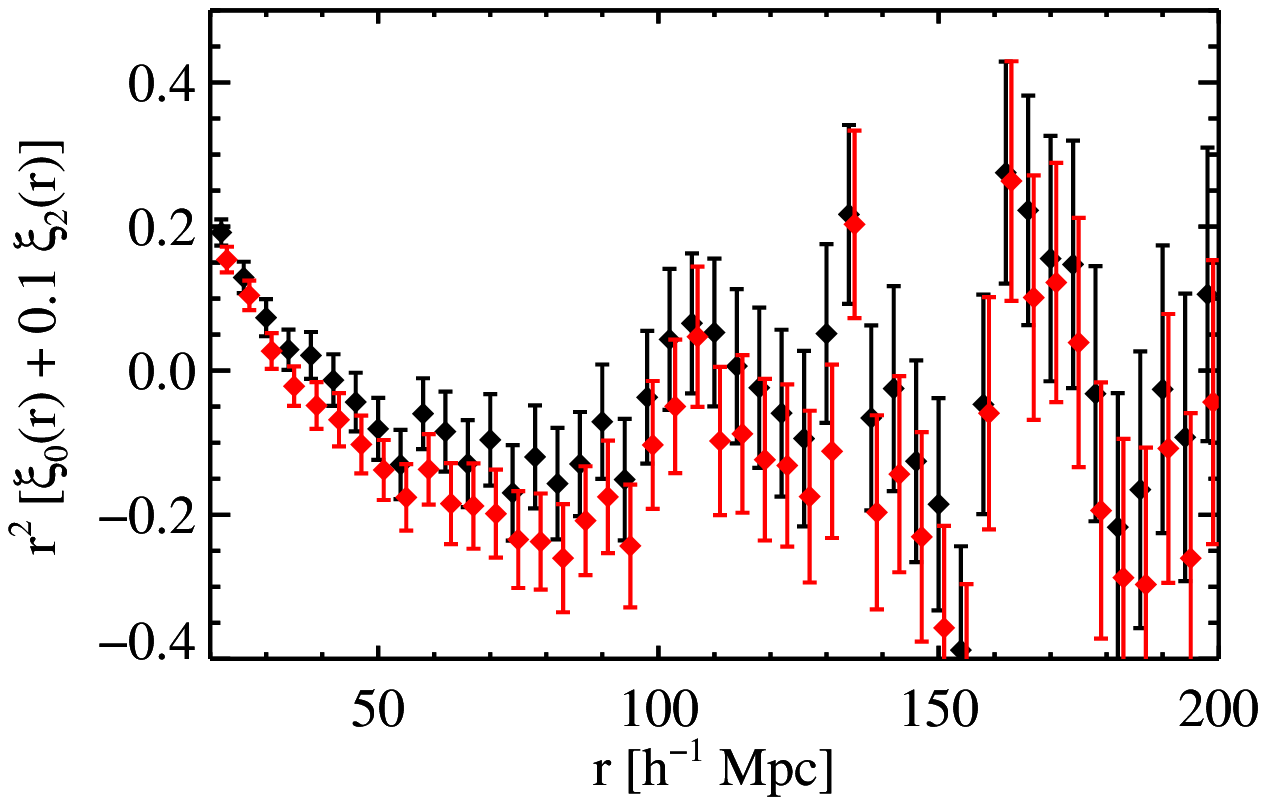}
\caption{Monopole (upper panel) and quadrupole (middle panel)
correlation functions
found by method 1 (red) and method 2 (black).
The bottom panel shows the combination $\xi_0+0.1\xi_2$ 
found by method 1 (red)  and method 2 (black).
}
\label{fig:monoquadm1m2}
\end{figure}

Figure \ref{fig:monoquadm1m2} displays the monopole and quadrupole
signals found by the two methods.
The two methods are slightly offset from one another, but the
peak structure is very similar.
Figure \ref{fig:monoquadm1m2} also shows the combination $\xi_0 + 0.1\xi_2$
which, because of the small monopole-quadrupole anti-correlation 
(section \ref{covsubsec}), is a better-determined quantity.
The peak structure seen in figure \ref{ximubins1} is also present
in these figures.

Because of the continuum estimation procedure 
(sections \ref{contfitsec1} and \ref{contfitsec2}),
we can expect that the monopole and quadrupole shown in 
figure \ref{fig:monoquadm1m2}
are deformed with respect to the true monopole and quadrupole.
The most important difference is that the measured monopole is negative
for $60~\hMpc<r<100~\hMpc$ while the true $\Lambda$CDM monopole remains
positive for all $r<130~\hMpc$.
The origin of the  deformation in the continuum estimate
is demonstrated in appendix \ref{mocksec} where
both the true and estimated continuum can be used to derive
the correlation function (figure \ref{fig:m1m2}).
As expected, the deformation is a slowly varying function
of $r$ so neither the position of the BAO peak nor its amplitude
above the slowly varying part of the  correlation function
are significantly affected.

\subsection{Covariance of the monopole and quadrupole}
\label{covsubsec}

In order to determine the significance of the peak we must
estimate the covariance matrix of the 
monopole and quadrupole.
If the fluctuations $\delta_i$ in
equation (\ref{eq:thexi})
in different pixels were uncorrelated, the variance of $\xi_A$ would simply
be the weighted products of the fluctuation variances. This yields a result
that is $\sim30\%$ smaller than the true correlation variance that we
compute below. The reason is, of course, that the 
$\delta$-pairs are correlated, either from LSS or from correlations
induced by instrumental effects or continuum subtraction;
this effect reduces  the effective number of pairs and introduces
correlations between $(r,\mu)$ bins.

Rather than determine the full covariance matrix
for $\xi(r,\mu)$, we determined directly 
the covariance matrix for
$\xi_0(r)$ and $\xi_2(r)$
by standard techniques of dividing the full quasar sample into
subsamples according to position on the sky.
In particular we used the sub-sampling technique described below. 
We also tried a bootstrap technique \citep{bootstrap} 
consisting of substituting the 
entire set of $N$ subdivisions of the data by $N$ of these subdivisions 
chosen at random (with replacement) to obtain a ``bootstrap'' sample. 
The covariances are then measured from the ensemble of bootstrap samples. 
Both techniques give consistent results.

The adopted covariance matrix for the monopole and quadrupole
uses the sub-sampling technique.
We divide the data into angular sectors and calculate a 
correlation function in each sector. Pairs of pixels belonging to 
different sectors contribute only to the sector of the pixel with lower 
right ascension. We investigated two different divisions of the sky data: 
defining 800 (contiguous but disjoint) sectors of similar solid angle, 
and taking the plates as defining the sectors (this latter version does 
not lead to disjoint sectors). The two ways of dividing the data 
lead to similar 
covariance matrices.

Each sector $s$ in 
each division 
of the data provides
a measurement of $\xi_s(r,\mu)$ that can be used to derive
a monopole and quadrupole, $\xi_{\ell s}(r)$, $(\ell=0,2)$.
The covariance of the whole BOSS sample can then be estimated
from the weighted and rescaled covariances for each
sector:
\begin{eqnarray}
\sqrt{W(r)W(r^\prime)}
\mathrm{Cov}[\hat{\xi}_\ell(r),\hat{\xi}_{\ell'}(r^\prime)]
\hspace*{30mm}
\nonumber \\
\hspace*{10mm}
 =\left\langle
\sqrt{W_s(r)W_s(r^\prime)}
\left[
\hat{\xi}_{\ell s}(r)\hat{\xi}_{\ell' s}(r^\prime)
-\overline{\xi}_\ell(r)\overline{\xi}_{\ell'}(r^\prime)
\right]
\, \right\rangle
\;.
\label{eq:covariance}
\end{eqnarray}
The average denoted by $\langle\,\rangle$ is the simple
average over sectors, while $\overline{\xi}_\ell(r)$ denotes
the correlation function  measured for the whole BOSS sample. 
The $W_s(r)$ are the summed pixel-pair weights for the radial
bin $r$ for the sector $s$ 
and $W(r)$ is the same sum for the whole BOSS sample.

\begin{figure}[ht]
\includegraphics[width=\columnwidth]{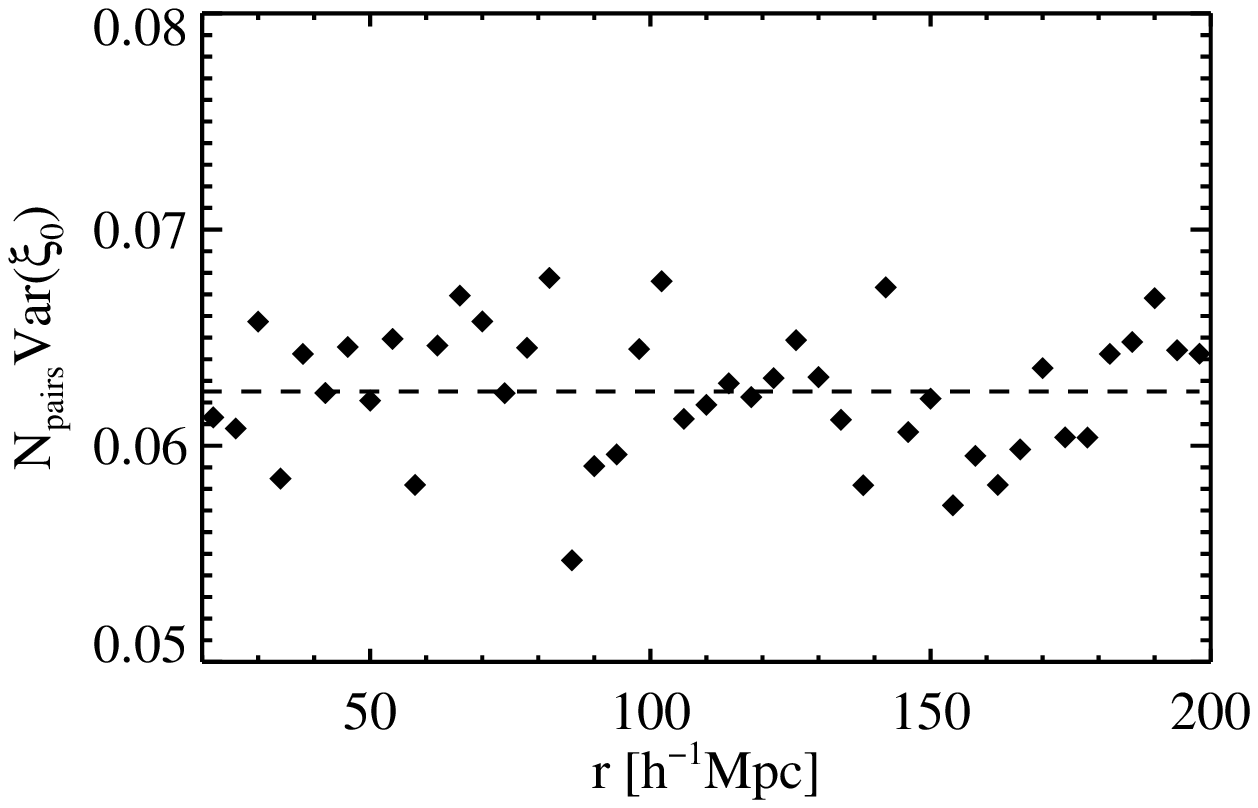}
\includegraphics[width=\columnwidth]{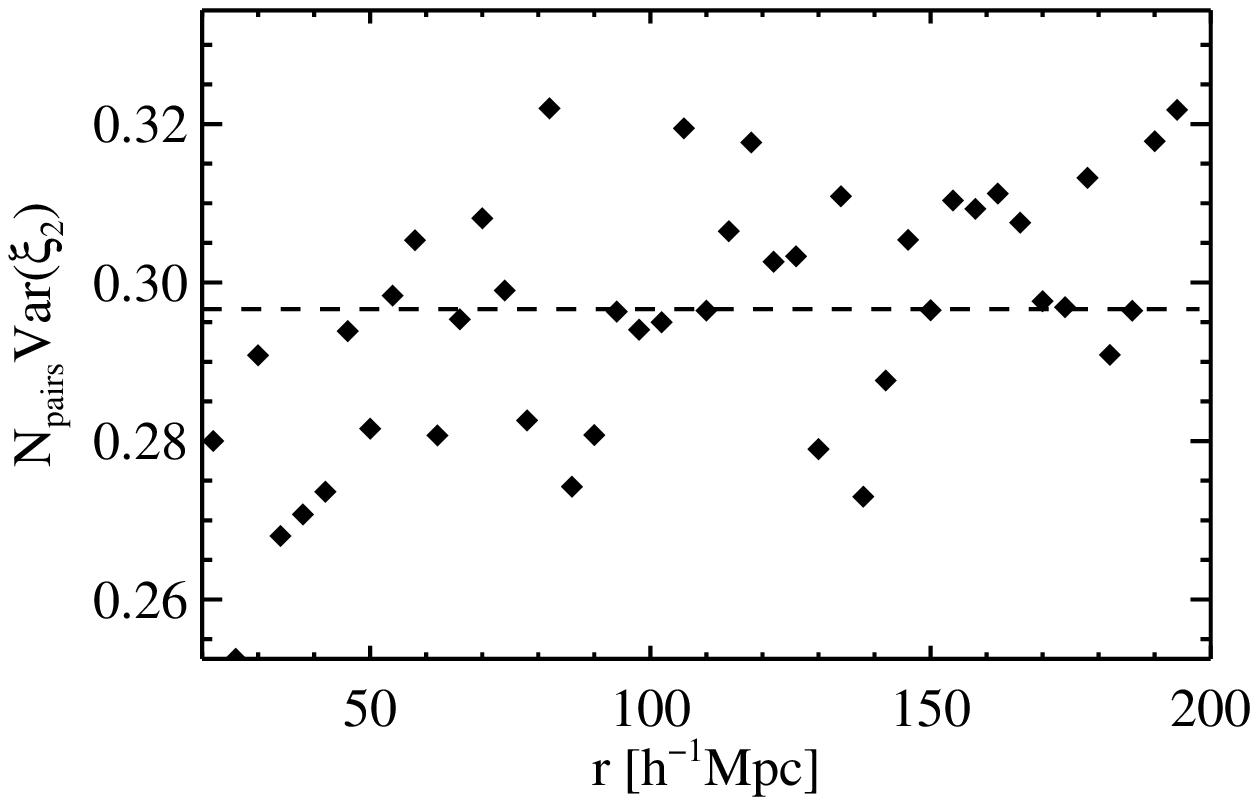}
\includegraphics[width=\columnwidth]{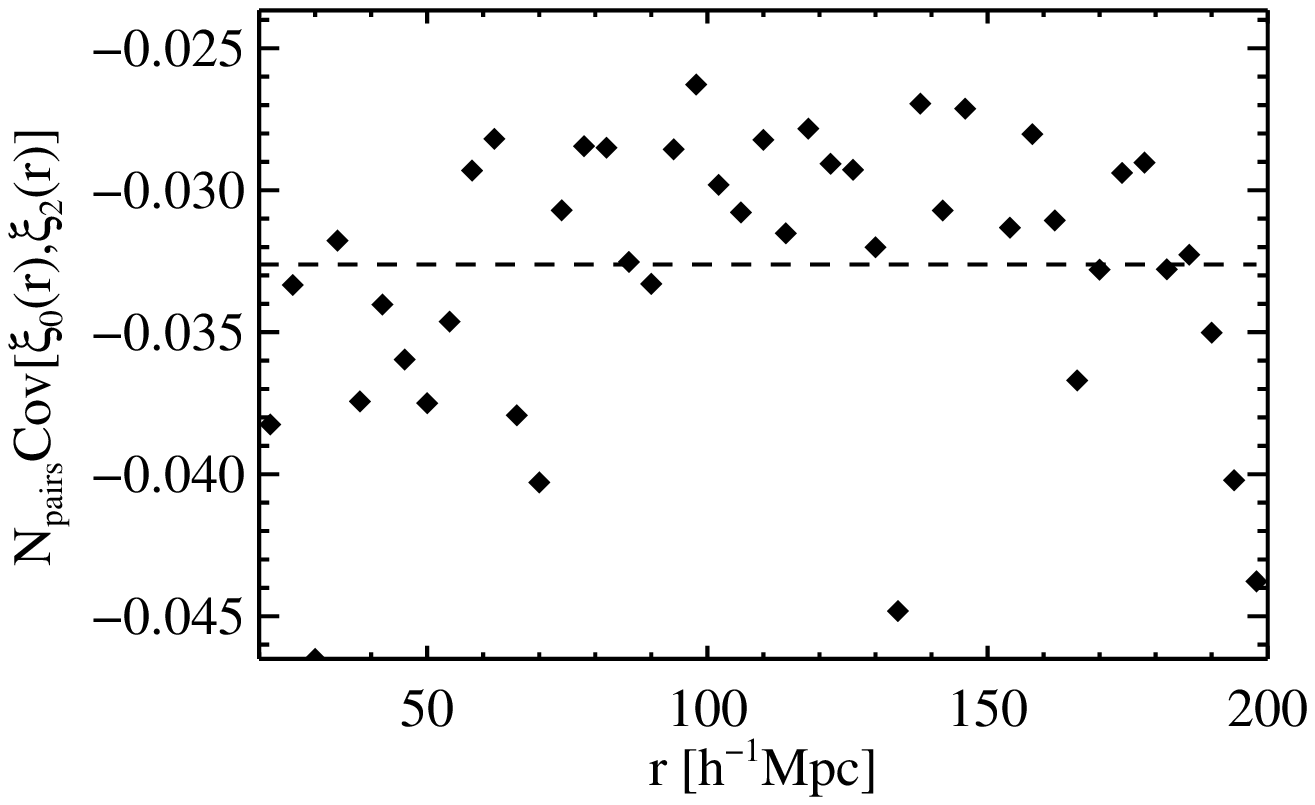}
\caption{
The r-dependence of the product of the
monopole (top) and quadrupole (middle) variances
and the number of pairs in the $r$-bin.
The bottom panel shows the product for 
the monopole-quadrupole covariance ($r=r^\prime$ elements).
The dotted lines show the means for $r>20\,\hMpc$.
}
\label{fig:varmonoquad}
\end{figure}

The most important terms in the covariance matrix are the
$r=r^\prime$ terms, i.e.  the
monopole and quadrupole variances.
They are shown in 
figure \ref{fig:varmonoquad} as a function of $r$.
In the figure, they are  multiplied by the number $N$
of pixel pairs in the $r$-bin.
The product is nearly independent of $r$, as expected
for a variance nearly equal to the pixel variance divided by $N$.
For the monopole, the variances are only about 30\% higher
than what one would calculate naively assuming
uncorrelated pixels and equation (\ref{pixvariance}).
Figure \ref{fig:varmonoquad} also displays the monopole-quadrupole
covariance times number of pairs, which also is nearly
independent of $r$.

\begin{figure}[ht]
\includegraphics[width=\columnwidth]{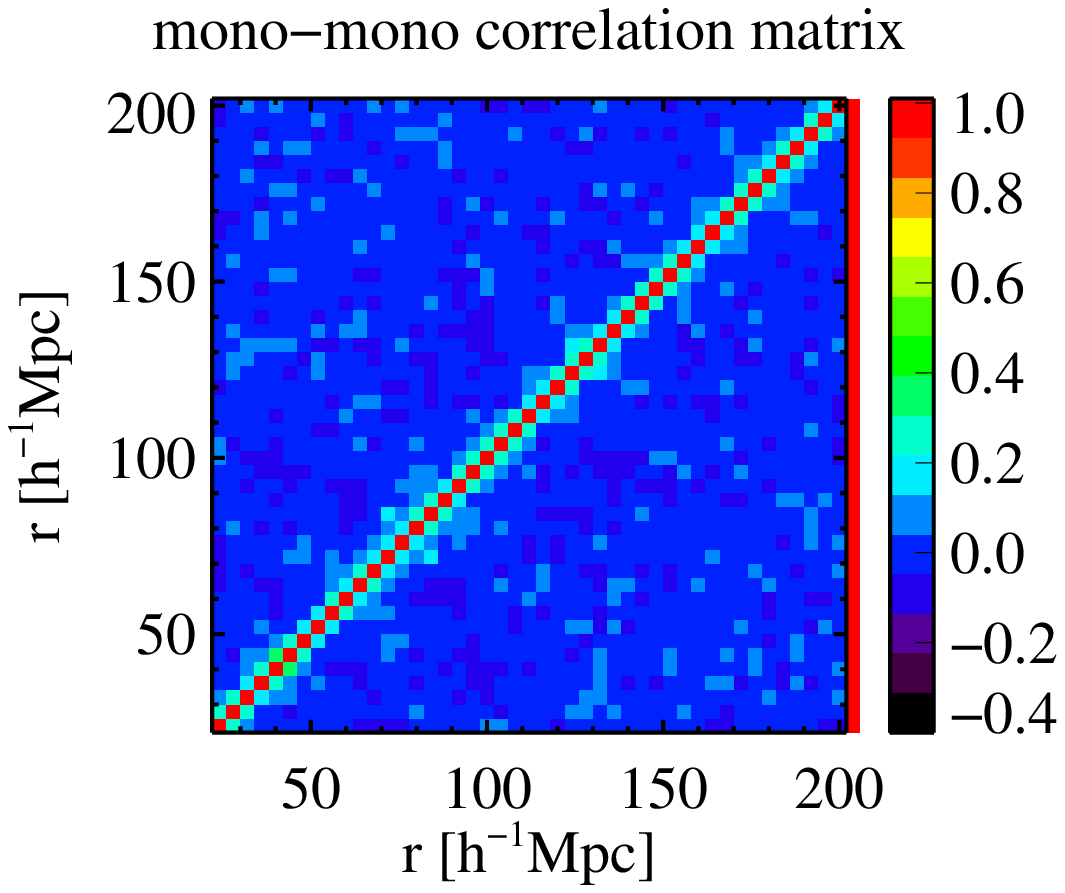}
\includegraphics[width=\columnwidth]{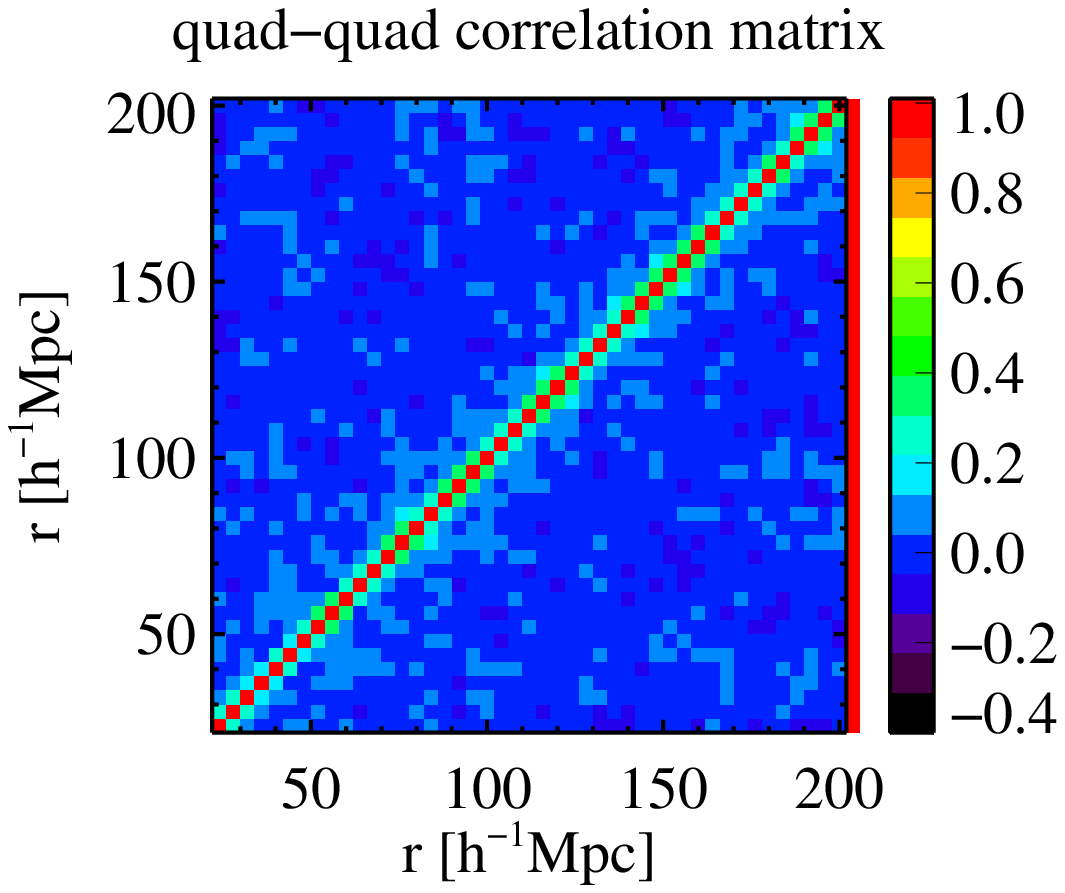}
\includegraphics[width=\columnwidth]{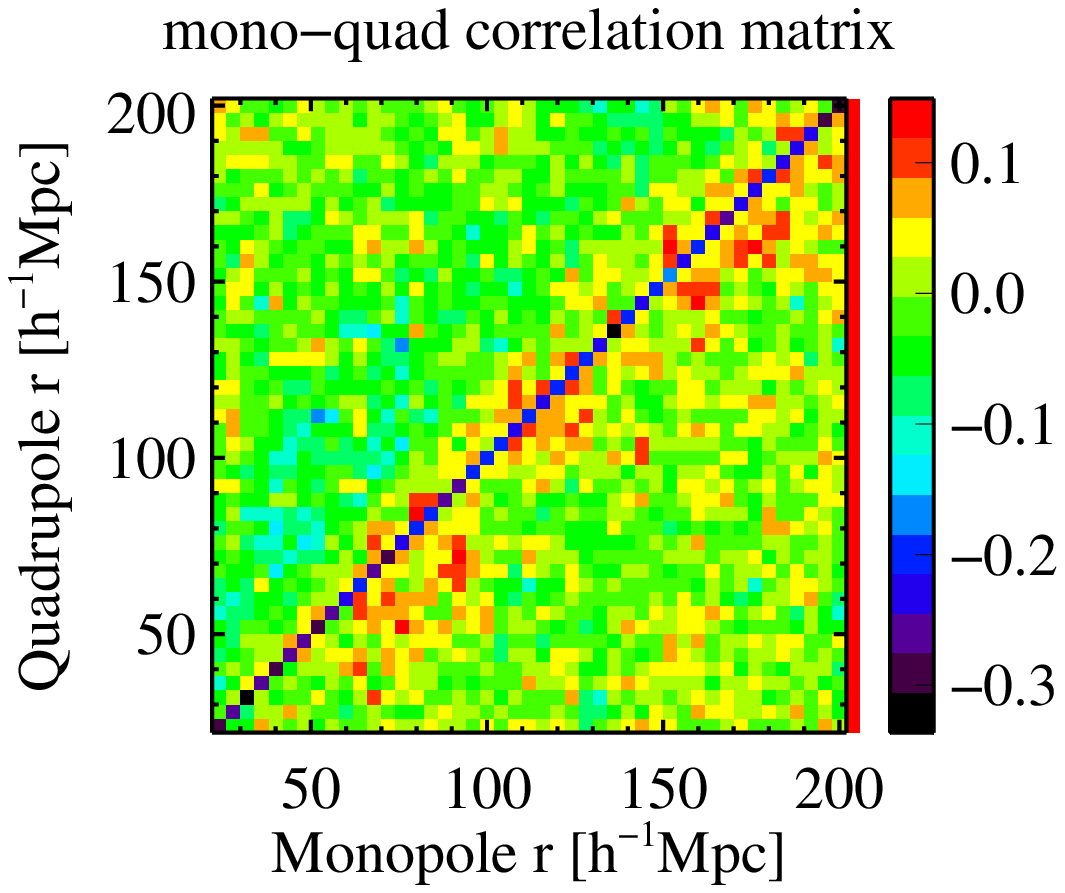}
\caption{
The monopole and quadrupole covariance matrix.
The monopole-monopole and quadrupole-quadrupole elements
are normalized to the variance:  $C_{ij}/\sqrt{C_{ii}C_{jj}}$.
The monopole-quadrupole elements are normalized to the
mean of the quadrupole and monopole variances.
The first off-diagonal elements of the monopole-monopole and
quadrupole-quadrupole elements are $\sim20\%$ of the diagonal elements.
The diagonal elements of the quadrupole-monopole covariance are
$\sim-0.2$ times the geometric 
mean of the monopole and quadrupole variances.
}
\label{fig:covmq}
\end{figure}

Figure \ref{fig:covmq} 
displays 
the monopole-monopole and quadrupole-quadrupole
covariances.
Nearest-neighbor covariances are of order 20\%.
Figure \ref{fig:covmq} also shows 
monopole-quadrupole covariance.

\begin{figure}[ht]
\includegraphics[width=\columnwidth]{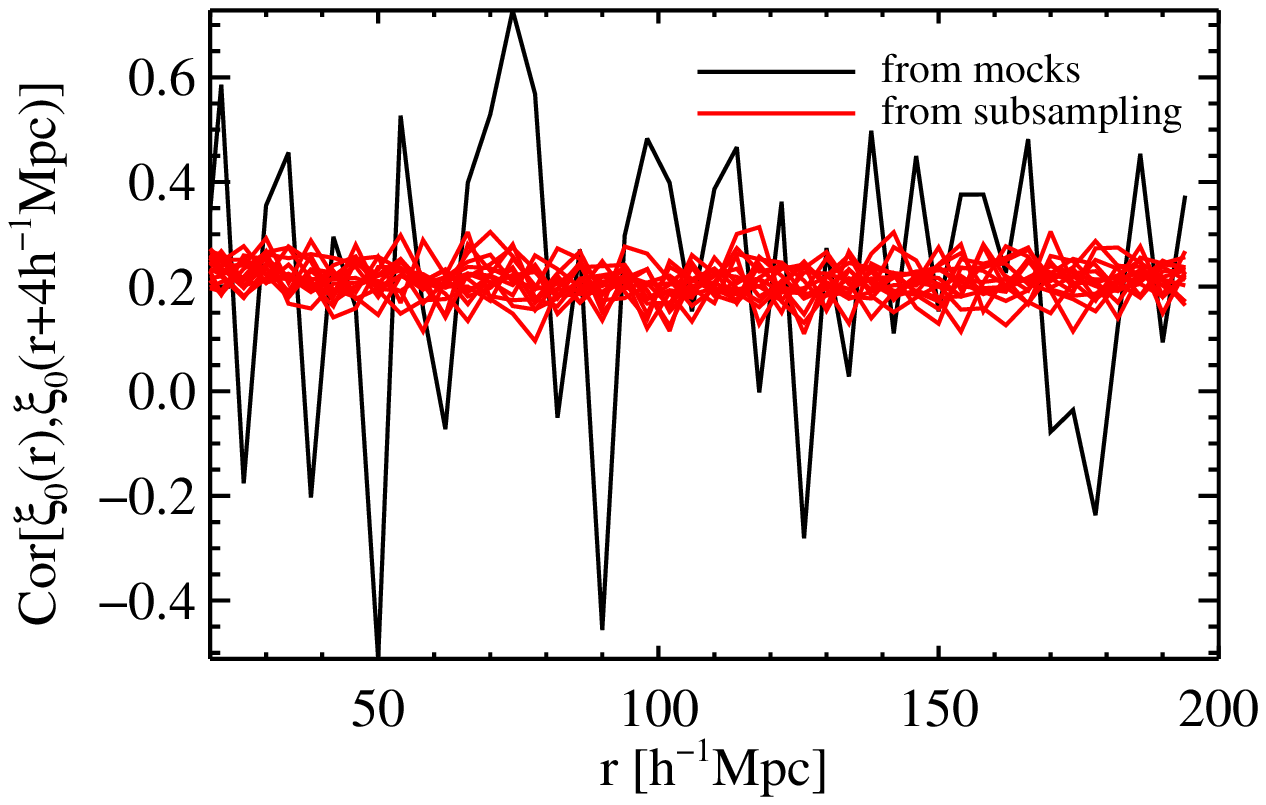}
\includegraphics[width=\columnwidth]{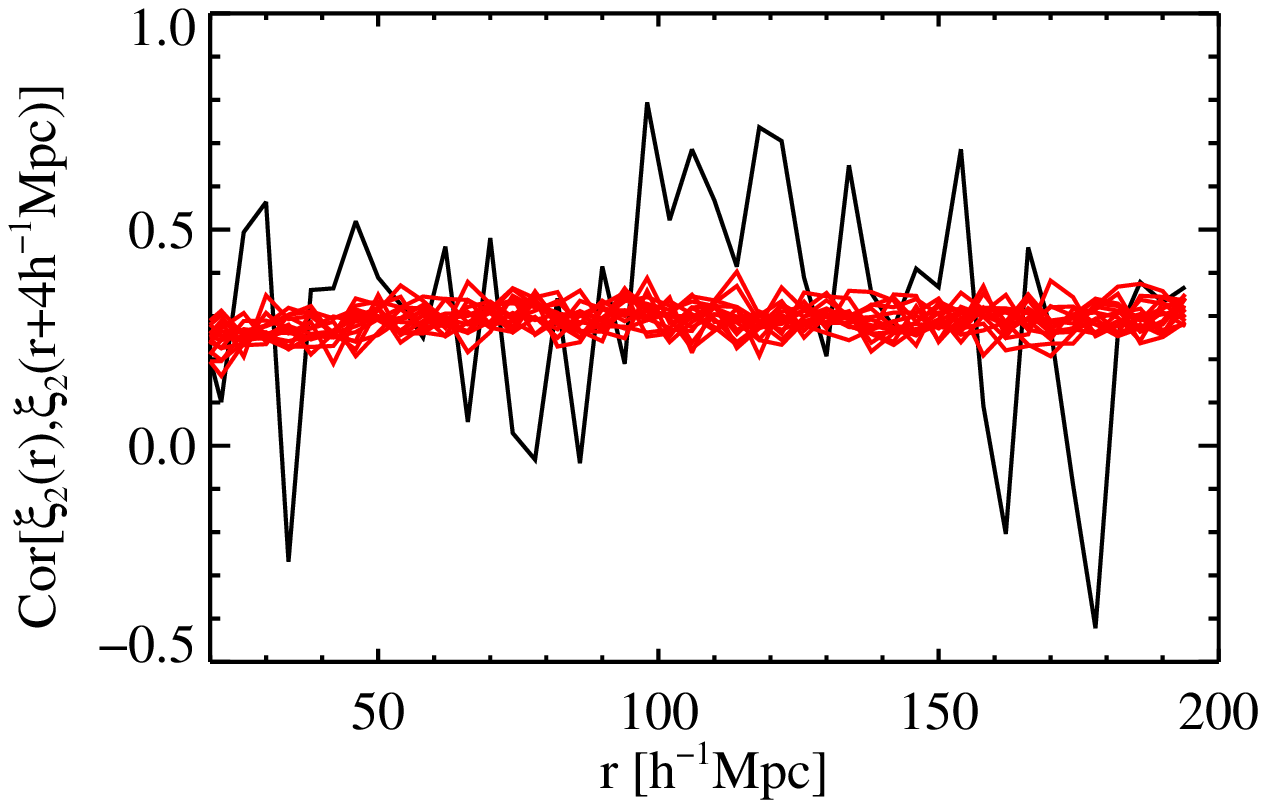}
\includegraphics[width=\columnwidth]{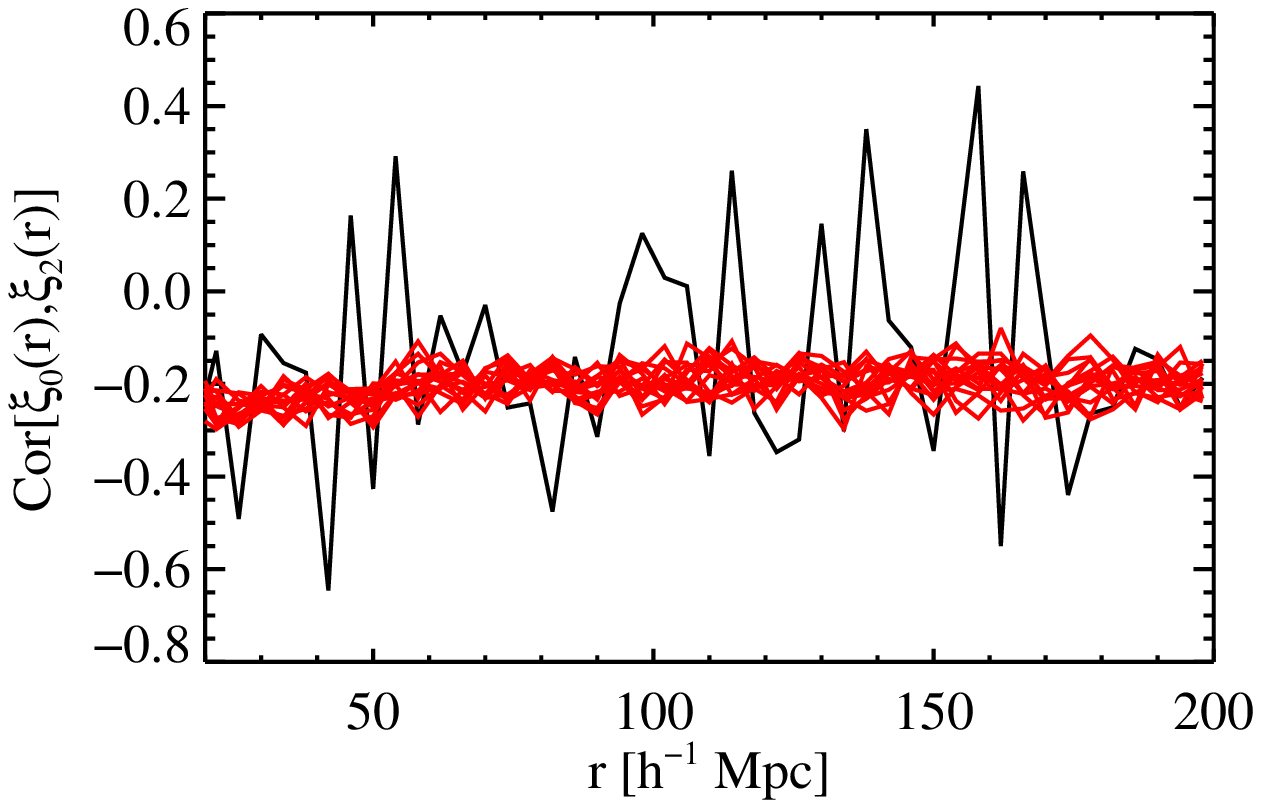}
\caption{Verification of the off-diagonal elements
of the covariance matrix with the 15 sets of mock spectra. 
The black lines show correlations derived from the dispersion
of the 15 measurements and the red lines show the correlations
expected from the covariance matrix calculated by sub-sampling.
The top and middle panels show the correlation between neighboring bins 
for monopole and quadrupole respectively. The bottom panel the correlation
between monopole and quadrupole measured at the same distance bin.
}
\label{mockcov}
\end{figure}

We used the 15 sets of mock spectra to test
our method for calculating the covariance matrix.  
From the 15 measurements of $\xi_\ell(r)$ one can
calculate the average values of $\xi_\ell(r)\xi_{\ell^\prime}(r^\prime)$
and compare them with those expected from the covariance
matrix.
Figures \ref{mockcov} shows this comparison
for the monopole and quadrupole variance, the monopole and quadrupole
covariances between neighboring r-bins and the monopole-quadrupole
covariance. The agreement is satisfactory.

\subsection{Detection significance of the  BAO peak}

In this section, we estimate the
significance of our detection of a BAO peak at $105~\hMpc$.
At the statistical power of the present data, it is
clear that the peak significance will depend to some extent
on how we treat the
so-called ``broadband'' correlation function on which the peak
is superimposed.
In particular, the significance
will depend strongly on the $r$-range over which the correlation functions
are fitted.  
To the extent that the BAO peak is known to be present in the
matter correlation function and that the \Lya absorption is known to trace
matter,  the actual significance is of limited interest for
cosmology.
Of greater interest is the uncertainty in the derived cosmological
parameter constraints (section \ref{cosmosec})
which will be  non-linear reflections of the peak significance
derived here.

A detection of the BAO peak requires comparing the quality of a 
fit with no peak (the null hypothesis) to that of a  fit
with a peak. Typically, this exercise would 
be performed by choosing a test statistic, such as the $\chi^2$,  
computing the distribution of this quality indicator from a 
large number of peak-less simulations and looking at the consistency of 
the data with this distribution. Since our mock data sets are quite 
computationally expensive and only a handful are available, we chose a 
different approach.

\begin{figure}[ht]
\includegraphics[width=\columnwidth]{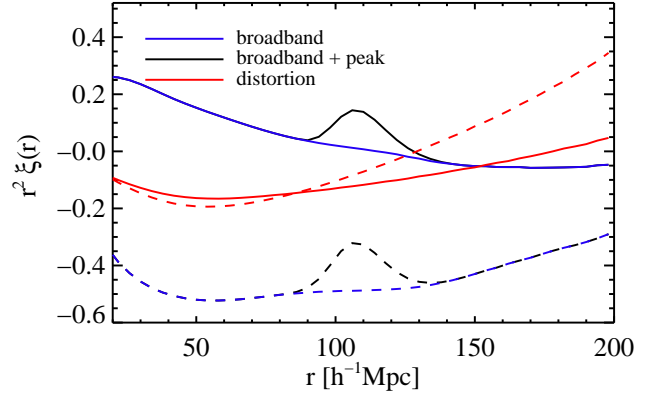}
\caption{
The fitting functions 
used for the determination of the peak detection significance:
$r^2\xi^{bb}(r)$ (blue), 
$r^2[\xi^{peak}(r)+\xi^{bb}(r)]$ (black) 
and $r^2\xi^{dist}(r)$ (red) for the monopole (solid lines) 
and quadrupole (dashed lines).
}
\label{threefunctions}
\end{figure}

Our detection approach uses the following expression to fit the 
observed monopole and quadrupole.
\begin{equation}
\xi_\ell(r) = 
  B_\ell\xi_\ell^{BB}(r) 
+ C_\ell\xi^\mathrm{peak}_\ell(r) 
+ A_\ell\xi_\ell^\mathrm{dist}(r) 
\label{eq:detection_fit}
\end{equation}
where 
$\xi_\ell^{BB}$ is a 
broadband term to describe the LSS correlation function in the
absence of a peak,
$\xi^\mathrm{peak}_\ell$ is a peak term,  
and
$\xi_\ell^\mathrm{dist}$ is a ``distortion'' term used
to model the effects of continuum subtraction.
The broadband term is derived from the fiducial
$\Lambda$CDM cosmology defined by
the parameters in equation (\ref{fiducialdef}).
It is obtained by fitting the shape of 
the fiducial correlation function with an 8-node 
spline function \emph{masking the region of the peak} 
$(80~\hMpc < r < 120~\hMpc)$. The peak term is the difference 
between the theoretical correlation function and the broadband term. 
Finally, the distortion term is calculated from simulations,  
as the difference in 
the monopole or quadrupole 
measured using the true continuum 
and that measured from fitting the 
continuum as described in appendix \ref{mocksec}. 
The three components are shown in figure \ref{threefunctions}.

Expression  (\ref{eq:detection_fit})
contains three parameters each for the monopole and quadrupole 
(so six in total). 
We have performed fits 
leaving all six parameters free and fits where we
fix the ratio $C_2/C_0$
to be equal to its nominal value used to generate
our mock spectra (the value given by assuming
a  ``redshift distortion parameter''
$\beta=1.4$, see appendix \ref{mocksec}).
We define the test statistic as the $\chi^2$ difference between fitting  
equation (\ref{eq:detection_fit}) simultaneously to monopole and 
quadrupole by fixing $C_0$ to zero (a ``peak-less'' four or five-parameter fit) 
and fitting for $C_0$ (a  five or six-parameter fit).  In 
our detection fits we do not fit for the BAO position but  fix it to the 
theoretical prediction. The distribution for this test statistic 
(``$\Delta\chi^2_\mathrm{det}$'') under the null hypothesis is a 
$\chi^2$ distribution with one degree of freedom. 
The significance is then given by $\sigma=(\Delta\chi^2_\mathrm{det})^{1/2}$.

\begin{figure}[ht]
\includegraphics[width=\columnwidth]{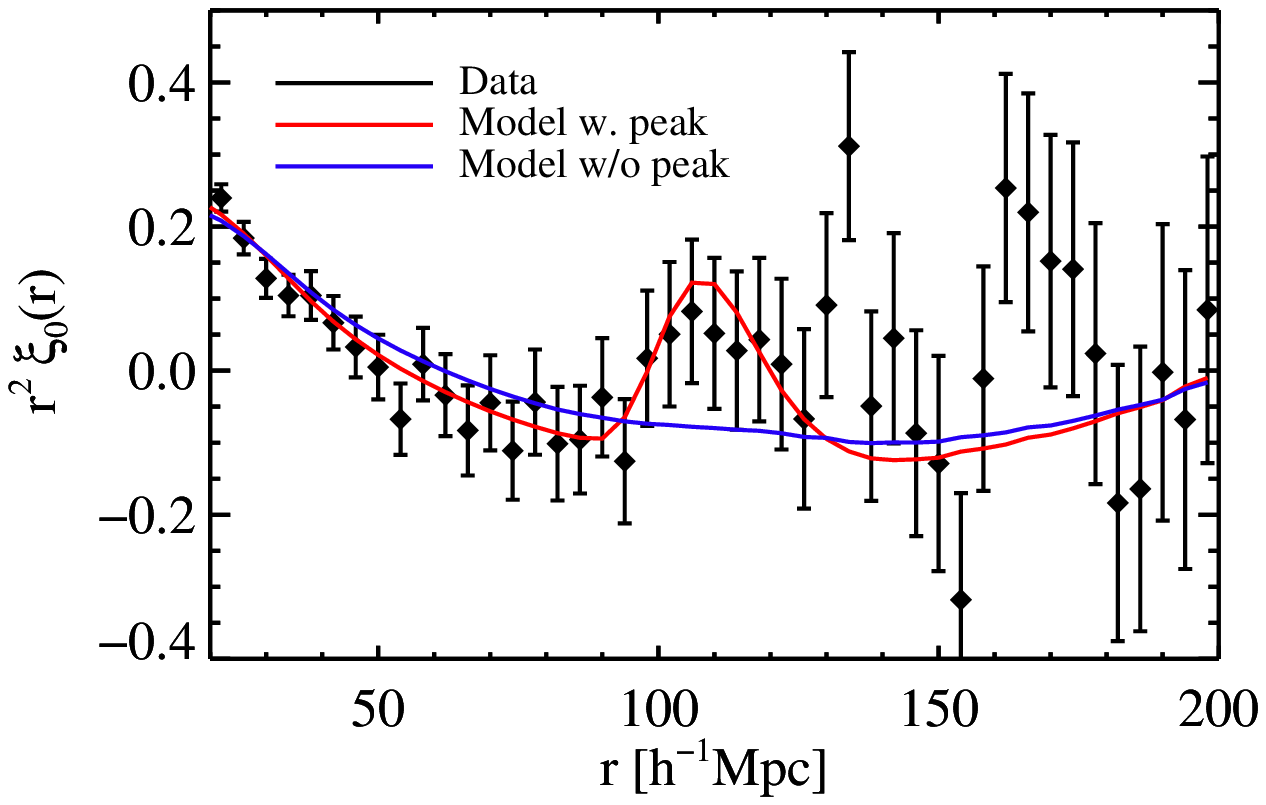}
\includegraphics[width=\columnwidth]{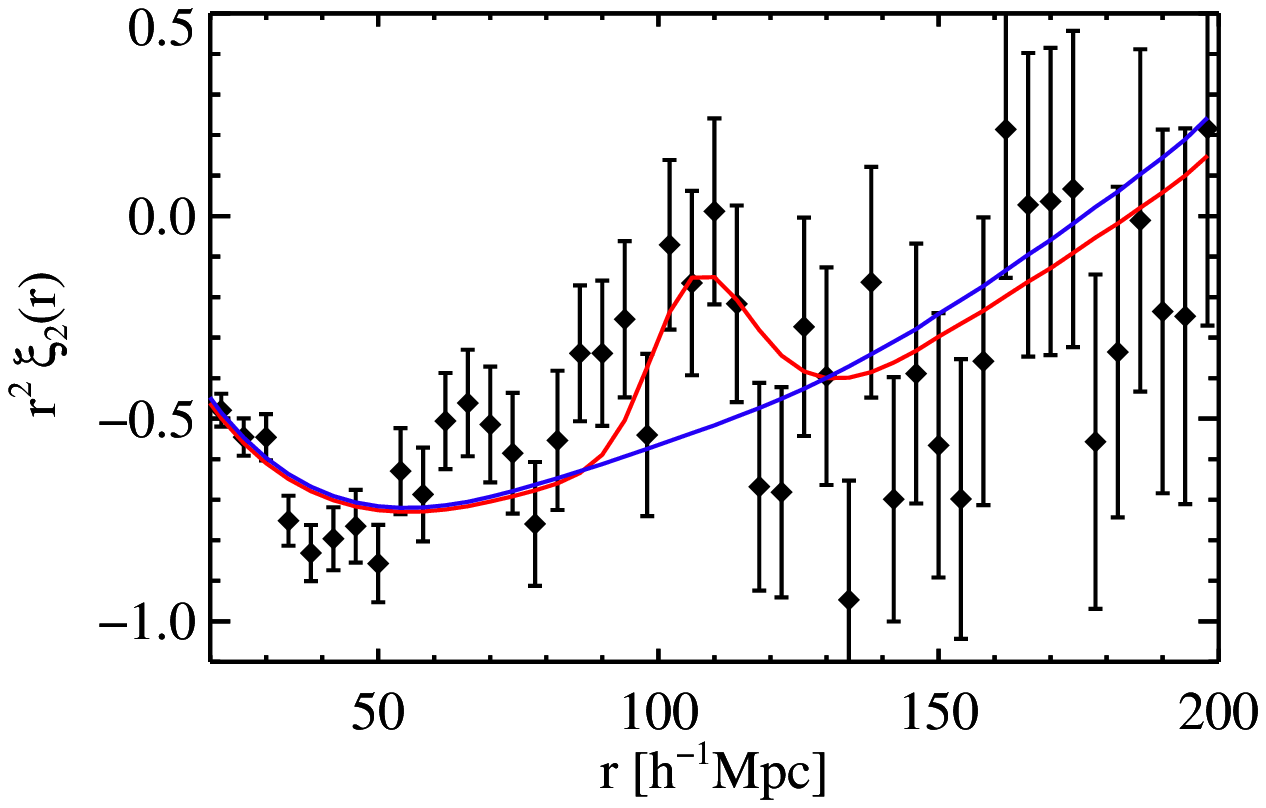}
\caption{Monopole and quadrupole fits with a BAO peak (red line)
and without a BAO peak (blue line, Method 2).
The fitting range is $r_{\rm min}<r<200~\hMpc$ with $r_{\rm min}=20~\hMpc$.
}
\label{fig:detection_plots}
\end{figure}

Figure \ref{fig:detection_plots} shows the fits to monopole (top panel) 
and quadrupole (bottom panel) and the corresponding fits with and without
peaks and fixing $\beta=1.4$. 
For method 2,
we obtain $\chi^2/DOF=93.7/85$ (111.8/86) with (without) 
a peak,  giving $\Delta\chi^2=18.1$
for a detection significance of $4.2\sigma$.
For method 1, 
we obtain $\chi^2/DOF=93.2/85$ (102.2/86)
for a significance of $3.0\sigma$.
Allowing $\beta$ to be a free parameter gives essentially the same
detection significances.

The detection significance of $\sim4\sigma$ is typical of that which
we found in the 15 sets of mock spectra.  For the mocks, the
significances ranged from 0 to $6\sigma$ with a mean of $3.5\sigma$.

\begin{figure}[ht]
\includegraphics[width=\columnwidth]{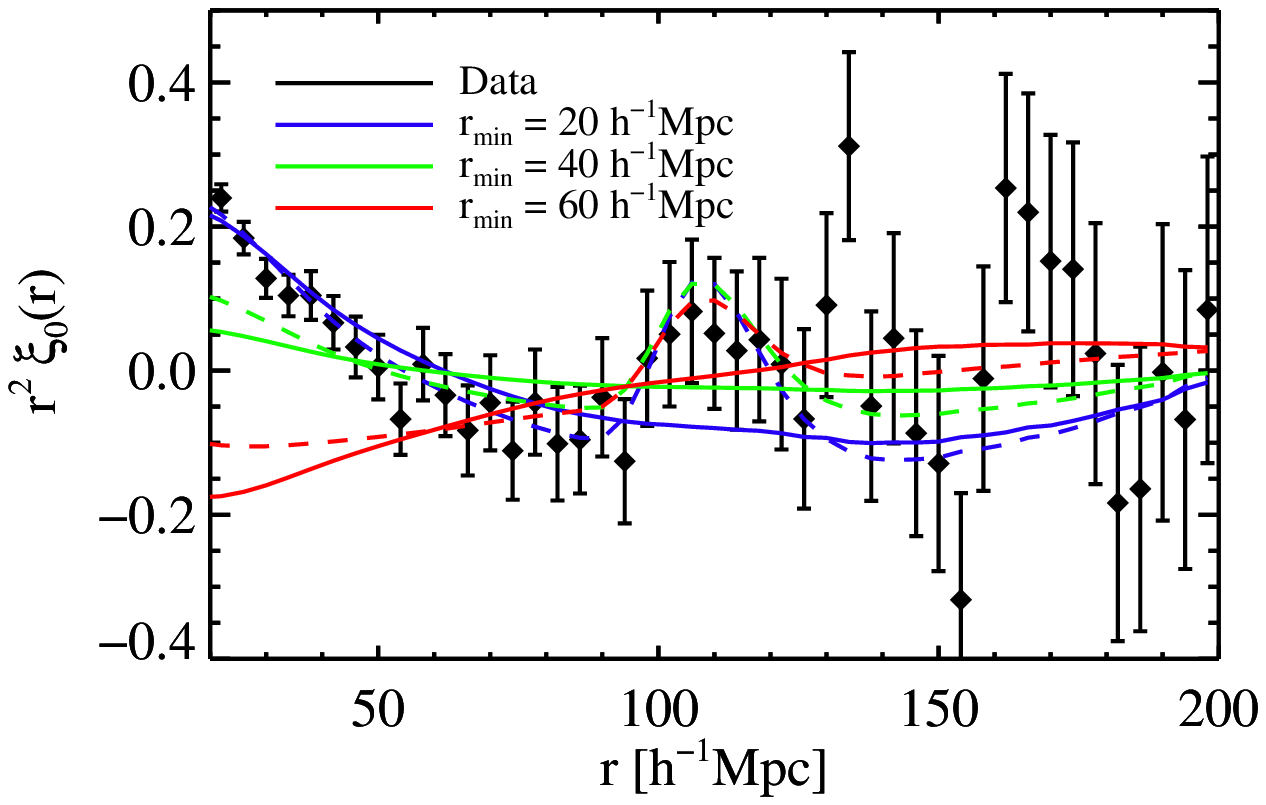}
\includegraphics[width=\columnwidth]{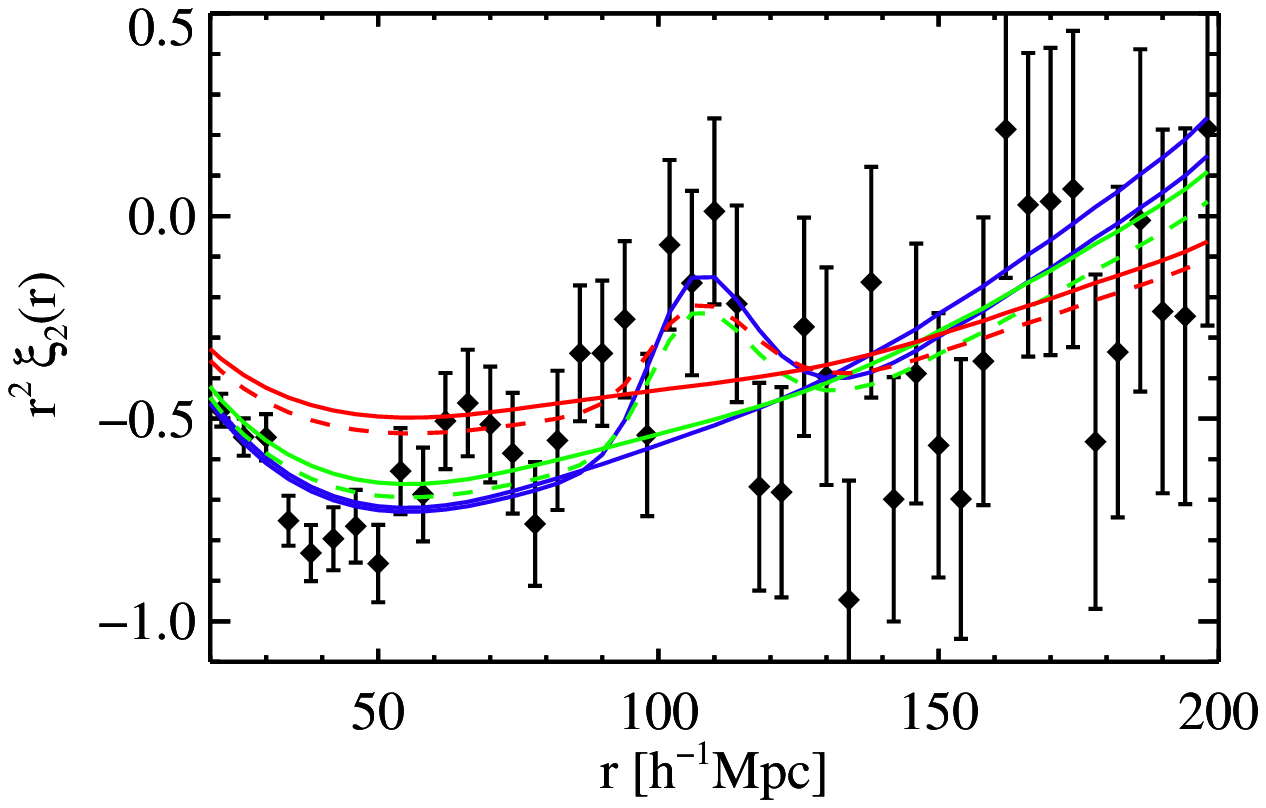}
\caption{Same as figure \ref{fig:detection_plots} except for
different fitting ranges: blue, green and red curves are for fits with 
$r_{\rm min}=20~\hMpc$,
$40~\hMpc$ and $60~\hMpc$ respectively 
(Method 2).
The solid lines for fits without a BAO peak and the dashed lines
with a peak.
}
\label{fig:detection_rcut}
\end{figure}

Our significance depends strongly on the fitting range.
For a lower boundary of the range of 
$r_\mathrm{min}=20,40, 60~\hMpc$ we obtain a significance 
of $\sigma=4.2$, 3.2 and 2.3, respectively (method 2). 
The reason for this result is illustrated 
in figure \ref{fig:detection_rcut}, where the results of the fits
with and without peaks
are compared to data for different values 
of $r_\mathrm{min}=20, 40, 60~\hMpc$. Reducing the fitting range poses 
less stringent constraints on the distortion and broadband terms,
thus allowing some of the peak to be attributed to the broadband.
In particular, the statistically insignificant bump in the 
quadrupole at $\sim65~\hMpc$ causes the fitted broadband
to increase as  $r_\mathrm{min}$ is increased to $60~\hMpc$, decreasing
the amplitude of the BAO peak.
For the monopole, the $r_\mathrm{min}=60~\hMpc$ fit predicts a
positive slope for $\xi_0(r)$ that decreases the amplitude
of the peak but predicts a
$\xi_0(r<50~\hMpc)$ to be much less than what is measured.

\section{Cosmology with the BAO peak}
\label{cosmosec}

The observed position of the BAO peak 
in $\xi(r,\mu)$ is determined by two sets of
cosmological parameters:  the ``true'' parameters
and the ``fiducial'' parameters.
Nature uses the true cosmology to create correlations
at the true sound horizon, $r_s$.  The true cosmology 
transforms physical separations between \Lya absorbers
into angles on the sky and redshift differences:
$\theta_{BAO}=r_s/\da(z)(1+z)$ and $\Delta z_{BAO}=r_sH(z)/c$.
We, on the other hand,  
use a  ``fiducial'' cosmology (defined by equation \ref{fiducialdef})
to transform angular and redshift
differences to local distances at the redshift in question to reconstruct
$\xi(r,\mu)$.
If the fiducial cosmology is the true cosmology, 
the reconstructed
peaks will be at the calculated fiducial sound horizon, $r_{s,f}$.  
Limits on the difference between the fiducial and reconstructed
peak position can be used to constrain the differences between
the fiducial and true cosmological models.

\subsection{The peak position}

The  use of  incorrect fiducial $\da$, $H$ and $r_s$
leads to shifts in the BAO peak position in the transverse and radial
directions by the multiplicative factors  $\alpha_t$ and $\alpha_r$:
\begin{equation}
\alpha_t \equiv \frac{\da(z)/r_s}{D_{A,f}/r_{s,f}}
\end{equation}
\begin{equation}
\alpha_r\equiv
\frac{H_f(z)r_{s,f}}{H(z)r_s}
\end{equation}
where the subscript $f$ refers to the fiducial model.
Following \citet{xu2012}, we will use a 
fitting function, $\tilde{\xi}_\ell(r)$,
for the monopole and quadrupole that
follows the expected peak position
as a function of $(\alpha_t,\alpha_r)$:
\begin{equation}
\tilde{\xi}_\ell (r)=
\hat{\xi}_\ell(r,\alpha_t,\alpha_r,b,\beta)
+ A_\ell(r)
\hspace*{5mm}\ell=0,2 \;.
\label{fitfunction}
\end{equation}
Here, the two functions $\hat{\xi}_\ell$,
derived from the power spectrum given in appendix \ref{mocksec}, 
describe the underlying mass correlation function, 
the linear bias $b$ and redshift distortion parameter $\beta$,  
and the movement
of the BAO peak for $\alpha_t,\alpha_r\ne1$.
The functions $A_\ell$ take into account distortions,
as described below.

For $\alpha_t=\alpha_r\equiv\alpha_{iso}$, 
there is a simple isotropic scaling of the coordinates
by $\alpha_{iso}$ and 
$\hat{\xi}(r)$ is given  by
$\hat{\xi}_\ell(r,\alpha_{iso})=f_\ell(b,\beta)\xi_{\ell,f}(\alpha_{iso} r)$,
where $\xi_{\ell,f}$
are the fiducial monopole and quadrupole and
the normalizations $f_\ell$ are the functions of
the bias and redshift-distortion parameter
given by \citet{hamilton}.
For  $\alpha_t\ne\alpha_r$, \citet{xu2012} found
an approximate
formula for $\hat{\xi}(r)$ 
that was good
in the limit $|\alpha_t-\alpha_r|\ll 1$.
We take
the more direct route of numerically expanding
$\xi_f(\alpha_tr_t,\alpha_rr_r)$ in Legendre polynomials, $P_\ell(\mu)$,
to directly calculate the $\hat{\xi}_\ell(r,\alpha_t,\alpha_r)$.

The functions $A_\ell(r)$ describe broadband distortions
due to continuum subtraction and the fact that the broadband
correlation function is not expected to change
in the same way as the BAO peak position when one deviates
from the fiducial model.
They correspond to the term
$A_\ell\xi_\ell^\mathrm{dist}$ in equation (\ref{eq:detection_fit}).
We have used two forms to represent $A_\ell(r)$:
\begin{equation}
A^{(1)}_\ell(r) = \frac{a_{\ell}}{r^2}+\frac{b_{\ell}}{r} +c_{\ell}
\label{nuissance1}
\end{equation}
and
\begin{equation}
A^{(2)}_\ell(r) = \frac{a_{\ell}}{r^2}+\frac{b_{\ell}}{r} +c_{\ell}
+\frac{d_{\ell}}{\sqrt{r}} \;.
\label{nuissance2}
\end{equation}
The observed monopole and quadrupole can then be fit to (\ref{fitfunction})
with free parameters $\alpha_t$, $\alpha_r$, bias, $\beta$,
and the nuisance parameters ($a_\ell , b_\ell , c_\ell$ and $d_\ell$).

We first fixed  $(\alpha_t,\alpha_r)=(1,1)$ to determine if we 
find reasonable values of $(b,\beta)$.
These two parameters are highly degenerate since both the quadrupole
and monopole have amplitudes that are proportional to $b^2$ times
polynomials in $\beta$. A well-determined combination is
$b(1+\beta)$, for which we find a value $0.38\pm0.07$; 
this is in agreement
with $b(1+\beta)=0.336\pm0.012$ found at $r\sim40~\hMpc$
by \citet{Slosar}.
The larger error of our fit reflects the substantial freedom
we have introduced with our distortion function.

\begin{figure}
\includegraphics[width=\columnwidth]{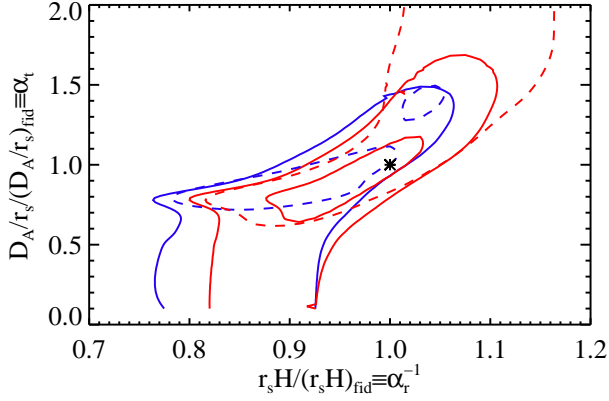}
\caption{
The contours for $(\da/r_s, r_sH)$
obtained by fitting the monopole and quadrupole to 
(\ref{fitfunction}).
The
broadband distortions are eqn. (\ref{nuissance1}, dashed lines)
or  (\ref{nuissance2}, solid lines).
The blue lines are for method 1 and the red lines for method 2.
All contours are for $\Delta\chi^2=4$  except for the interior solid red
contour which is for $\Delta\chi^2=1$.
}
\label{DAHcontours}
\end{figure}

We next freed all parameters to constrain $(\alpha_t,\alpha_r)$.
The contours for the two methods and two broadbands
are shown in figure \ref{DAHcontours} and the
$\chi^2$ for the fiducial and best-fit models 
are given  in table \ref{dlytable}.
The broadband term in equation (\ref{nuissance2}) fits the data better than
that in equation (\ref{nuissance1}) both for the fiducial parameters
and for the best fit.  
For broadband in equation (\ref{nuissance2}), the
$\chi^2$ for the fiducial model is acceptable for 
both methods:  
$\chi^2/DOF=85.0/80$ for method 1 and
$\chi^2/DOF=71.5/80$ for method 2.

The contours in the figure are elongated 
along the direction for which the BAO peak position stays
approximately fixed at large $\mu$ (near the radial direction, where the
observations are most sensitive). 
The best constrained combination of $\da$ and $H$  
of the form $(\da^\zeta H^{\zeta-1}/r_s)$
turns out to have $\zeta\sim0.2$.
This low value of $\zeta$ reflects the fact that
we are mostly sensitive to the BAO peak in the radial direction. 
At the one standard-deviation level, the precision on
this combination is about 4\%.
However, even this combination is sensitive to the
tails in the contours.
A more robust indicator of the statistical accuracy
of the peak-position determination comes from fits
imposing $\alpha_t=\alpha_r\equiv\alpha_{iso}$, as has
generally been done in  previous BAO studies with the
exception of \citet{chuang2012} and \citet{xu2012}.
This constraint does not correspond to any particular
class of cosmological models.
It does however eliminate the tails in the contours
in a way that is similar to the imposition of outside
data sets.
The two methods and broadbands give consistent
results, as seen in table \ref{dlytable}.

\begin{table*}
\begin{center}
\caption{Results with the
the two methods and two broadbands (equations \ref{nuissance1} and
\ref{nuissance2}).
Columns 2 and 3 give the $\chi^2$ for the fiducial model and for the
model with the minimum $\chi^2$.
Column 4 gives the best fit for $\alpha_{iso}$ with the constraint
($\alpha_t=\alpha_r\equiv\alpha_{iso}$).
Column 5 gives $Hr_s/[Hr_s]_{fid}$ with the $2\sigma$ limits in parentheses.  
Column 6 gives the $Hr_s/[Hr_s]_{fid}$  
deduced by combining our data with that of WMAP7 \citep{wmap7ref}
(see section \ref{hofzsec}).
}
\begin{tabular}{l|c|c|c|c|c}
 method \& & 
$\chi^2_{fid}/DOF$   &  
$\chi^2_{min}/DOF$   & 
$\alpha_{iso}$  &
$Hr_s/[Hr_s]_{fid}$  &
$Hr_s/[Hr_s]_{fid}$  \\
broadband & & & & & (with WMAP7) \\
\hline \hline 
& & & & & \\[-2mm]
Method 1 (\ref{nuissance2})  & 
85.0/80  & 
84.6/78 & 
$1.035\pm0.035$  &
$0.876\pm0.049$ $(^{+0.188}_{-0.111})$ &
$ 0.983\pm 0.035 $     \\ [2mm]
Method 2 (\ref{nuissance2})  & 
71.5/80  & 
71.4/78 & 
$1.010\pm 0.025$    &
$0.954\pm0.077$ $(^{+0.152}_{-0.154})$ &
$ 1.000\pm 0.036 $    \\ [2mm]
Method 1 (\ref{nuissance1})  & 
104.3/82 & 
99.9/80 & 
$1.027\pm0.031$     &
$0.869\pm0.044$ $(^{+0.185}_{-0.084})$ &
$ 0.988 \pm 0.034 $    \\ [2mm]
Method 2 (\ref{nuissance1})  & 
88.4/82  & 
87.7/80 & 
$1.004\pm0.024$    &
$0.994\pm0.111$ $(^{+0.166}_{ -0.178})$ &
$ 1.006\pm0.032 $    
\label{dlytable}
\end{tabular}
\end{center}
\end{table*}%

\begin{figure}
\includegraphics[width=\columnwidth]{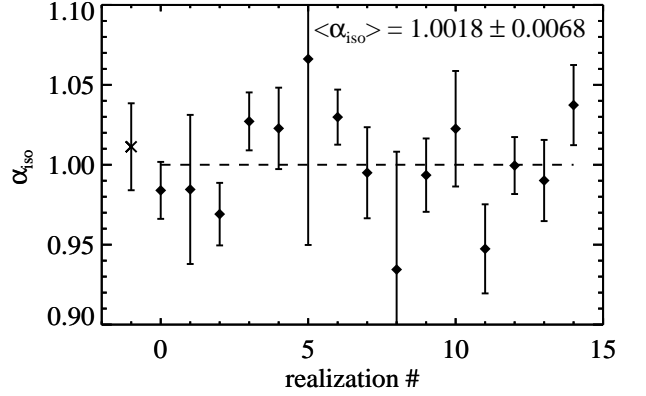}
\caption{
The measurements of 
$\alpha_{iso}$  ($=\alpha_t=\alpha_r$)
for
the 15 sets of mock spectra and for the data (realization=-1).
The large errors for realization 5 and 8 are due to the very
low significance of the BAO peak found on these two sets.
}
\label{alphamagicmocks}
\end{figure}

We used the sets of mock spectra to search for biases
in our measurement of  
$\alpha_{iso}$.
The mean value reconstructed for this quantity on individual
mocks is $1.002\pm0.007$,
suggesting
that there are no significant biases in the determination
of the BAO scale.
Figure \ref{alphamagicmocks} shows the
values and errors for the individual mocks along with that
for the data.
Both the measured value and its uncertainty for the data is
typical of that found for individual sets of mock spectra.

\begin{figure}
\includegraphics[width=\columnwidth]{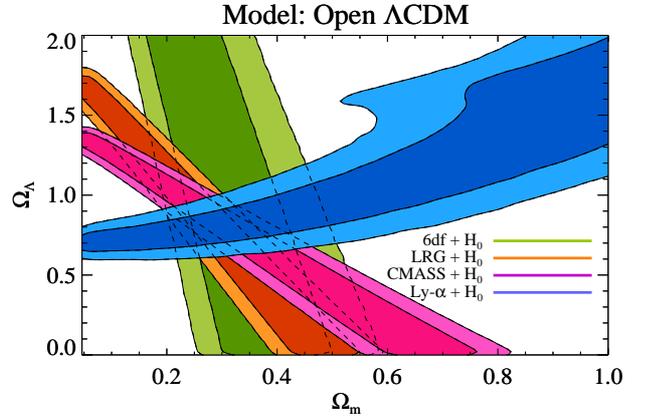}
\caption{
Constraints on the matter and dark-energy density parameters
$( \om , \ol )$ assuming a dark-energy pressure-density
ratio $w=-1$.
The blue regions are the one and two standard deviation
constraints derived from our
contours in figure \ref{DAHcontours} 
(method 2, broadband \ref{nuissance2})
combined with a measurement of 
$H_0$ \citep{riess2011}.
Also shown are one and two standard deviation
contours from lower redshift measurements
of $\dv/r_s$ (also combined with $H_0$) at 
$z=0.11$ [6dF: \citet{6df}], 
$z=0.35$ [LRG: \citet{percival2010}] and
$z=0.57$ [CMASS: \citet{bossgal}].
All constraints use a WMAP7 \citep{wmap7ref} 
prior on the baryon-to-photon ratio
$\eta$ but do not otherwise incorporate CMB results.
}
\label{omolfig}
\end{figure}

\begin{figure}[ht]
\includegraphics[width=\columnwidth]{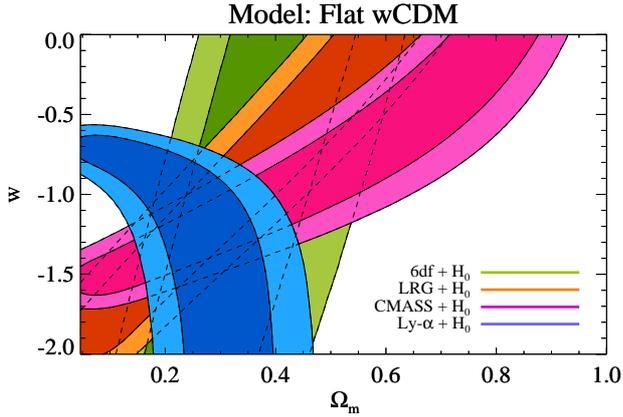}
\caption{As in figure \ref{omolfig} with constraints on
the matter density parameter, $\om$, and dark-energy pressure-density
ratio $w$ assuming $\om+\ol=1$.
}
\label{omwfig}
\end{figure}

\subsection{Constraints on cosmological models}

Our constraints on $(\da/r_s,Hr_s)$ can be used to constrain
the cosmological parameters.
In a $\Lambda$CDM cosmology,
apart from the pre-factors of $H_0$ that cancel, 
$\da/r_s$ and $Hr_s$
evaluated at $z=2.3$
depend primarily on $\om$ through $r_s$ and on $\ol$ which, with $\om$,
determines $\da$ and $H$.
The sound horizon
also depends on $H_0$ 
(required to derive $\Omega_\gamma$ from $T_{CMB}$), 
on the effective number of neutrino species $N_\nu$ 
(required to derive the radiation density from the photon density),
and on the baryon-to-photon number ratio, $\eta$
(required for the speed of sound). 

Figure \ref{omolfig} shows the $\Lambda$CDM constraints
on $(\om ,\ol)$  derived from 
the contours in figure \ref{DAHcontours} 
combined with the most recent measurement
of $H_0$ \citep{riess2011}.
We use the contours for method 2 and the broadband of 
equation \ref{nuissance2} which gives better fits to the data  than
the other method and broadband.
The contours also assume $N_\nu=3$ 
and the WMAP7 value of $\eta$ \citep{wmap7ref}.
Also shown are constraints from BAO measurements of $\dv/r_s$
\citep{percival2010,bossgal,6df}.

The \Lya contours are nicely orthogonal to the
lower redshift $\dv/r_s$ measurements, reinforcing the requirement
of dark energy from BAO data.
In fact, our measurement is the only BAO measurement that
by itself requires dark energy: $\ol>0.5$.
This is because at $z=2.3$ the universe is strongly matter 
dominated and the $\sqrt{\om}$ factor in $H$ partially cancels
the $1/\sqrt{\om}$ in $r_s$, enhancing the importance
of the $\ol$ dependence of $H$.

Figure \ref{omwfig} shows the constraints
on $(\om ,w$; where $w$ is the dark-energy pressure-density ratio)
assuming a flat universe: $\Omega_k=0$.
Our result is the only BAO measurement that by itself
requires negative $w$.  Our limit $w<-0.6$ requires
matter domination at $z=2.3$.
\begin{equation}
\frac{\rho_{de}(z=2.3)}{\rho_m(z=2.3)} < 0.3 \left(
\frac{\ol/\om}{0.73/0.27} 
\right) \;.
\label{wlimit}
\end{equation}

\subsection{Constraints on $H(z)$}
\label{hofzsec}
 
The contours in figure \ref{DAHcontours} 
give the measurements
of $Hr_s$ given in table \ref{dlytable}.
A measurement of the expansion rate deep in the matter-dominated
epoch can be used to demonstrate the deceleration of the expansion
at that time.
Unfortunately, our data are not yet precise enough to do this.
To make a
more precise measurement of $H(z=2.3)$, 
we must add further constraints to eliminate the long
tails in figure \ref{DAHcontours}.
These tails correspond to models where
$1/H(z=2.3)$ is increased (resp. decreased) with respect to the fiducial value
while $\da(z=2.3)$ is decreased  (resp. increased).
For flat models, this would imply a change in the mean of $1/H$
(averaged up to $z=2.3$) that is opposite to that of the change in
$1/H(z=2.3)$, which requires  a functional form $H(z)$ 
that strongly differs from the fiducial case.
It is possible to construct models with this property by introducing
significant non-zero curvature.

\begin{figure}
\includegraphics[width=\columnwidth]{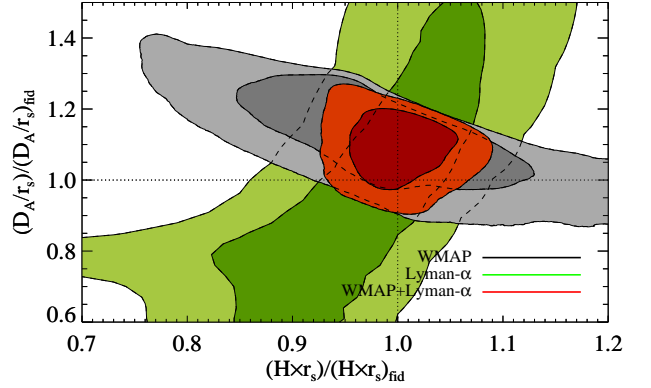}
\caption{
Constraints on  $(\da/r_s, r_sH)_{z=2.3}$ within the framework
of OwOwaCDM models.
The green contours are our $1\sigma$ and $2\sigma$
constraints using method 2 and broadband
(\ref{nuissance2}).
The gray contours are the  $1\sigma$ and $2\sigma$ constraints from
WMAP7 \citep{wmap7ref}.
The red contours show the combined constraints.
}
\label{DAHcontourscombo}
\end{figure}

Because of the importance of curvature, the tails are
eliminated once WMAP7 constraints \citep{wmap7ref} are included.
This is done in figure \ref{DAHcontourscombo} within the
framework of non-flat models
where the dark-energy pressure-density ratio, $w(z)$, is
determined by two parameters, $w_0$ and $w_a$: $w(z)=w_0+w_az/(1+z)$.
As expected, the WMAP7 results in this framework constrain $\da$ and $1/H$
to migrate in roughly the same direction as one moves away from
the fiducial model.
Combining WMAP7 constraints with ours 
gives the values of $H(z=2.3)r_s$ given in the last column
of table \ref{dlytable}.
For what follows, we adopt the mean of methods 1 and 2 that use
the more flexible broadband of equation (\ref{nuissance2}):
\begin{equation}
\frac{H(z=2.3)r_s}{[H(z=2.3)r_s]_{fid}}= 0.992\pm0.035 \;.
\end{equation}
The precision on $H$ is now  sufficient to study the
redshift evolution of $H(z)$.

The fiducial model has
$r_{s}=152.76~$Mpc 
and $H(z=2.3)=3.23H_0$, $H_0=70 {\rm km\,s^{-1}Mpc^{-1}}$.
These results produce
\begin{equation}
\frac{H(z=2.3)r_s}{1+z}= (1.036\pm0.036)\times 10^4~{\rm km\,s^{-1}} \;.
\end{equation}
or equivalently
\begin{equation}
\frac{H(z=2.3)}{1+z}= (67.8\pm2.4){\rm km\,s^{-1}Mpc^{-1}} 
\,\left(\frac{152.76~{\rm Mpc}}{r_s}\right) \;.
\end{equation}

\begin{table}
\begin{center}
\caption{ Recent measurements of $H(z)/(1+z)$.
The BAO-based measurements use $r_s=152.76~{\rm Mpc}$ as
the standard of length and are shown as the filled circles
in figure \ref{adotfig}.  The quoted uncertainties in $H(z)$ do not include
uncertainties in $r_s$ which are expected to be negligible 
,$\approx 1\%$ \citep{wmap7ref}.
The measurements of \citet{wigglezsn} use supernova
data and therefore measure $H(z)$ relative to $H_0$.
We quote the results they obtain without assuming
a flat universe and plot them as the open green circles
in figure \ref{adotfig} assuming $h=0.7$.
}
\begin{tabular}{l|c|c|c}
 $z$   &  $H(z)/(1+z)$   &method & reference \\
    &  ${\rm km\,s^{-1}Mpc^{-1}}$ &   &  \\
\hline \hline
2.3 & $66.5\pm7.4$ & BAO  &this work \\
2.3 & $67.8\pm2.4$ & BAO+WMAP7 &this work \\
0.35 & $60.8\pm3.6$ & BAO &\citet{chuang2012} \\
0.35 & $62.5\pm5.2$ & BAO &\citet{xu2012} \\
0.57 & $58.8\pm2.9$ & BAO + AP & \citet{reid2012} \\
0.44 & $57.4\pm5.4$ & BAO + AP &\citet{wigglezhz} \\
0.60 & $54.9\pm3.8$ & &\\
0.73 & $56.2\pm4.0$ & &\\
& & & \\
0.2 & $(1.11\pm0.17)H_0$ & AP + SN &\citet{wigglezsn} \\
0.4 & $(0.83\pm0.13)H_0$ & &\\
0.6 & $(0.81\pm0.08)H_0$ & &\\
0.8 & $(0.83\pm0.10)H_0$ & &\\
& &   & \\
0 & $73.8\pm2.5$ & &\citet{riess2011} \\
\label{hofztable}
\end{tabular}
\end{center}
\end{table}%

\begin{figure}[ht]
\includegraphics[width=\columnwidth]{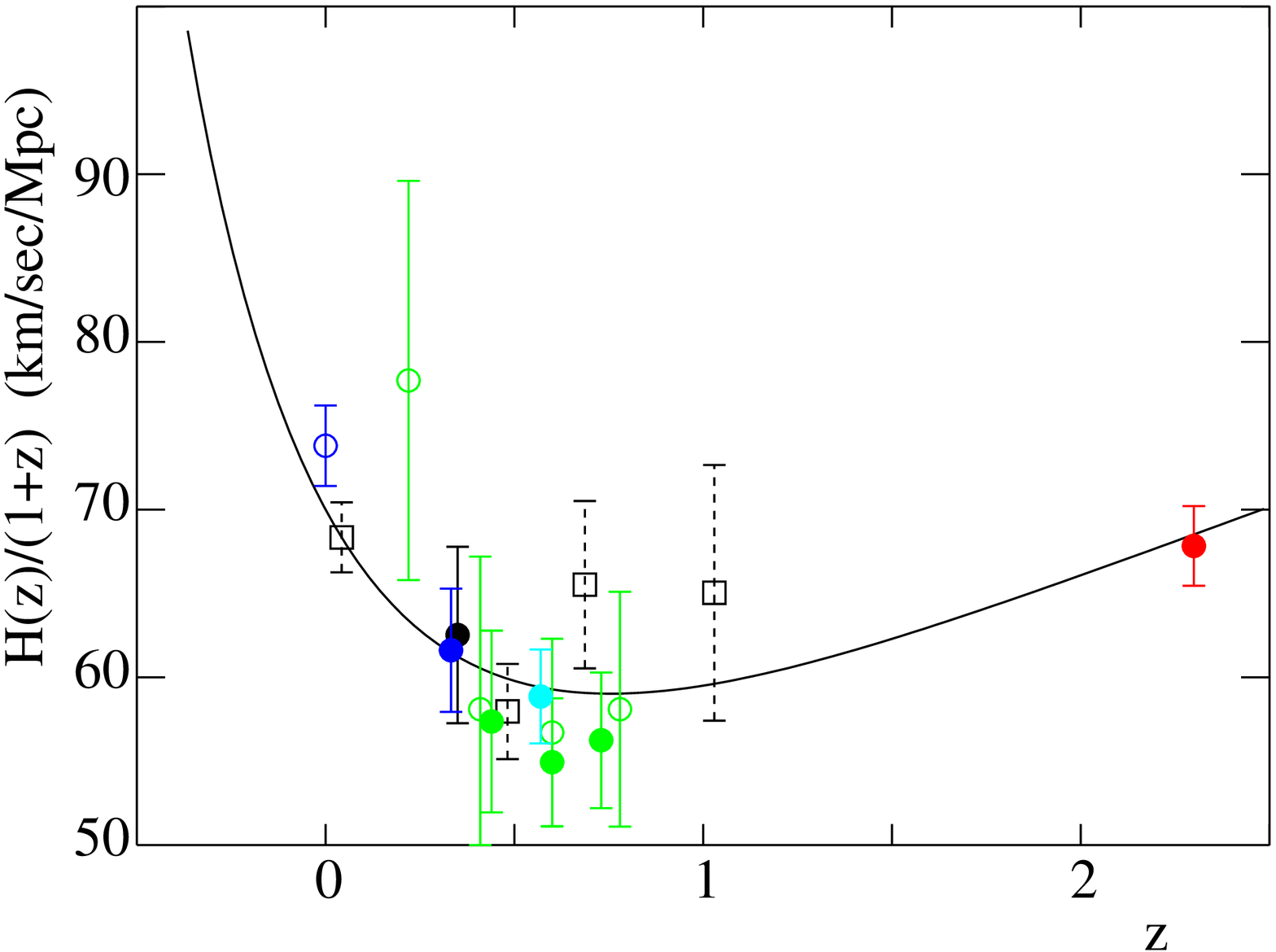}
\caption{
Measurements of $H(z)/(1+z)\;vs\;z$ demonstrating
the acceleration of the expansion for $z<0.8$ and deceleration
for $z>0.8$.
The BAO-based measurements are the filled circles:
[this work: red],
[\citet{xu2012}: black]
[\citet{chuang2012}: blue],
[\citet{reid2012}, cyan],
and  
[\citet{wigglezhz}: green].
The open green circles are from WiggleZ \citep{wigglezsn}
Alcock-Paczynski data 
combined with supernova data yielding $H(z)/H_0$ (without
the flatness assumption) plotted here assuming
$H_0=70{\rm km\,s^{-1}Mpc^{-1}}$.
The open blue circle is the $H_0$ measurement of \citet{riess2011}. 
The open black squares with dashed error bars show
the results of \citet{riesshofz} which were derived by differentiating
the SNIa Hubble diagram and assuming spatial flatness.
(For visual clarity, the \citet{riesshofz} point at $z=0.43$ has been
shifted to $z=0.48$.)
The line is the $\Lambda$CDM prediction for $(h,\om ,\ol)$
$=(0.7,0.27,0,73)$.
}
\label{adotfig}
\end{figure}

This number can be compared with the measurements 
of $H(z)$ at lower redshift
shown  in table \ref{hofztable} and figure \ref{adotfig}.
Other than those of $H_0$, the measurements that we use
can be divided into two classes:  those (like ours) that use
$r_s$ as the standard of length and those that use $c/H_0$ as the
standard of length.

The comparison with our measurement is simplest 
with BAO-based
measurements that use $r_s$ 
as the standard of length and therefore measure $H(z)r_s$ 
(as is done here).   
The first attempt at such a measurement was made by \cite{gaztanaga09},
a result debated in subsequent papers by \cite{me09},
\cite{yoome10}, \cite{kazin10}, and \cite{cabre11}.
Here, we use four more recent measurements.
\citet{chuang2012} and \citet{xu2012} studied the SDSS DR7 
LRG sample and decomposed
the BAO peak into radial and angular components, thus extracting
directly $Hr_s$ and $\da/r_s$.
\citet{wigglezhz} and \citet{reid2012} took a more indirect route. 
They first used the angle-averaged
peak position to derive $\dv(z)/r_s=((1+z)^2\da^2czH^{-1}/r_s$.
They then studied the Alcock-Paczynski effect on the broadband
galaxy correlation function to determine
$\da(z)H(z)$.  Combining the two measurements yielded $H(z)r_s$.

It is evident from comparing our $H(z)$ measurement (filled red circle
in figure \ref{adotfig}) 
to the other BAO-based measurements (other filled circles)
that $H(z)/(1+z)$ decreases between $z=2.3$ and $z=0.35-0.8$.
To demonstrate deceleration quantitatively, we fit the eight
BAO-based values of $H(z)$ in table 2 
to the $o\Lambda$CDM form 
$H(z)=H_0(\ol + \om(1+z)^3 + (1-\ol-\om)(1+z)^2)^{1/2}$.
Marginalizing over $\ol$ and $H_0$ we find
\begin{equation}
\frac{[H(z)/(1+z)]_{z=2.3}}{[H(z)/(1+z)]_{z=0.5}}
= 1.17\pm0.05 \;,
\end{equation}
clearly indicating deceleration between $z=2.3$ and $z=0.5$.
This measurement is in good agreement with the fiducial value of 1.146.
We emphasize that this result is independent of $r_s$,
assuming only that the BAO-peak position
is redshift-independent in comoving coordinates.
The result also does not assume spatial flatness.

To map the expansion rate 
over the full range
$0<z<2.3$,
we must adopt the fiducial value of $r_s$ and compare
the resulting $H(z)$ with $H_0$ and with other BAO-free
measurements.
Besides the $H_0$ measurement of \citet{riess2011}, we use
the WiggleZ analysis combining their Alcock-Paczynski data
with distant supernova data from the Union-2 compilation
\citep{union2ref}.
The supernova analysis does not use the poorly known mean
SNIa luminosity, so the SNIa Hubble diagram gives 
the luminosity distance in units of $H_0^{-1}$, $\dl(z)H_0$.
Combining this result with the Alcock-Paczynski measurement of $\da(z)H(z)$
yields $H(z)/H_0$.  The values are given in table \ref{hofztable}.

We fit all the data in table \ref{hofztable} 
(filled and open circles in figure \ref{adotfig})
to the $\Lambda$CDM
form of $H(z)$.
This yields an estimate of  the redshift of minimum $H(z)/(1+z)$ 
\begin{equation}
z_{d-a}= 0.82 \pm 0.08
\end{equation}
which compares well with the fiducial value:
$z_{d-a}=(2\ol/\om )^{1/3}-1=0.755$.

In this analysis, we have not used two other sources 
of information on $H(z)$ at high redshift.
The first use
high-redshift type Ia supernovae to probe the era where the 
universe transitions
from deceleration to acceleration (e.g., \citet{riess2004,riesshofz}).
The data of \citet{riesshofz} (plotted as the open squares
in figure \ref{adotfig})) yielded useful measurements up to $z\sim1.1$.  
However, this data yields
constraints on $H(z)$ 
that are weaker than those of BAO-based methods
because of the need to differentiate the
distance-redshift relation.  
Moreover, these inferences of $H(z)$ assume spatial flatness.
Fitting the SNe data to a model 
with an evolving deceleration parameter $q(z) = q_0+(dq/dz)_0 z$ and assuming
flatness,
\citet{riesshofz} and \citet{riess2004}
were able to demonstrate that $(dq/dz)_0 > 0$, i.e.\ a negative
3rd-derivative of $a(t)$.  However, we point out that in a more general
$q(z)$ model, the demonstration that $dq/dz>0$ at low redshift is not
equivalent to a demonstration that $\ddot{a}$
becomes negative in the past.

Another approach to
determining $H(z)$ uses the evolution of stellar populations as
a clock to infer $dt/dz$ \citep{stellarclocks,moresco12}.
This method yields results that are consistent with $\Lambda CDM$ expectations,
but the uncertainties (statistical and systematic) are larger than
those of the determinations in Table 2, so we have not plotted
them in figure \ref{adotfig}.

\section{Conclusions}
\label{conclusionssec}
  
In this paper, we have presented the first observation
of the BAO peak using the \Lya forest.
It represents both the first BAO detection 
deep in the matter dominated epoch and
the first to use a tracer of mass that is not galactic.
The results are consistent with concordance $\Lambda$CDM,
and require, by themselves, the existence of dark energy.
Combined with CMB constraints, we deduce the expansion
rate at $z=2.3$ and demonstrate directly the sequence
of deceleration and acceleration expected in dark-energy
dominated cosmologies.
These results have been confirmed with higher precision by \citet{method3}
using the same underlying DR9 data set but more aggressive data cuts
and a more nearly optimal statistical method.

    BOSS continues to acquire data and will eventually produce a quasar
sample three times larger than DR9.  
We can thus expect improved precision in our measurements of 
distances and expansion rates, leading to improved constraints 
on cosmological parameters.  The \Lya forest
may well be the most practical method for obtaining precise
$D_A(z)$ and $H(z)$ measurements at $z>2$, thanks to the large number
of independent density measurements per quasar.  It is reassuring
that the first sample large enough to yield a detection of 
BAO produces a signal in good agreement with expectations.  In the
context of BAO dark energy constraints, high redshift measurements
are especially valuable for breaking the degeneracy between 
curvature and the equation of state history  More generally,
however, by probing an epoch largely inaccessible to other methods,
BAO in the \Lya forest have the potential to reveal surprises,
which could provide critical insights into the origin of cosmic acceleration.

\begin{acknowledgements}

We thank Carlos Allende Prieto, Ashley Ross and Uros Seljak 
for stimulating
discussions and Adam Riess for providing the data points of
\citet{riesshofz} in figure \ref{adotfig}.

Funding for SDSS-III has been provided by the Alfred P. Sloan
Foundation, the Participating Institutions, the National Science
Foundation, and the U.S. Department of Energy Office of Science.
The SDSS-III web site is http://www.sdss3.org/.

The French Participation Group of SDSS-III was supported by the
Agence Nationale de la
Recherche under contract ANR-08-BLAN-0222.

The research leading to these results has received funding 
from the European Union Seventh Framework Programme (FP7/2007-2013) 
under grant agreement no. [PIIF-GA-2011-301665].

SDSS-III is managed by the Astrophysical Research Consortium for the
Participating Institutions of the SDSS-III Collaboration including the
University of Arizona,
the Brazilian Participation Group,
Brookhaven National Laboratory,
University of Cambridge,
Carnegie Mellon University,
University of Florida,
the French Participation Group,
the German Participation Group,
Harvard University,
the Instituto de Astrofisica de Canarias,
the Michigan State/Notre Dame/JINA Participation Group,
Johns Hopkins University,
Lawrence Berkeley National Laboratory,
Max Planck Institute for Astrophysics,
Max Planck Institute for Extraterrestrial Physics,
New Mexico State University,
New York University,
Ohio State University,
Pennsylvania State University,
University of Portsmouth,
Princeton University,
the Spanish Participation Group,
University of Tokyo,
University of Utah,
Vanderbilt University,
University of Virginia
University of Washington,
and Yale University.

\end{acknowledgements}

\appendix

\section{Mock quasar spectra}
\label{mocksec}

We have produced mock spectra in order to tune the
analysis procedure and to study statistical uncertainties
and systematic 
effects in the measured correlation function.

In some galaxy clustering studies (e.g. \citet{bossgal}) 
the covariance matrix of the measured correlation function is obtained from
mock data sets. In this case, it is crucial to have very realistic mocks with
the right statistics.

In order to do so, we would need to generate several realizations of 
hydrodynamical simulations, with a large enough box to cover the whole survey
(several $\rm Gpc^3$) and at the same time have a good enough resolution to 
resolve the Jeans mass of the gas (tenths of $\rm kpc$). This type of 
simulations are not possible to generate with current technology, but luckily
in this study the covariance matrix is obtained from the data itself, and 
the mock data sets are only used to test our analysis and to study possible
systematic effects.

In the last few years there have been several methods proposed to generate
simplified mock Lyman-$\alpha$ surveys by combining Gaussian fields and 
nonlinear transformations of the field 
\citep{legoff,greig,mockref}.
In this study we used a set of mocks generated using
the process described in \citet{mockref}, the same method used in the 
first publication of the Lyman-$\alpha$ correlation function from BOSS 
\citep{Slosar}.

The mock quasars were generated at the angular positions and
redshifts of the BOSS quasars.  The unabsorbed spectra (continua)
of the quasars were generated using the Principal
Component Analysis (PCA) eigenspectra of \citet{qsopca},
with amplitudes for each eigenspectrum randomly drawn from Gaussian
distributions with sigma equal to the corresponding eigenvalues as published
in  \citet{suzuki06}
table 1.
A detailed description will be provided by \citet{expandedmocks},
accompanying a public release of the mock catalogs.

We generated the field of transmitted flux fraction, $F$, that have
a $\Lambda$CDM power spectrum 
with the fiducial parameters
\begin{displaymath}
(\om ,\Omega_{\Lambda},\Omega_bh^2,h,\sigma_8,n_s)_{\rm fid}
\end{displaymath}
\begin{equation}
\hspace*{20mm}=(0.27,0.73,0.0227,0.7,0.8,0.97)
\label{fiducialdef}
\end{equation}
where $h=H_0/100~{\rm km\,s^{-1}Mpc^{-1}}$.
These values produce a fiducial sound horizon of
\begin{equation}
r_{s,fid}=152.76~{\rm Mpc} \;.
\end{equation}

Here, we use the parametrized fitting formula introduced by
\citet{macdo2003} to fit the results of the power spectrum
from several numerical simulations,
\begin{equation}
 P_F(k,\mu_k) = b_{\delta}^2 (1+\beta \mu_k^2)^2 P_L(k) D_F(k,\mu_k) ~,
\label{ips}
\end{equation}
where $\mu_k=k_{\parallel}/k$ is the cosine of the angle between \vec{k}
and the line of sight,
$b_{\delta}$ is the density bias parameter,
$\beta$ is the
redshift distortion parameter, 
$P_L(k)$ is the
linear matter power spectrum, 
and $D_F(k,\mu_k)$ is a non-linear term
that approaches unity at small $k$. This form of $P_F$ is the expected
one at small $k$ in linear theory, and provides a good fit to the
3-D \Lya observations reported in \citet{Slosar}.
We do not generate a density and a velocity field, but 
directly create the \Lya forest absorption field instead, with the
redshift distortions being directly introduced in the input power
spectrum model of equation (\ref{ips}), with the parameter $\beta$
that measures the strength of the redshift distortion.

To model the evolution of the forest with redshift, 
$b_\delta$ varies with redshift according to 
$b_\delta=0.14[(1+z)/3.25]^{1.9}$ \citep{macdo2006}. 
The redshift distortion parameter is given a fixed value of 
$\beta_F=1.4$. The non-linear correction factor 
$D(k,\mu_k)$ is taken from \cite{macdo2003}.
The flux field was constructed 
by generating
Gaussian random fields $g$ with an appropriately chosen 
power spectrum \citep{mockref}
to which  the log-normal transformation $F = \exp(-a e^{\upsilon g})$
is applied 
\citep{lognormal,bi92,gnedinhui96}.
Here $a$ and $\upsilon$ are free parameters chosen to
reproduce the flux variance and mean transmitted flux fraction
\citep{macdo2006}.

DLA's were added to the spectra according to the 
procedure described in \citet{dlasmocks}.

Finally, the spectra were modified to include the effects of 
the BOSS spectrograph point spread function (PSF), 
readout noise, photon noise, and flux miscalibration.

Fifteen independent realizations of the BOSS data were 
produced and analyzed with the same procedures as those
for the real data.

\begin{figure}[ht]
\includegraphics[width=\columnwidth]{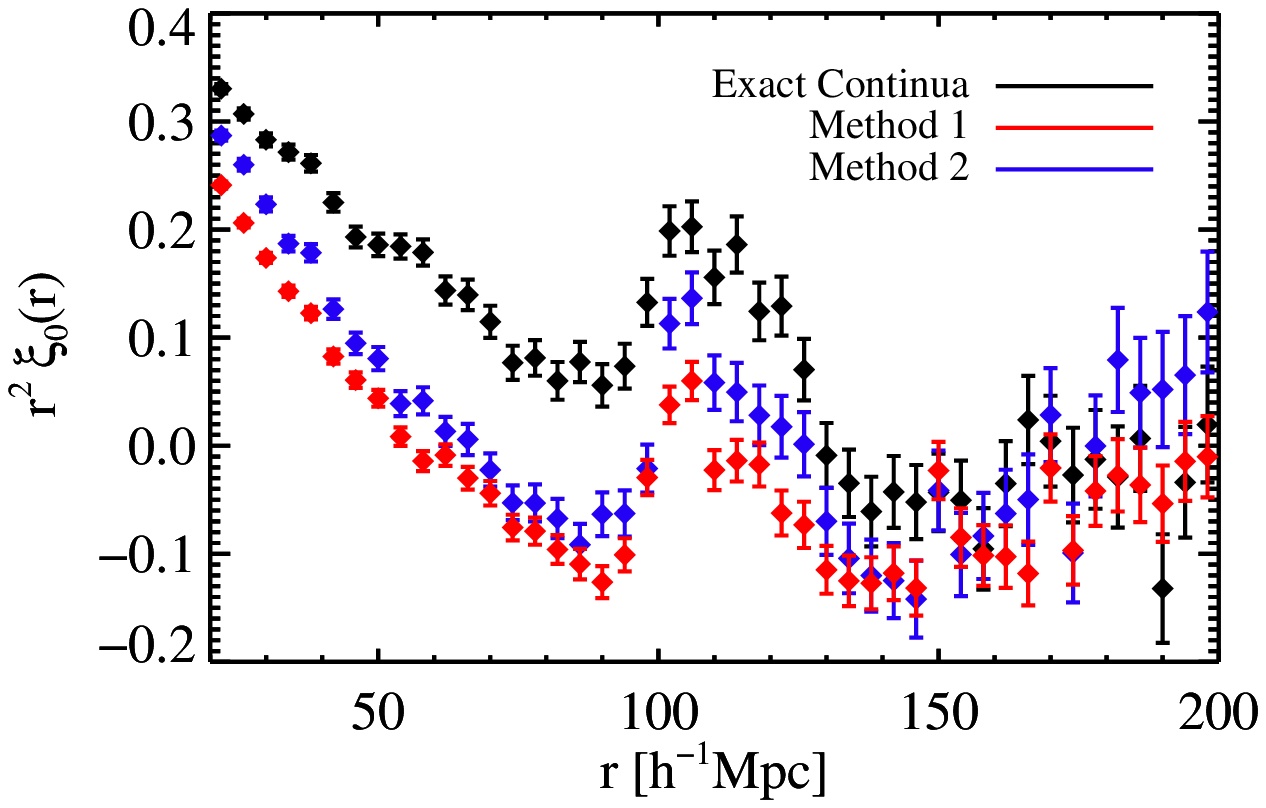}
\includegraphics[width=\columnwidth]{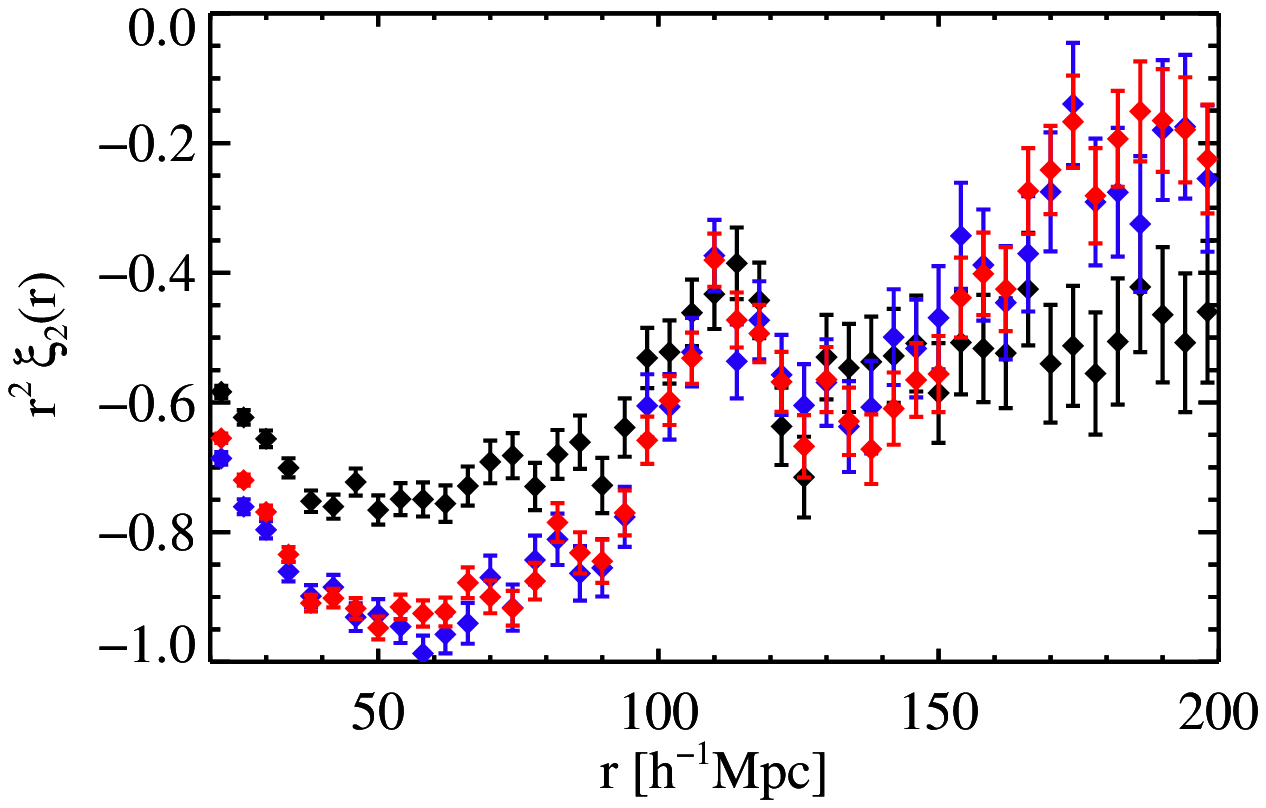}
\caption{
The effect of the continuum estimation procedure on the
correlation function found with the 
mock spectra.
The black dots are the
average of monopole and quadrupole obtained with the 15 sets using the 
exact continua.  The blue (red) dots
show those obtained with the continuum estimation
of  method 1 (method 2) as described in 
section \ref{contfitsec1}  (\ref{contfitsec2}).
}
\label{fig:m1m2}
\end{figure}

\begin{figure}[ht]
\includegraphics[width=\columnwidth]{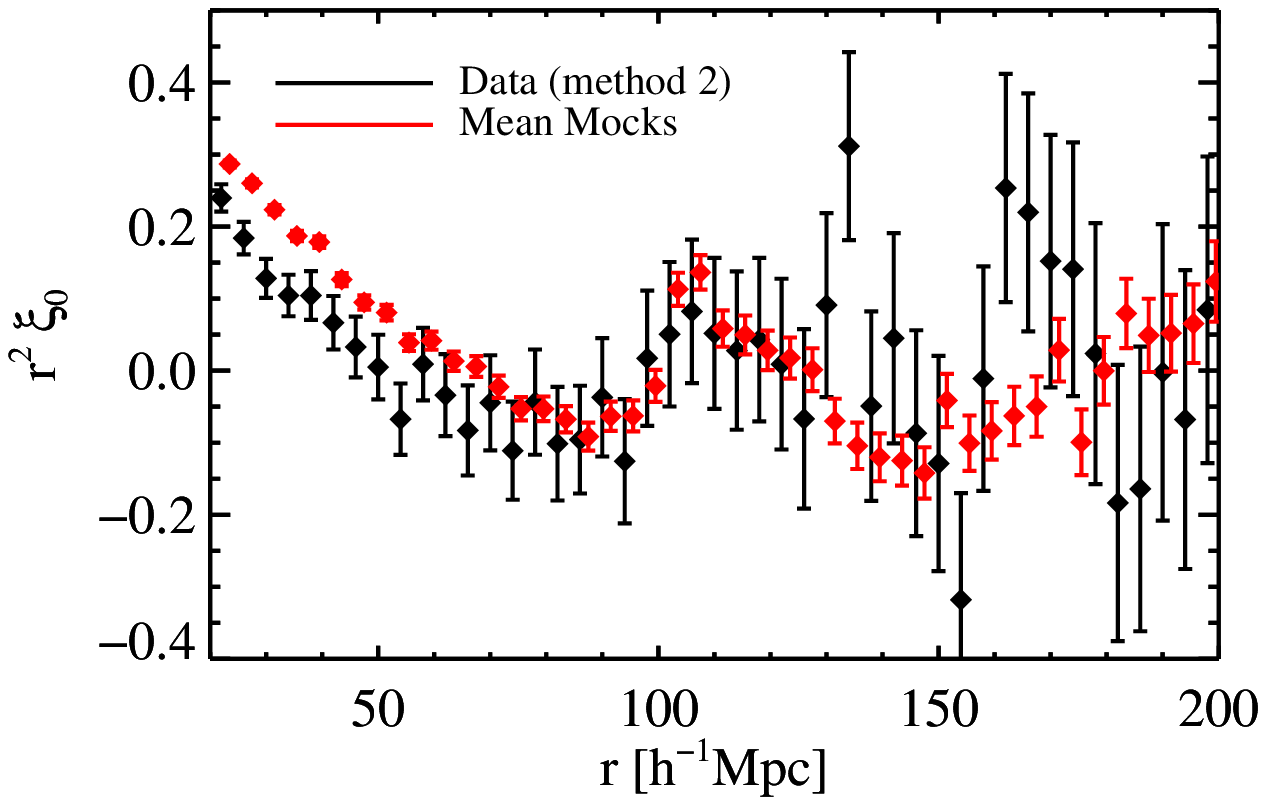}
\includegraphics[width=\columnwidth]{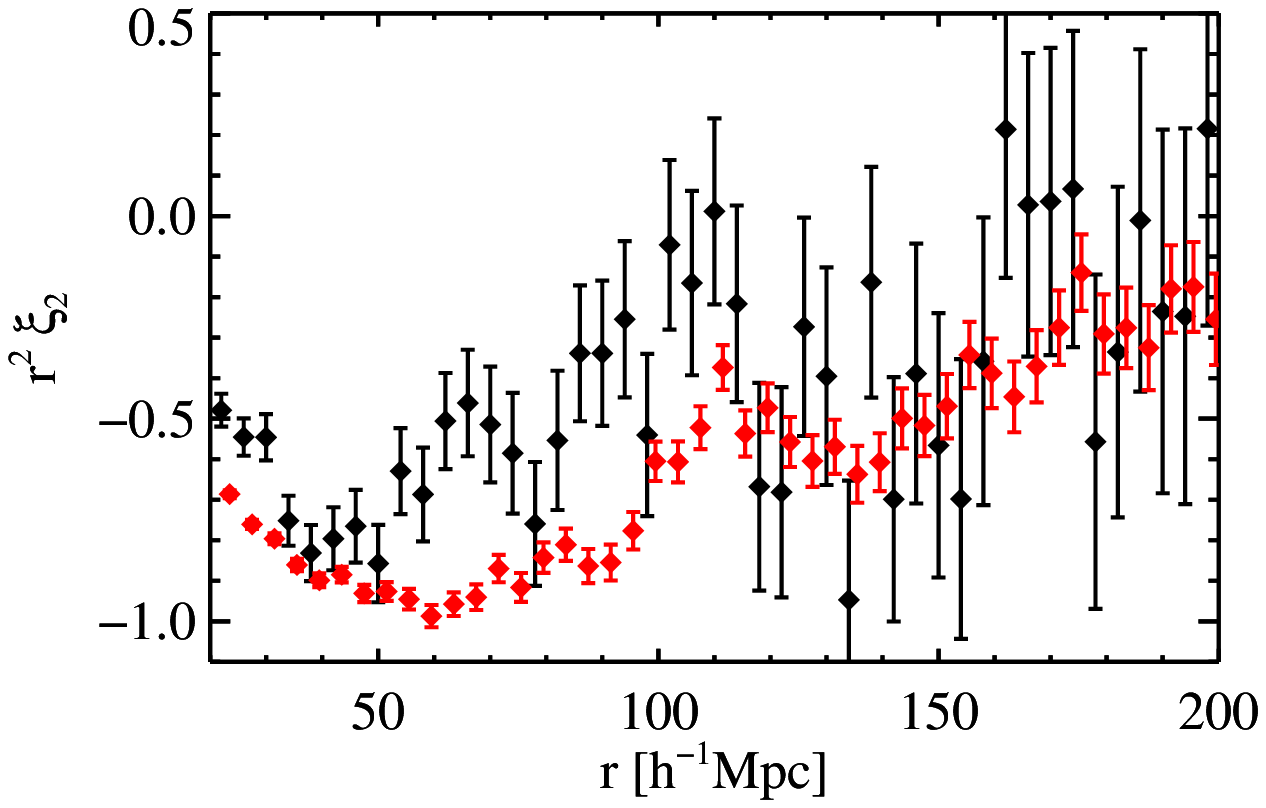}
\caption{
Comparison of the correlation function for the mock spectra and
that for the data.
The red dots show the mean of the 15 sets of mock spectra and the
black dots show the data.
}
\label{fig:monoquad}
\end{figure}

We used the mock spectra to understand how our analysis procedure
modifies the correlation function.
Figure \ref{fig:m1m2}  shows the average over 15 mocks of
the reconstructed quadrupole
and monopole using methods 1 and 2 
(sections \ref{contfitsec1} and \ref{contfitsec2}) 
and that reconstructed with 
the true continuum.
The monopole and quadrupole for the two methods have a general
shape that follows that found with the true continuum including
the position of the BAO peak.
However,
both methods produce a monopole that becomes negative 
for $60~\hMpc<r<100~\hMpc$ while the true monopole remains
positive for all $r<130~\hMpc$.
As discussed in section \ref{contfitsec2},
this result is due to the continuum estimation of the two methods
which introduced negative correlations.
For both methods, however, the BAO peak remains visible with
a deviation above  the ``broadband'' correlation function  that
is hardly affected by the distortion.

Figure \ref{fig:monoquad} 
presents
$\xi_0(r)$ and $\xi_2(r)$ found with the data, along with the mean
of 15 mocks.  
The figure demonstrates that 
our mocks do not perfectly reproduce
the data.
In particular, for $r<80~\hMpc$, the monopole is underestimated and
the quadrupole overestimated.
Since we use only peak positions to extract
cosmological constraints, we
only use the mocks qualitatively
to search for  possible systematic
problems in extracting the peak position.

\section{Results for a Fiducial BOSS \Lya Forest Sample}
\label{kgsample}

\begin{figure}[ht]
\includegraphics[width=\columnwidth]{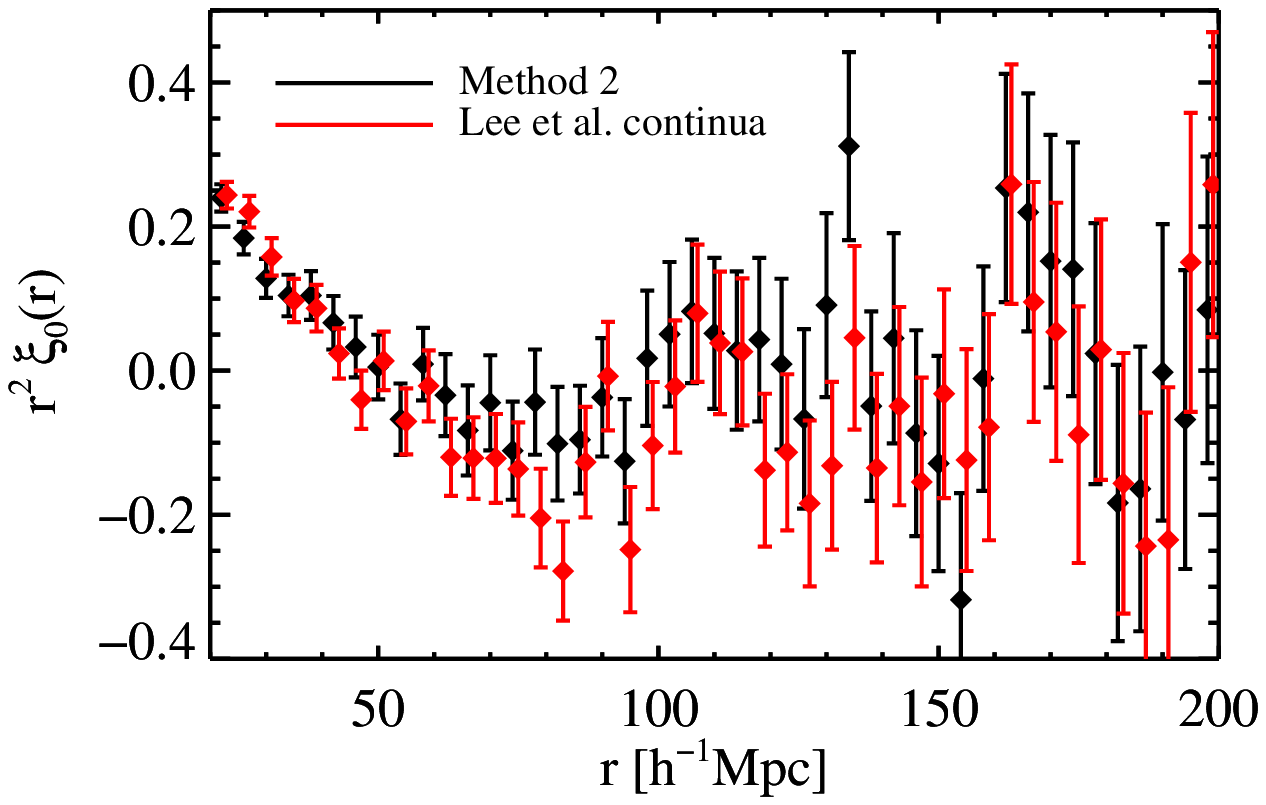}
\includegraphics[width=\columnwidth]{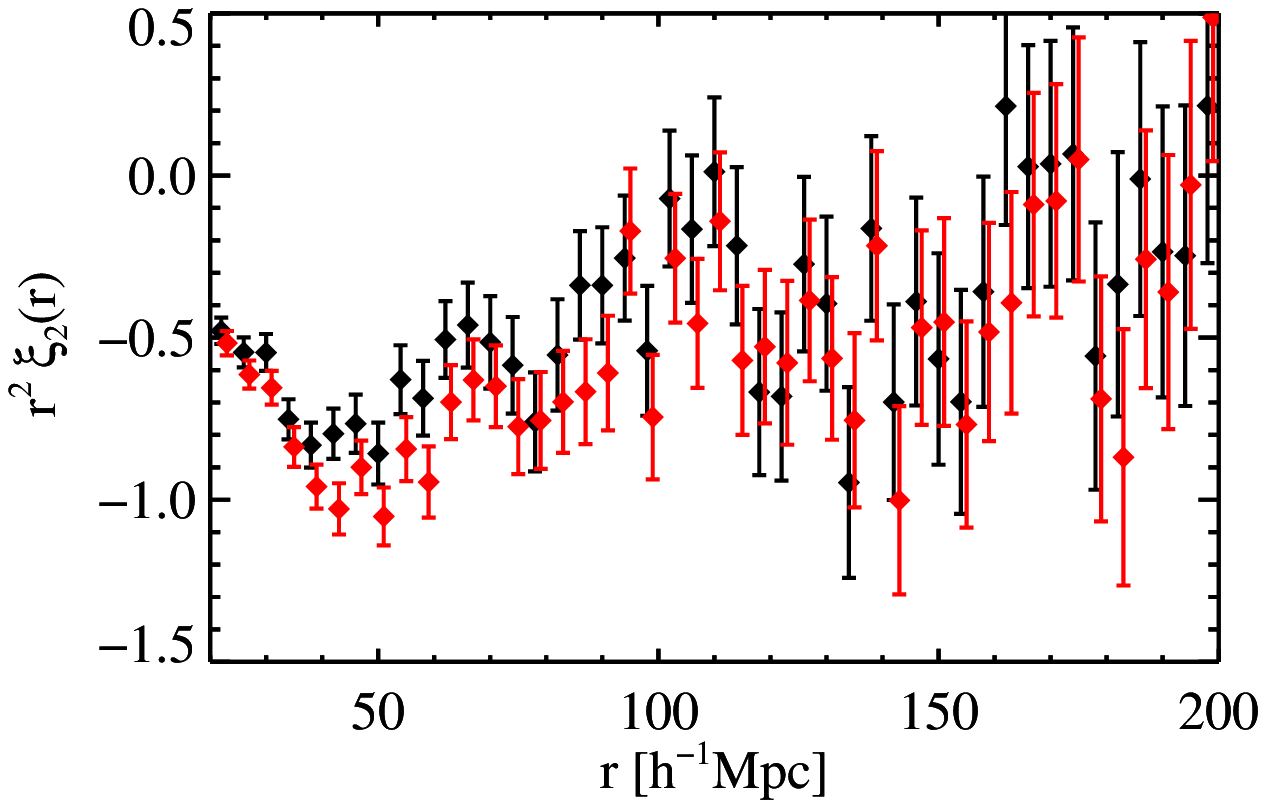}
\caption{
Comparison of the monopole and quadrupole correlation functions
for the sample used here (black dots) and
for the sample and continua of \citet{lee13} (red dots).
}
\label{KGfig}
\end{figure}

The spectra analyzed here are all available through SDSS DR9
\citep{dr9ref}, and the DR9 quasar catalog is described by
\cite{dr9qso}.  A \Lya forest analysis requires many detailed
choices about data selection and continuum determination.
To aid community analyses and comparison of results from different groups, 
\citet{lee13} has presented
a fiducial BOSS \Lya forest sample that uses constrained PCA
continuum determination \citep{lee12} and reasonable
choices of masks for DLAs, BALs, and data reduction artifacts.
Both the data selection and the continuum determination differ 
from those used here.  Figure \ref{KGfig} compares the Method 2 correlation
function from this paper's analysis to that obtained by applying
the Method 2 weights and correlation measurement code directly
to the continuum-normalized spectra of the fiducial Lee et al. sample.
The good agreement in this figure, together with the good agreement
between our Method 1 and Method 2 results, demonstrates the
robustness of the BAO measurement, and the more general correlation
function measurement, to the BOSS DR9 data.

\begin{thebibliography}{}

\bibitem[Abazajian et al.(2009)]{abazajian09}Abazajian K.N., J.K. Adelman-McCarthy, M.S. Ag\"ueros et al. 2009, \apjs, 182, 543

\bibitem[Ahn et al.(2012)]{dr9ref} Ahn, C.P., R. Alexandroff, C. Allende Prieto  et al. 2012, \apjs, 203, 21
% arXiv:1207.7137

\bibitem[Alcock \& Paczynski(1979)]{apeffect}Alcock, C. \& B. Paczynski 1979, Nature, 281, 358

\bibitem[Amanullah et al.(2010)]{union2ref}Amanullah, R. et al. 2010, \apj, 716, 712

\bibitem[Anderson et al.(2012)]{bossgal}Anderson, L., E. Aubourg, S. Bailey et al. 2012, \mnras, 427, 3435
%arXiv:1203:6594 

\bibitem[Bailey et al.(in preparation)]{expandedmocks}Bailey, S. et al. 2012, in preparation

\bibitem[Becker et al.(1995)]{firstref}Becker, R.H., R.L. White \& D.J. Helfand 1995, \apj, 450, 559B

\bibitem[Beutler et al.(2011)]{6df}Beutler F. et al. 2011, \mnras, 416, 3017

\bibitem[Bi et al.(1992)]{bi92}Bi, H.G., G. B\"orner \& Y.Chu 1992, \aap, 266, 1

\bibitem[Blake et al.(2011a)]{wigglez}Blake C. et al. 2011, \mnras, 415, 2892

\bibitem[Blake et al.(2011b)]{wigglezsn}Blake C. et al. 2011, \mnras, 418, 1725

\bibitem[Blake et al.(2012)]{wigglezhz}Blake C. et al. 2012, \mnras, 425, 405
 

\bibitem[Blanton et al.(2003)]{Blanton:2001yk}Blanton, M.R., H.~Lin, R.~H.~Luptonn et al. 2003, \aj, 125, 2276

\bibitem[Bolton et al.(2012)]{pipeline}Bolton A. et al. 2012, \aj, 144, 144

\bibitem[Bovy et al.(2011)]{coreref}Bovy, J. et al. 2011, \apj, 729, 141
\bibitem[Bovy et al.(2012)]{bovy2012}Bovy, J. et al. 2012, \apj, 749, 41

\bibitem[Cabr\'e \& Gazta\~naga(2011)]{cabre11}Cabr\'e, A. \& E. Gazta\`naga 2011, \mnras, 412, L98

\bibitem[Caucci et al.(2008)]{caucci08}Caucci, S., S. Colombi, C. Pichon et al. 2008, \mnras, 386, 211

\bibitem[Cen et al.(1994)]{cen94}Cen R., Miralda-Escud\'e J., Ostriker J. P., Rauch M. 1994, \apj, 437, L9


\bibitem[Chuang \& Wang(2012)]{chuang2012}Chuang, Chia-Hsun \& Yun Wang 2012, \mnras, 426, 226

\bibitem[Cole et al.(2005)]{2dfbao}Cole, S. et al. 2005, \mnras, 362, 505

\bibitem[Coles \& Jones(1991)]{lognormal}Coles, P. \& B. Jones. 1991, \mnras,248, 1

\bibitem[Croft et al.(1997)]{croft97}Croft, R. et al. 1997, \apj, 488, 532


\bibitem[Croft et al.(1998)]{croft98}Croft, R. et al. 1998, \apj, 495, 44


\bibitem[Croft et al.(1999)]{croft99}Croft, R.A.C., et al. 1999, \apj, 520, 1 

\bibitem[Croft et al.(2002)]{croft02}Croft, R.A.C., D. Weinberg,M. Bolte, et al. 2002, \apj, 581, 20 

\bibitem[Dawson et al.(2013)]{bossoverview} Dawson, K., D. Schlegel, C. Ahn  et al. 2013, \aj, 145, 10
% arXiv:1208.0022



\bibitem[de Bernardis et al.(2000)]{boomerang}de Bernardis, P., P. A. R. Ade, J. J. Bock, et al. 2000, \nat, 404, 955 


\bibitem[Eisenstein et al.(2005)]{sdss1bao}  Eisenstein D. J., Zehavi, I., Hogg, D.W. et al. 2005,  \apj, 633, 560

\bibitem[Eisenstein et al. (2007)]{eisenstein07}Eisenstein, D.J. 2007, \apj, 664, 675

\bibitem[Eisenstein et al.(2011)]{Eisenstein:2011sa} Eisenstein, D.J. et al. 2011, \aj, 142, 72


\bibitem[e.g. Efron \& Gong(1983)]{bootstrap}Efron, B. \& G. Gong 1983, The American Statistician, 37, 36
  

\bibitem[Font-Ribera et al.(2012a)]{mockref}Font-Ribera, A., P. McDonald \& J. Miralda Escud\'e 2012, \jcap, 01, 001

\bibitem[Font-Ribera et al.(2012b)]{dlasmocks} Font-Ribera, A. \& Miralda-Escudé, J. 2012, \jcap, 07, 028

\bibitem[Fukugita et al.(1996)]{fukugita96} Fukugita, M., Ichikawa, T., Gunn, J.~E., et al. 1996, \aj, 111, 1748

\bibitem[Gallerani et al.(2011)]{gallerani11}Gallerani, S., F. S. Kitaura \& A. Ferrara 2011, \mnras, 413L, 6

\bibitem[Gazta\~{n}aga et al.(2009)]{gaztanaga09}Gazta\~{n}aga, E., A. Cabr\'e \& L. Hui(2009), \mnras, 399, 1663

\bibitem[Gnedin \& Hui(1996)]{gnedinhui96} Gnedin, N. \& L. Hui 1996, \apj, 472 L73 

\bibitem[Greig et al.(2011)]{greig}Greig, B., J. Bolton, J. Wyithe \& B. Stuart 2011, \mnras, 418, 1980

\bibitem[Gunn et al.(1998)]{gunn98}Gunn, J.E., et al. 1998, \aj, 116, 3040 
%(SDSS Camera)

\bibitem[Gunn et al.(2006)]{gunn06}Gunn, J.E., et al. 2006, \aj, 131, 2332 
%(SDSS Telescope)



\bibitem[Hamilton(1992)]{hamilton}Hamilton, A.J.S. 1992, \apj, 385L, 5

\bibitem[Hernquist et al.(1996)]{hernquist96}Hernquist L., Katz N., Weinberg D. H.,  Miralda-Escud\'e J. 1996, \apj, 457, L51

\bibitem[Kazin et al.(2010)]{kazin10}Kazin, E. A., M. R. Blanton, R. Scoccimarro et al. 2010, \apj, 719, 1032

\bibitem[Kirkpatrick et al.(2011)]{Kirkpatrick:2011gq} Kirkpatrick,J.A., D.~J.~Schlegel, N.~P.~Ross, et al., 2011, \aj, 743, 125 

\bibitem[Kitaura et al.(2012)]{kitaura12}Kitaura, F.-S., S. Gallerani \& A. Ferrara 2012, \mnras, 420, 61

\bibitem[Komatsu et al.(2011)]{wmap7ref}Komaatsu, E. et al. 2011, \apjs, 192, 18

\bibitem[Lawrence et al.(2007)]{ukidssref}Lawrence, A., S.J. Warren, O.Almaini et al. 2007, \mnras, 379, 1599

\bibitem[Le Goff et al.(2011)]{legoff}Le Goff, J. M. et al. 2011, \aap, 534, 135

\bibitem[Lee et al.(2012a)]{lee12}
{Lee}, K.-G., {Suzuki}, N., \& {Spergel}, D.~N. 2012a, \aj, 143, 51

\bibitem[Lee et al.(2012b)]{lee13} Lee, K.-G., Bailey, S., 
Bartsch, L.~E., et al.\ 2012b, arXiv:1211.5146 

\bibitem[Martin et al.(2005)]{galexref}Martin, D.C., J. Fanson, D. Schiminovich et al. 2005, \apj, 619, L1

\bibitem[McDonald et al.(2000)]{macdo2000}McDonald, P. et al. 2000, \apj, 543, 1

\bibitem[McDonald (2003)]{macdo2003}McDonald, P. 2003, \apj, 585, 34

\bibitem[McDonald et al.(2006)]{macdo2006}McDonald, P., U. Seljak, D.J. Schlegel et al. 2006, \apjs, 163, 80

\bibitem[McDonald \& Eisenstein(2007)]{McDonald:2006qs} McDonald, P. \& D.~Eisenstein 2007, \prd, 76, 063009 

\bibitem[McQuinn \& White(2011)]{mcquinnwhiteref}McQuinn, M. \& M. White 2011, \mnras, 415, 2257

\bibitem[Mehta et al.(2012)]{mehtaref}Mehta, K.T., A. Cuesta, X. Xu et al. 2012, \mnras, 427, 2168
% arXiv:1202.0092

\bibitem[Miralda-Escud\'e et al.(1996)]{miralda96}Miralda-Escud\'e J., Cen R., Ostriker J. P., Rauch M. 1996, \apj, 471, 582

\bibitem[Miralda-Escud\'e(2009)]{me09}Miralda-Escud\'e J. (2009), arXiv:0901.1219


\bibitem[Moresco et al.(2012)]{moresco12}Moresco, M., A. Cimatti, R.Jimenez et al. 2012, \jcap, 006, 1208

\bibitem[Noterdaeme et al.(2012)]{noterdaeme}Noterdaeme, P., P. Petitjean, W.C. Carithers et al. 2012, \aap, 547, L1
  
\bibitem[Nusser \& Haehnelt(1999)]{nusser99}Nusser, A. \& M. Haehnelt 1999, \mnras, 303 179

\bibitem[Padmanabhan et al.(2008)]{padmanabhan08}Padmanabhan, N., Schlegel, D.~J., Finkbeiner, D.~P., et al. 2008, \apj, 674, 1217

\bibitem[Padmanabhan et al.(2009)]{padmanabhan09}Padmanabhan, N. et al. 2009, \prd, 79, 3523

\bibitem[Padmanabhan et al.(2012)]{padmanabhan12}Padmanabhan, N., X. Xu, D.J. Eisenstein et al. 2012, \mnras, 427, 2132
% arXiv:1202.0090

\bibitem[Palanque-Delabrouille et al.(2011)]{PalanqueDelabrouille:2010wi}Palanque-Delabrouille, N., C.~.Y\`eche, A.~D.~Myers  et al. 2011, \aap, 530, 122

\bibitem[P\^aris et al.(2011)]{paris2011}P\^aris, I., Petitjean, P., Rollinde, et al. 2011, \aap, 530, 50


\bibitem[P\^aris et al.(2012)]{dr9qso}P\^aris, I., P. Petitjean, E. Aubourg et al. 2012, \aap, 548, 66
%arXiv:1210.5166

\bibitem[Percival et al.(2010)]{percival2010} Percival, W.J., B.A. Reid, D.J.Eisenstein et al. 2010, \mnras, 401, 2148


\bibitem[Petitjean et al.(1995)]{petitjean95}Petitjean P., M\"ucket J. P.,  Kates R. E.  1995, \aap, 295, L9

\bibitem[Pichon et al.(2001)]{pichon01}Pichon, C., J. L. Vergely, E. Rollinde et al. 2001, \mnras, 326, 597

\bibitem[Pier et al.(2003)]{pier03} Pier, J.~R., Munn, J.~A., Hindsley, R.~B., et al. 2003, \aj, 125, 1559

\bibitem[Pieri et al.(2010)]{metalref}Pieri, M. M. et al. 2010, \apj, 724, L69

\bibitem[Reid et al.(2012)]{reid2012}Reid, B.A., L. Samushia, M.White et al. 2012, \mnras, 426, 2719
% arXiv:1203.6641  

\bibitem[Richards et al.(2009)]{richards09}Richards, G.T. et al. 2009, \apjs, 180, 67

\bibitem[Riess et al.(2004)]{riess2004}Riess,A.G., L.-G. Strolger, J. Tonry et al. 2004, \apj, 607, 665

\bibitem[Riess et al.(2007)]{riesshofz}Riess,A.G., L.-G. Strolger, S. Casertano et al. 2007, \apj, 659, 98

\bibitem[Riess et al.(2011)]{riess2011}Riess,A.G., L. Macri, S. Casertano et al. 2011, \apj, 730, 119

\bibitem[Ross et al.(2012)]{ross2012}Ross, N.P. et al. 2012, \apjs 199, 3

\bibitem[Seo et al.(2012)]{seo}Seo, H.-J., S. Ho, M. White et al. 2012, \apj, 761, 13 
% arXiv:1201.2172

\bibitem[Slosar et al.(2011)]{Slosar}Slosar, A. et al. 2011, \jcap, 09, 001

\bibitem[Slosar et al.(2013)]{method3}Slosar, A. et al.(2013), arXiv:1301.3459

\bibitem[Smee et al.(2012)]{bossspectrometer}Smee, S., J.E.Gunn, A. Uomoto et al. 2012, arXiv:1208.2233

\bibitem[Smith et al.(2002)]{smith02} Smith, J.~A., Tucker, D.~L., Kent, S., et al. 2002, \aj, 123, 2121


\bibitem[Stern et al.(2010)]{stellarclocks}Stern, D.. et al. 2010, \jcap,  02, 008

\bibitem[Stoughton et al.(2002)]{stoughton02} Stoughton, C., Lupton, R.~H., Bernardi, M., et al. 2002, \aj, 123, 485 


\bibitem[Suzuki et al.(2005)]{qsopca}Suzuki, N., D. Tytler, D.Kirkman et al. 2005, \apj, 618, 592

\bibitem[Suzuki(2006)]{suzuki06}Suzuki, N. 2006, \apjs, 163, 110 

\bibitem[Theuns et al.(1998)]{theuns98}Theuns T., Leonard A., Efstathiou, G.,  Pearce F.R.,  Thomas, P.A. 1998, \mnras, 301, 478

\bibitem[Viel et al.(2004)]{viel04}Viel, M., M.G. Haehnelt \& V. Springel 2004, \mnras, 354, 684

\bibitem[Weinberg et al.(1998)]{weinberg98}Weinberg, D.H. et al. 1998, ASPC, 148, 21
%http://adsabs.harvard.edu/abs/1998ASPC..148...21W

\bibitem[Weinberg et al.(2012)]{baotheory}Weinberg, D.H., J. J. Mortonson, D. J. Eisenstein et al. 2012, arXiv:1201.2434


\bibitem[White(2003)]{white2003}White, M. 2003, The Davis Meeting On Cosmic Inflation. 2003 March 22-25, Davis CA., p.18
%http://adsabs.harvard.edu/abs/2003dmci.confE..18W

\bibitem[Worseck \& Wisotzki(2006)]{fluxflucref}Worseck, G. \& L. Wisotzki 2006, \aap, 450, 495; A. Lidz et al. (2010) \apj 718, 199


\bibitem[Xu et al.(2012)]{xu2012}Xu, X., Cuesta, A.J., N. Padmanabhan et al. 2012, arXiv:1206.6732


  
\bibitem[Yeche et al.(2009)]{Yeche:2009yh}Y\`eche, C.,  P.~Petitjean, J.~Rich,  et al. 2009, \aap, 523, A14

\bibitem[Yoo \& Miralda-Escud\'e(2010)]{yoome10}Yoo, J. \& J. Miralda-Escud\'e 2010, \prd, 82, 3527
	

\bibitem[York et al.(2000)]{York:2000gk}York, D. G. {\it et al.}  (SDSS Collaboration) 2000, \aj, 120, 1579


\bibitem[Zhang et al.(1995)]{yu95}Zhang, Y., Anninos P.,  Norman M. L. 1995, \apj,  453, L57




  






\end{thebibliography}
\end{document}